\documentclass[numberedappendix, appendixfloats]{emulateapj}
\usepackage{hyperref}
\usepackage{longtable, graphicx, subfigure, epsfig, natbib}
\usepackage{threeparttablex}
\usepackage[fleqn]{amsmath}
\usepackage{lmodern}
\usepackage[T1]{fontenc}
\usepackage{url}
\usepackage{soul}
\usepackage{todonotes}
\newcounter{column_number}
\newcommand{\Msun}{\ifmmode {M_{\odot}}\else{M$_{\odot}$}\fi}
\newcommand{\Rsun}{\ifmmode {R_{\odot}}\else${R_{\odot}}$\fi}
\newcommand{\lapprox }{{\lower0.8ex\hbox{$\buildrel <\over\sim$}}}
\newcommand{\gapprox }{{\lower0.8ex\hbox{$\buildrel >\over\sim$}}}

\def\asec{\ifmmode^{\prime\prime}\else$^{\prime\prime}$\fi}

\def\Ha{$\mathrm{H}\alpha$}

\def\Prot{$P_{\mathrm{rot}}$}
\def\Pmem{$P_{\mathrm{mem}}$}
\def\amin{\ifmmode^{\prime}\else$^{\prime}$\fi}
\def\asec{\ifmmode^{\prime\prime}\else$^{\prime\prime}$\fi}

\usepackage{aas_macros}
\usepackage{natbib}
\slugcomment{Accepted to {\it ApJ} April 5, 2018}
\shorttitle{NGC 752: Membership, Rotation, and Activity}
\shortauthors{Ag\"ueros et al.}

\begin{document}

\title{A New Look at an Old Cluster: The Membership, Rotation, and Magnetic Activity of Low-Mass Stars in the 1.3-Gyr-Old Open Cluster NGC 752}

\author{M.~A.~Ag{\"u}eros\altaffilmark{1}, 
E.~C.~Bowsher\altaffilmark{1},
J.~J.~Bochanski\altaffilmark{2},
P.~A.~Cargile\altaffilmark{3},
K.~R.~Covey\altaffilmark{4},
S.~T.~Douglas\altaffilmark{1},
A.~Kraus\altaffilmark{5},
A.~Kundert\altaffilmark{6},
N.~M.~Law\altaffilmark{7},
A.~Ahmadi\altaffilmark{8},
H.~G.~Arce\altaffilmark{9}} 

\altaffiltext{1}{Department of Astronomy, Columbia University, 550 West 120th Street, New York, NY 10027, USA; \email{marcel@atro.columbia.edu}}
\altaffiltext{2}{Department of Chemistry, Biochemistry, and Physics, Rider University, 2083 Lawrenceville Road,
Lawrenceville, NJ 08648, USA}
\altaffiltext{3}{Harvard-Smithsonian Center for Astrophysics, 60 Garden Street, Cambridge, MA 02138, USA}
\altaffiltext{4}{Department of Physics and Astronomy, Western Washington University, Bellingham, WA 98225, USA }
\altaffiltext{5}{Department of Astronomy, University of Texas at Austin, 2515 Speedway, Stop C1400, Austin, TX 78712, USA}
\altaffiltext{6}{Department of Astronomy, University of Wisconsin-Madison, Madison, WI 53706, USA}
\altaffiltext{7}{Department of Physics and Astronomy, University of North Carolina, Chapel Hill, NC 27599, USA}
\altaffiltext{8}{University of Calgary, 2500 University Dr.~NW, Calgary, Alberta T2N 1N4, Canada}
\altaffiltext{9}{Department of Astronomy, Yale University, New Haven, CT 06520, USA}

\begin{abstract}  
The nearby open cluster NGC 752 presents a rare opportunity to study stellar properties at ages $>$1~Gyr. However, constructing a membership catalog for it is challenging; most surveys have been limited to identifying its giants and dwarf members earlier than mid-K. We supplement past membership catalogs with candidates selected with updated photometric and proper-motion criteria, generating a list of 258 members, a $>$50\% increase over previous catalogs. Using a Bayesian framework to fit MESA Isochrones \& Stellar Tracks evolutionary models to literature photometry and the Tycho-\textit{Gaia} Astrometric Solution data available for 59 cluster members, we infer the age of, and distance to, NGC 752: $1.34\pm0.06$~Gyr and $438_{-6}^{+8}$~pc. We also report the results of our optical monitoring of the cluster using the Palomar Transient Factory. We obtain rotation periods for 12 K and M cluster members, the first periods measured for such low-mass stars with a well-constrained age $>$1 Gyr. We compare these new periods to data from the younger clusters Praesepe and NGC 6811, and to a theoretical model for angular-momentum loss, to examine stellar spin down for low-mass stars over their first 1.3 Gyr. While on average NGC 752 stars are rotating more slowly than their younger counterparts, the difference is not significant. Finally, we use our spectroscopic observations to measure \Ha\ for cluster stars, finding that members earlier than $\approx$M2 are magnetically inactive, as expected at this age. Forthcoming {\it Gaia} data should solidify and extend the membership of NGC 752 to lower masses, thereby increasing its importance for studies of low-mass stars.
\end{abstract}


\section{Introduction}\label{intro}
A star's age is one of its most fundamental parameters. It is also, for low-mass,  main-sequence field stars, notoriously difficult to measure accurately. Over the past decade, a number of authors have proposed age-rotation and age-magnetic activity relations as tools for determining ages for $\lapprox$1~\Msun\ stars \citep[e.g.,][]{mamajek2008, barnes2010, reiners2012, Matt2015}. While measurements for small samples of solar-type stars with precise, $>$1 Gyr ages derived from isochrone fits \citep{Meibom2011,Meibom2015} or asteroseismology \citep{Angus2015,vanSaders2016} exist, by and large these relations for lower-mass stars have been calibrated using observations of the coeval, $<<$1-Gyr-old populations in nearby open clusters \citep[e.g., Praesepe, the Hyades, and the Pleiades;][]{agueros11, Douglas2014, covey16}.

The failure of {\it Kepler}'s second reaction wheel and the mission's rebirth as {\it K2} \citep{howell2014} was an opportunity to measure new rotation periods (\Prot) for members of open clusters along the ecliptic. The result has been a notable increase in our understanding of the rotational behavior of $\lapprox$1~\Msun\ stars in the linchpin clusters listed above \citep{Douglas2016, Rebull2016a,Rebull2016b,stauffer2016, Douglas2017}. 

Unfortunately, with the exception of the $\approx$3-Gyr-old Ruprecht 147 \citep[][]{Curtis2013}, a target of {\it K2}'s Campaign 7, none of the clusters surveyed by {\it Kepler} or {\it K2} is sufficiently old and close to enable the \Prot\ measurements needed to extend our understanding of the rotational evolution of low-mass stars to ages $>$1~Gyr. Targeted studies of older clusters remain critical for understanding the nature and evolution of low-mass stars. 

NGC 752 (01$^{\rm h}$58$^{\rm m}$, $+$37$^\circ$ 52$\amin$), discovered by Caroline Herschel in 1783, could become a benchmark cluster for studying stellar rotation and activity at 1-2~Gyr. While nearby for a cluster of its age \citep[$(m - M)_o\approx8$;][]{daniel1994}, NGC 752 has received relatively little attention, in part because of how difficult it has been to establish a high-confidence membership catalog for the cluster. Surveys such as that of \citet{daniel1994} were limited to identifying cluster giants and main-sequence members earlier than mid-K, mostly due to a lack of proper motion (PM) data for fainter stars.

Later-type NGC 752 members can now be identified using all-sky photometric and astrometric surveys. As was demonstrated by \citet[][hereafter KH07]{adam2007}, combining data from e.g., the Sloan Digital Sky Survey \citep[SDSS;][]{york00}, the Two Micron All Sky Survey \citep[2MASS;][]{2mass}, and the third U.S.~Naval Observatory CCD Astrograph Catalog \citep[UCAC3;][]{zacharias2010}, can yield precise PMs with standard deviations $\sigma \approx$ 3~mas~yr$^{-1}$, spectrophotometric distances accurate to within about 10\%, and spectral types (SpTs) accurate to within about 1 subclass. Even for sparse and slow-moving clusters such as Coma Berenices, these surveys reveal the low-mass stellar populations that eluded previous searches.

\begin{deluxetable}{lccc}[!t]
\tablewidth{0pt}
\tabletypesize{\scriptsize}
\tablecaption{Comparison of the Main NGC 752 Membership Catalogs \label{memstats}}
\tablehead{
\colhead{} & \colhead{}  & \colhead{Non-}   & \colhead{SpT}   \\
\colhead{Catalog} &  \colhead{Members}  & \colhead{Members}  & \colhead{Range\tablenotemark{$\ddagger$}}     
}
\startdata
\citet{daniel1994}	& 157\tablenotemark{$\dagger$} 	& 98	 	& \nodata	 \\
\citet{mermilliod1998}	& 17\tablenotemark{$\dagger$} 	& 13	 	& \nodata	 \\ 
This work 		& 258 	& \nodata 	& F0-M4 	 
\enddata
\tablenotetext{$\dagger$}{Includes both probable and possible members.}
\tablenotetext{$\ddagger$}{\citet{daniel1994} and \citet{mermilliod1998} do not provide spectral types for their stars. The \cite{daniel1994} stars are F-K dwarfs and K-type red giants; the \citet{mermilliod1998} stars are all red giants. See Figure~\ref{cmd}.}
\tablecomments{Fourteen of the \citet{mermilliod1998} probable members are also probable members in \cite{daniel1994}. \citet{mermilliod1998} identify an additional two possible members that do not appear in the \citet{daniel1994} catalog, and reclassify one \citet{daniel1994} non-member as a probable member.}
\label{mem_sum}
\end{deluxetable}

In Section~\ref{members}, we summarize previous work on NGC 752's membership before providing an improved and expanded membership catalog for the cluster. We use this new catalog to derive a more accurate age for and distance to the cluster in Section~\ref{props}. In Section~\ref{rot}, we describe our Palomar Transient Factory \citep[PTF;][]{nick2009,rau2009} observations of NGC 752 and use the resulting data to measure \Prot\ for 12 K and M cluster members. In Section~\ref{spec}, we describe our spectroscopic campaign to characterize chromospheric activity in this cluster. We place these results in context in Section~\ref{disc} to constrain the evolution of low-mass stars up to 1.3 Gyr. We conclude in Section~\ref{concl}. 

\begin{figure}[t]
\centerline{\includegraphics[width=.97\columnwidth, angle = 90]{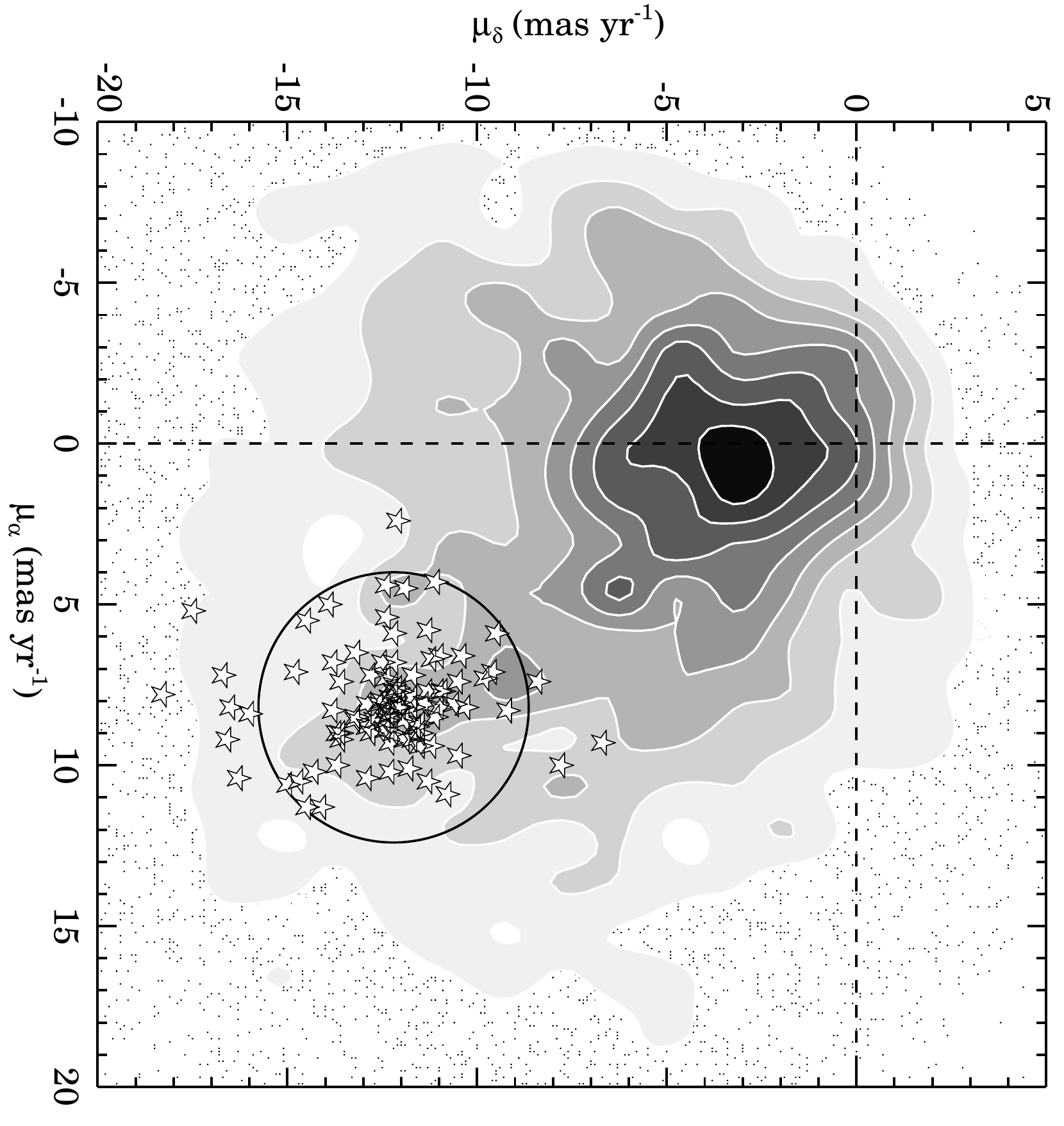}}
\caption{Proper-motion distribution for $\gapprox$$16,000$ stars $<$4$^\circ$ from the center of NGC 752 and with DMs between 6.5 and 8.5~mag. The stars are the 105 probable and possible cluster members identified by \citet{daniel1994} for which we measure proper motions. The circle is the 2$\sigma$ limit for a M0 cluster member.} 
\label{mem_id}
\end{figure}

\section{Consolidating and Expanding NGC 752's Membership}\label{members}
\subsection{Consolidating membership data from the literature}\label{lit}
We began by compiling membership information for NGC 752 from \citet{daniel1994} and \citet{mermilliod1998}. \citet{daniel1994} provided the most comprehensive membership catalog for the cluster, derived from previous PM and radial velocity (RV) studies and new RV measurements. This catalog is divided into three membership levels: probable member, possible member, and probable non-member. A star's membership is determined from its PM; \citet{daniel1994} give the results of \citet{plat1991} priority in the case of conflicts in the literature. The membership status was adjusted if there was strong evidence for non-membership based on the RV measurements made by \citet{psh1988} for 19 stars or by \citet{daniel1994} for 79 stars. The final catalog of 255 stars contains 109 probable members, 48 possible members, and 98 probable non-members. 

\citet{mermilliod1998} conducted an 18-year RV survey of NGC 752's red giants. The resulting catalog of 30 stars includes 15 probable members, two possible members, and 13 non-members. 

There is significant overlap between the catalogs: 22 of the 30 \citet{mermilliod1998} stars are in \citet{daniel1994}. The only significant difference concerns Platais 172, classified as a non-member by \citet{daniel1994} and as a probable member by \citet{mermilliod1998}. We therefore adopt the \cite{daniel1994} catalog as the bedrock of our membership catalog, adding Platais 172 and two possible members identified by \cite{mermilliod1998} that were not studied by \cite{daniel1994}. 

We also had access to RV measurements for 123 candidate cluster members.  These include RVs published by \citet[][]{daniel1994} for 92 stars \citep[including 19 RVs from][]{psh1988}, as well as measurements for 76 stars shared with us by C.~Pilachowski \citep[45 of which also have RVs published in][]{daniel1994}. For each of these 76 stars, $\approx$15 spectra were obtained as part of a long-term monitoring campaign with the Hydra spectrograph on the WIYN 3.5-m telescope, Kitt Peak, AZ.\footnote{The WIYN Observatory is a joint facility of the University of Wisconsin-Madison, Indiana University, Yale University, and the National Optical Astronomy Observatory.} The RVs were derived from spectra of the Mg b triplet (5167, 5173, 5184~\AA) obtained using the bench-mounted spectrograph with the blue fiber cables. To provide the highest possible precision, the same fibers were placed on the same stars for every observation. A subset of, on average, eight non-variable stars with known RVs were used to establish the zero point for each frame. With this approach, it was possible to obtain relative precision from run to run and night to night of 200~m~s$^{-1}$ for an individual star (C.~Pilachowski, pers.~comm.). 

We examined a number of other studies of NGC 752 in order to identify other candidate cluster members. However, these usually relied on the \citet{daniel1994} membership catalog \citep[e.g.,][]{sestito2004, giardino2008,barta2011} and did not include new PM or RV data, so we did not take them into account when making membership determinations.

\subsection{Identifying new candidate members}\label{tech}
Past surveys of NGC 752 found many FGK members, but only small numbers of late-K and M dwarfs. These low-mass members span much of the dynamic range of our PTF observations, and correctly identifying them is critical for interpreting the results of our rotational monitoring program. We therefore used the techniques first described in KH07 to add new candidate, low-mass members to the catalog described above.

Our candidate selection pipeline used astrometric and photometric data from 2MASS and UCAC3.\footnote{This analysis was undertaken before the release of UCAC4 and UCAC5.} Since NGC 752 and most of the surrounding area do not have SDSS coverage, we adapted our spectral-energy-distribution (SED) fitting procedure from KH07 to use USNO-B1.0 photometry \citep{monet2003}; see the Appendix for details and Table~\ref{SEDs} for the SED template magnitudes in the USNO-B1.0 filters. We combined the astrometric measurements to calculate PMs and the photometric measurements to calculate spectrophotometric distances and photometric spectral types (SpTs) for objects within 4$^\circ$ of the cluster center.

For our astrometric analysis, we fitted the absolute positions reported in each catalog with a linear solution in RA and DEC, $\sigma$-clipping at 3$\sigma$ to remove potentially erroneous measurements. For our photometric analysis, we fitted all available photometry against a grid of SED models, where the photometric SpT of the best-fit model was adopted as the object's SpT, and the average difference between the absolute magnitudes of the template and the apparent magnitudes of the object was used to infer the distance modulus (DM) and hence distance. 

After measuring the PMs, SpTs, and DMs, we computed a membership probability \Pmem\ for each object using the methods described by \citet{sanders1971} and \citet{fran1989b}. We first cut our sample to include only objects with DMs between 6.5 and 8.5~mag, corresponding to $\approx$1.5 mag above and 0.5~mag below the mean cluster value of $\approx$8 \citep[][]{daniel1994}.\footnote{This is 0.2-0.3 mag brighter than more recent estimates for the cluster's DM; cf.~\citet{barta2007} and discussion in Section~\ref{props}.} We considered all other objects to be likely field stars and removed them from our catalog. Figure~\ref{mem_id} is the PM diagram for the $\gapprox$16,000 stars that meet this DM requirement, and shows quite clearly the difficulty in separating out NGC 752 members from field stars at this stage in our analysis. 

We divided the remaining objects by the inferred SpT, and fit their distribution on the sky and in the PM diagram with a model comprised of a cluster distribution (distributed as $e^{-r}$ in spatial position and Gaussian in PM, centered on the cluster's mean position and mean PM of (8, $-11$) mas yr$^{-1}$) and a field-star distribution (distributed constantly in position and as a bivariate Gaussian in PM, where the mean and $\sigma$ of the PM distribution was fit independently for each bin). A bivariate Gaussian was chosen for the field PM distribution because the cluster PM is low, and hence the traditional parametrization \citep[e.g.,][]{deacon2004} as an exponential (parallel to the cluster PM vector) and one-dimensional Gaussian (perpendicular to the cluster PM vector) breaks down. The PM diagram for field stars is largely dominated by dwarfs for most bins, but has a significant contribution from background giants for K stars, so a bivariate Gaussian allows the shape to vary between these extremes as needed.

\begin{figure}[th!]
\centerline{\includegraphics[width=1.0\columnwidth]{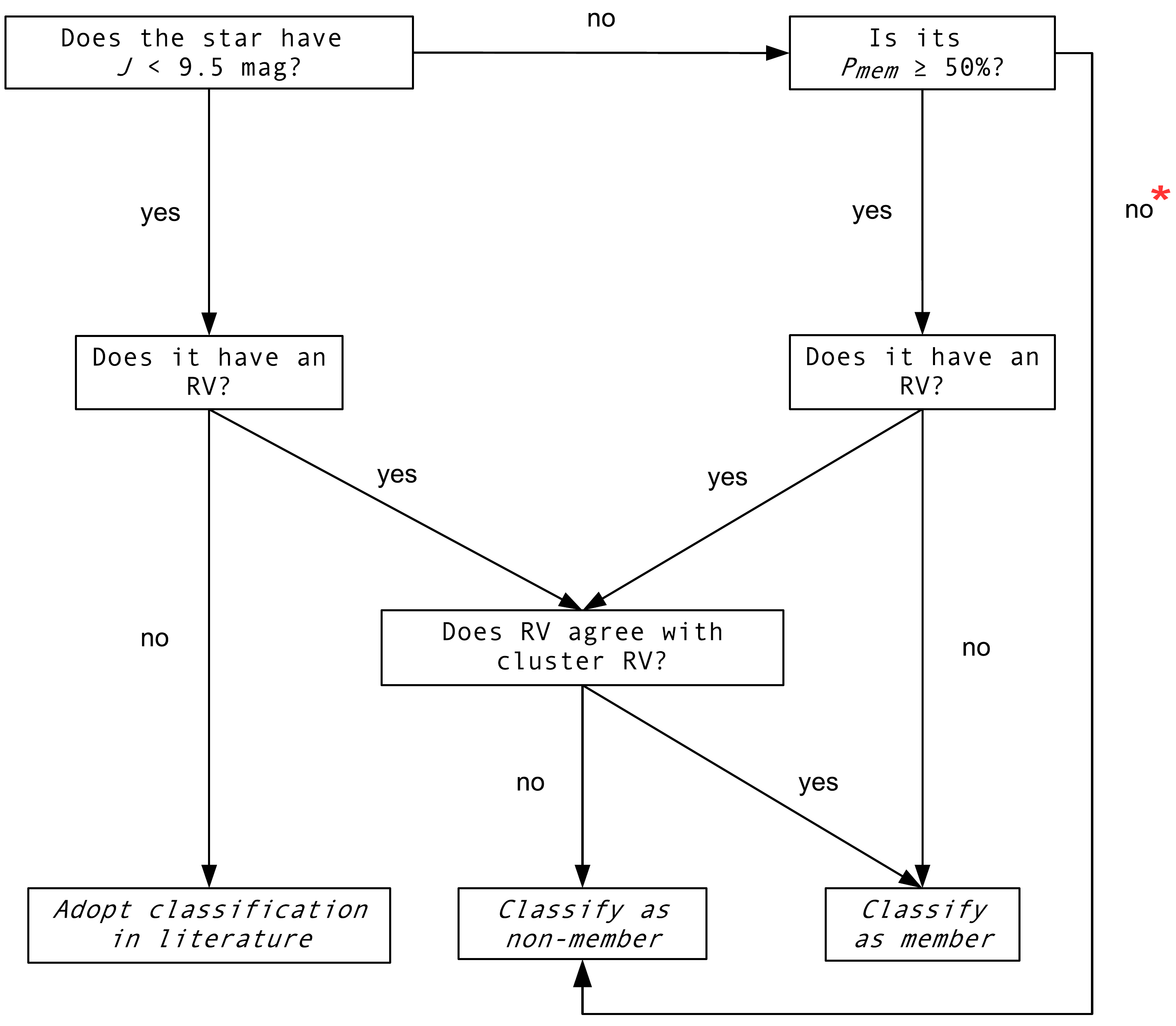}}
\caption{Constructing our membership catalog. The red asterisk on the far right indicates that not all stars with $J > 9.5$~mag and \Pmem~$< 50\%$ were rejected as members. Four with $10\% \leq$~\Pmem~$< 50\%$ and RVs consistent with the cluster's were included in our final catalog.}
\label{memb_flow}
\end{figure}

\begin{deluxetable*}{lcr@{$\pm$}lr@{$.$}lr}[!t] 
\tablewidth{0pt}
\tabletypesize{\scriptsize}
\tablecaption{Stars Identified as Members in the Literature and as Non-Members in This Work \label{demotedstats}}
\tablehead{
\colhead{2MASS ID} &  \colhead{Platais ID}  & \multicolumn{2}{c}{$J$}  & \multicolumn{2}{c}{\Pmem}   & \colhead{RV} \\ 
\colhead{} &  \colhead{}  & \multicolumn{2}{c}{(mag)}  & \multicolumn{2}{c}{(\%)}   &  \colhead{(km s$^{-1}$)} 
}
\startdata
01561395$+$3747048	&	477	&	9.78 & 0.02	&	99 & 8	&	8.17$^{\pm0.29}_{\pm0.43}$  \\ 
01564759$+$3724306	&	619	&	9.41 & 0.02	&	99 & 6	&	10.04$^{\pm1.11}_{\pm0.63}$  \\ 
01565304$+$3752094	&	641	&	9.37 & 0.02	& \multicolumn{2}{c}{\nodata} &	10.49$^{\pm1.37}_{\pm0.64}$ \\ 
01571034$+$3725552	&	722	&	12.18 & 0.02	&	86 & 9	&	$-$31.42$^{\pm0.40}$ \\ 
01571211$+$3759249	&	728	&	8.49 & 0.02	& \multicolumn{2}{c}{\nodata} 	&	9.46$^{\pm0.54}_{\pm0.46}$ \\ 
01572071$+$3751432	&	772	&	9.21 & 0.02	& \multicolumn{2}{c}{\nodata} 	&	9.08$^{\pm1.67}_{\pm0.60}$ \\ 
01573091$+$3754580	&	823	&	9.38 & 0.02	&	99 & 8	&	8.67$^{\pm1.27}_{\pm0.30}$ \\ 
01574395$+$3751421	&	888	&	9.56 & 0.02	&	99 & 9	&	11.28$^{\pm1.07}_{\pm0.55}$ \\ 
01581269$+$3734405	&	1008	&	10.19 & 0.02	&	99 & 8	&	9.84$^{\pm1.94}_{\pm0.50}$ \\ 
01550711$+$3732370	&	245	&	12.84 & 0.02	&	0 & 0	&	16.26$^{\pm0.23}_{\pm0.16}$ \\
01563444$+$3808495	&	563	&	12.02 & 0.02	&	7 & 2	&	$-$2.21$^{\pm0.37}_{\pm0.38}$ \\ 
\hline
01554473$+$3754428	&	361	&	12.44 & 0.02	&	4 & 5	&	\nodata	\\	
01560292$+$3736326	&	429	&	12.74 & 0.02	&	0 & 2	&	\nodata	\\	
01561369$+$3715569	&	475	&	11.78 & 0.02	&	43 & 6  & \nodata	\\	
01562944$+$3755147	&	542	&	13.21 & 0.02	&	0 & 0	&	\nodata	\\	
01565537$+$3804459	&	653	&	11.26 & 0.02	&	0 & 0	&	\nodata	\\	
01565614$+$3808161	&	655	&	11.71 & 0.02	&	0 & 0	&	\nodata	\\	
01570487$+$3807332	&	699	&	11.65 & 0.02	&	0 & 0	&	\nodata	\\	
01571468$+$3754109	&	748	&	12.11 & 0.02	&	0 & 0	&	\nodata	\\	
01572229$+$3736233	&	783	&	11.05 & 0.02	&	4 & 1	&	5.84$^{\pm0.49}_{\pm0.08}$ \\ 
01572297$+$3738215$\dagger$	&	786	&	11.70 & 0.02	&	0 & 1	&	7.57$^{\pm4.93}_{\pm0.95}$ \\ 
01573588$+$3758231	&	847	&	12.39 & 0.02	&	49 & 7	&	\nodata	\\	
01574443$+$3811067	&	889	&	11.65 & 0.02	&	0 & 0	&	\nodata	\\	
01575495$+$3720267	&	937	&	10.16 & 0.02	&	5 & 6	&	5.67$^{\pm0.38}_{\pm0.10}$ \\ 
01581337$+$3811413	&	1007	&	11.74 & 0.02	&	7 & 7	&	5.10$^{\pm0.17}_{\pm0.13}$ \\ 
01582839$+$3809303	&	1082	&	13.65 & 0.02	&	0 & 0	&	\nodata	\\	
01592608$+$3740398	&	1296	&	13.12 & 0.02	&	0 & 0	&	\nodata	\\	
01594206$+$3742114	&	1352	&	13.84 & 0.03	&	0 & 0	&	\nodata	\\	
01595680$+$3758104	&	1407	&	11.78 & 0.03	&	9 & 6	&	\nodata	\\	
01595738$+$3818094	&	1406	&	10.37 & 0.03	&	0 & 0	&	\nodata	\\	
02001767$+$3711032	&	1470	&	11.45 & 0.03	&	35 & 2 &	\nodata	\\	
02014168$+$3749290	&	1690	&	11.95 & 0.03	&	0 & 0	&	\nodata		
\enddata
\tablenotetext{$\dagger$}{Candidate binary}
\tablecomments{``Platais ID'' is the catalog number of the star in \citet{plat1991}. \Pmem\ is presented for all of the stars for which it was calculated, even though for those with $J < 9.5$ mag, its value (or absence) does not impact our membership decision as the classification of \citet{daniel1994} and \citet{mermilliod1998} takes precedence. The RV data are from the WIYN long-term monitoring campaign (C.~Pilachowshi, priv.~comm.). Most stars were observed multiple times on multiple nights, and an average RV and $\sigma$ were computed for each star on each night. The first (top) RV uncertainty corresponds to the $\sigma$ in the RV computed from the set of nightly average RVs (and is used for single stars, for which we define this $\sigma_1 <3$~km s$^{-1}$); the second (bottom) uncertainty is the average of the nightly $\sigma$ (and is used for binaries). The first 11 stars have RVs $>$2$\sigma$ from the cluster value of 5.5$\pm$0.6~km~s$^{-1}$ \citep{daniel1994}. The bottom 21 stars have \Pmem~$< 50\%$. Four of these have RVs consistent with membership but \Pmem~$< 10\%$ and are therefore excluded.}
\label{demotted_stats}
\end{deluxetable*}

\subsection{Producing an updated membership catalog}\label{new}
To assemble a definitive membership catalog, we began by combining the list of members and non-members assembled from the literature and the list of new candidate members constructed above. We then matched the stars in this merged catalog to 2MASS; only one likely member, with \Pmem~$= 89.5\%$, lacks a match because of confusion with a persistence artifact in the 2MASS image. We include it in our final catalog, but this star does not feature in our subsequent analysis.  

Because we rely on data from several photometric catalogs, all of which have a bright limit, we treated $J < 9.5$~mag stars differently than those fainter than this magnitude. Figure~\ref{memb_flow} is a decision tree illustrating how we constructed the membership catalog.

For the 154 stars with $J < 9.5$~mag, we used the information provided by \citet{daniel1994} and \citet{mermilliod1998} to assign initial membership status, identifying 41 probable and possible members and 113 non-members. For stars with $J > 9.5$, we selected the 212 with \Pmem~$\geq 50\%$ as candidate members.\footnote{For the 105 stars that are listed as probable and possible members in the literature and for which we calculated a \Pmem\ (including five that are brighter than $J = 9.5$ mag), the agreement is generally excellent: 87 (83\%) have \Pmem~$\geq 50\%$.} 

We further refined the membership status of the 123 candidate members with RV measurements obtained by \citet{daniel1994} and C.~Pilachowski (priv.~comm). For the 92 \citet{daniel1994} measurements we used the given RV uncertainties in making our comparison to the cluster value.\footnote{Two \citet{daniel1994} stars lack $\sigma$ values; for these we use the average $\sigma$ derived from the other \citet{daniel1994} RVs.} For the RVs that appear only in \citet{psh1988} we used the typical $\sigma$ quoted by these authors of 0.5 km~s$^{-1}$. 

Most of the stars observed as part of the WIYN long-term monitoring campaign were observed multiple times on multiple nights, and an average RV and $\sigma$ were computed for each star on each night. In addition, an uncertainty $\sigma_1$ corresponding to the $\sigma$ in the RV computed from the set of nightly average RVs and another $\sigma_2$ corresponding to the average of the nightly $\sigma$ were calculated. 

For one of these stars to be labeled a member, we required that its mean RV be within 2$\sigma_1$ of the cluster RV of $5.5$$\pm$$0.6$~km~s$^{-1}$ \citep{daniel1994}.\footnote{\citet{mermilliod1998} found   $4.68$$\pm$$0.11$~km~s$^{-1}$. For simplicity, we use the \citet{daniel1994} value for our RV tests.} This requirement resulted in the rejection of a number of stars that had been listed as members in the literature. For example, five stars with $J < 9.5$ mag, and four with $J > 9.5$ and \Pmem~$\geq 50\%$, are listed as members in the combined \citet{daniel1994} and \citet{mermilliod1998} catalog, but have RVs that are inconsistent with membership.

The 17 stars with $\sigma_1 > 3$ km s$^{-1}$ were labeled candidate binaries; 12 of these were identified as candidate binaries by \citet{daniel1994}. For the five new systems, we use $\sigma_2$ to test for the agreement with the cluster RV and classify four as members. The fifth, Platais 786, had been considered a member, but has a \Pmem~$= 0.1\%$, and we removed it from our membership catalog. 

We also checked stars with 10\% $\leq$~\Pmem~$<$ 50\% and RV measurements and identified four whose RVs are consistent with cluster membership. As a result, we added these stars to our final list of cluster members. Stars with RVs consistent with membership but \Pmem~$< 10\%$ were removed from our membership catalog. In addition to Platais 786, there are three stars formerly listed as members that are removed for this reason.  

\begin{deluxetable*}{lcr@{$\pm$}lr@{$\pm$}lcccccc}
\tablewidth{0pt}
\tabletypesize{\tiny}
\tablecaption{\Pmem\ Selected NGC 752 Members \label{tab:adam_members}}
\tablehead{
 \colhead{2MASS ID} &
 \colhead{Platais ID} &
 \multicolumn{2}{c}{$J$} &
 \multicolumn{2}{c}{$K$} &
 \colhead{Mass} &
 \colhead{SpT} &
 \colhead{DM} &
 \colhead{M$_{bol}$} &
 \colhead{Binary?\tablenotemark{a}} &
 \colhead{\Pmem}\\
 \colhead{} & 
 \colhead{} &
 \multicolumn{2}{c}{(mag)} & 
 \multicolumn{2}{c}{(mag)} & 
 \colhead{(\Msun)} & 
 \colhead{} & 
 \colhead{(mag)} & 
 \colhead{(mag)} &
 \colhead{} & 		
 \colhead{(\%)}
}
\startdata
01501676$+$3812369 & \nodata &  9.96 & 0.02 &  9.73 & 0.02 & 1.24 & F3.4 & 7.38$\pm$0.22 & 10.60$\pm$0.10 & \nodata & 54.6 \\
01523927$+$3822334 & \nodata & 10.79 & 0.02 & 10.53 & 0.02 & 1.06 & F7.4 & 7.58$\pm$0.20 & 11.70$\pm$0.08 & \nodata & 54.4 \\
01524348$+$3724497 & \nodata & 10.57 & 0.02 & 10.30 & 0.02 & 1.04 & F8.0 & 7.28$\pm$0.16 & 11.52$\pm$0.05 & \nodata & 93.7 \\
01524372$+$3808381 & \nodata & 11.59 & 0.02 & 11.12 & 0.02 & 0.82 & G9.3 & 7.17$\pm$0.04 & 12.73$\pm$0.05 & \nodata & 52.7 \\
01525891$+$3803515 & \nodata & 12.51 & 0.02 & 11.86 & 0.02 & 0.69 & K4.7 & 7.37$\pm$0.07 & 14.01$\pm$0.02 & \nodata & 52.8 \\
01531903$+$3759057 & \nodata & 11.75 & 0.02 & 11.41 & 0.02 & 0.96 & G3.5 & 8.02$\pm$0.07 & 12.76$\pm$0.04 & \nodata & 51.4 \\
01532120$+$3735162 & \nodata & 10.86 & 0.02 & 10.58 & 0.02 & 1.17 & F4.9 & 8.01$\pm$0.22 & 11.59$\pm$0.10 & \nodata & 97.7 \\
01533728$+$3724173 & \nodata & 11.45 & 0.02 & 11.11 & 0.03 & 0.82 & G8.3 & 7.19$\pm$0.08 & 12.55$\pm$0.11 & \nodata & 61.1 \\
01534317$+$3743224 & \nodata & 12.95 & 0.03 & 12.35 & 0.03 & 0.73 & K2.9 & 8.03$\pm$0.08 & 14.32$\pm$0.07 & \nodata & 57.4 \\
01535762$+$3756556 & \nodata & 14.56 & 0.04 & 13.67 & 0.04 & 0.51 & M1.0 & 8.23$\pm$0.14 & 16.20$\pm$0.04 & \nodata & 73.0 
\enddata
\tablenotetext{a}{Based on RV measurements published by \citet{daniel1994} (``D'') or \citet{mermilliod1998}, or collected by C.~Pilachowski (``P'').}
\tablecomments{The full version of this table is available online.}
\end{deluxetable*}

\begin{deluxetable*}{lccccccccc}
\tablewidth{0pt}
\tabletypesize{\tiny}
\tablecaption{Other NGC 752 Members \label{tab:not_adam_members}}
\tablehead{
 \colhead{2MASS ID} &
 \colhead{Platais ID} &
 \colhead{$J$} &
 \colhead{$K$} &
 \colhead{Mass} &
 \colhead{SpT} &
 \colhead{RV$_D$} &
 \colhead{RV$_P$} &
 \colhead{Binary?\tablenotemark{a}} \\
 \colhead{} & 
 \colhead{} &
 \colhead{(mag)} & 
 \colhead{(mag)} & 
 \colhead{(\Msun)} & 
 \colhead{} & 
 \colhead{(km s$^{-1}$)} & 
 \colhead{(km s$^{-1}$)} &
 \colhead{}
 }
\startdata
01510351$+$3746343 & \nodata & 7.35$\pm$0.02 & 6.70$\pm$0.03 & \nodata &\nodata & \nodata       & \nodata & \nodata \\
01513012$+$3735380 & \nodata & 7.34$\pm$0.02 & 6.76$\pm$0.02 & \nodata &\nodata & \nodata       & \nodata & \nodata \\
01543966$+$3811455 & \nodata & 7.13$\pm$0.01 & 6.52$\pm$0.02 & \nodata &\nodata & \nodata       & \nodata & \nodata \\
01551261$+$3750145 & \nodata & 7.82$\pm$0.02 & 7.22$\pm$0.02 & \nodata &\nodata & \nodata       & \nodata & \nodata \\
01551528$+$3750312 & \nodata & 7.80$\pm$0.02 & 7.19$\pm$0.02 & \nodata &\nodata & 4.50$\pm$0.50 & \nodata & \nodata \\
01552765$+$3759551 & \nodata & 7.62$\pm$0.02 & 7.03$\pm$0.02 & \nodata &\nodata & 4.90$\pm$0.30 & \nodata & \nodata \\
01552769$+$3734046 & 305     & 9.29$\pm$0.01 & 9.06$\pm$0.02 & \nodata & F5.6   & \nodata       & 6.25$^{\pm0.09}_{\pm0.19}$ & \nodata \\
01552928$+$3750262 & 313     & 9.00$\pm$0.02 & 8.69$\pm$0.02 & \nodata &\nodata & 4.60$\pm$0.50 & $-$11.0$^{\pm1.97}_{\pm36.7}$ & DP \\
01553936$+$3752525 & \nodata & 7.15$\pm$0.02 & 6.54$\pm$0.02 & \nodata &\nodata & \nodata       & \nodata & \nodata \\
01554239$+$3737546 & \nodata & 7.40$\pm$0.01 & 6.80$\pm$0.02 & \nodata &\nodata & \nodata       & \nodata & \nodata  
\enddata
\tablenotetext{a}{Based on RV measurements published by \citet{daniel1994} (``D'') or \citet{mermilliod1998}, or collected by C.~Pilachowski (``P'').}
\tablecomments{The full version of this table is available online.}
\end{deluxetable*}

Five stars identified as members in the literature are included in our final membership catalog but not in our  subsequent analysis. Four of these stars, Platais 654, 921, 952, and 1129, have poor SED fits ($\chi ^{2} > 3$) and therefore \Pmem~$< 50\%$. Still, all four fall on the cluster main sequence in $J$ versus $(J-K)$ and have PMs and RVs consistent with the expected values, so they are plausible members. The fifth star, Platais 684, is one of our newly identified candidate binaries: it has a good SED fit, photometry marginally consistent with membership (and therefore a low \Pmem), but RV variability that suggests it is a single-lined spectroscopic binary. Its nature needs be investigated further. 

In Table~\ref{mem_sum} we summarize the properties of our new catalog and compare it to those of \citet{daniel1994} and of \citet{mermilliod1998}. Our work has added 125 new stars to the cluster, reclassified five stars listed as non-members in the literature as cluster members, and extended NGC 752's membership to the mid-M stars. Conversely, we have removed 32 stars, or one-fifth of the merged \citet{daniel1994} and \citet{mermilliod1998} catalog, from the list of cluster members (see Table~\ref{demotedstats}).

\begin{figure}[ht!]
\centerline{\includegraphics[width=\columnwidth]{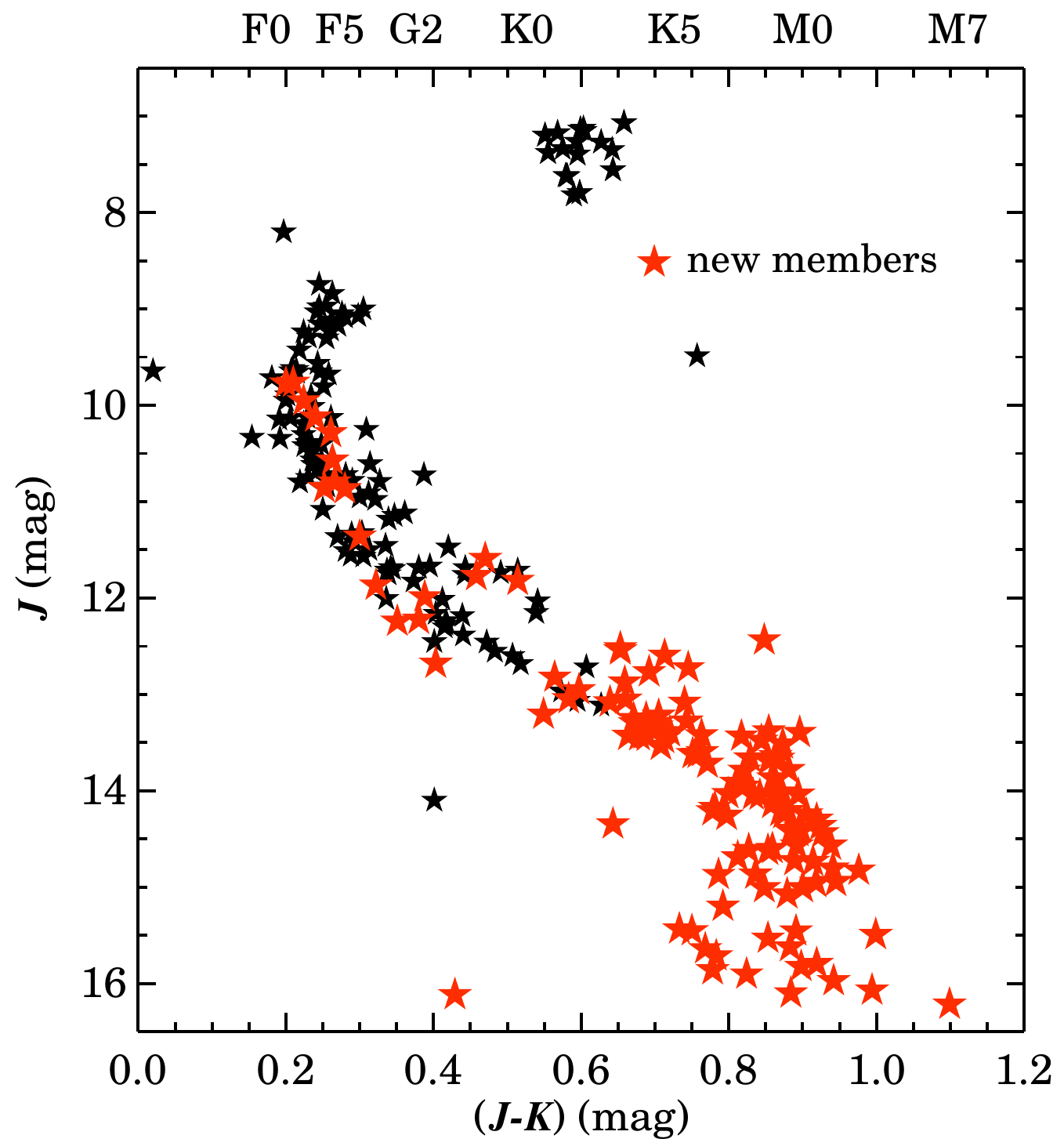}}
\caption{CMD for NGC 752. Members identified in the literature are in black; our new high-confidence members are in red. } 
\label{cmd}
\end{figure}

\begin{figure}[h!]
\centerline{\includegraphics[width=\columnwidth]{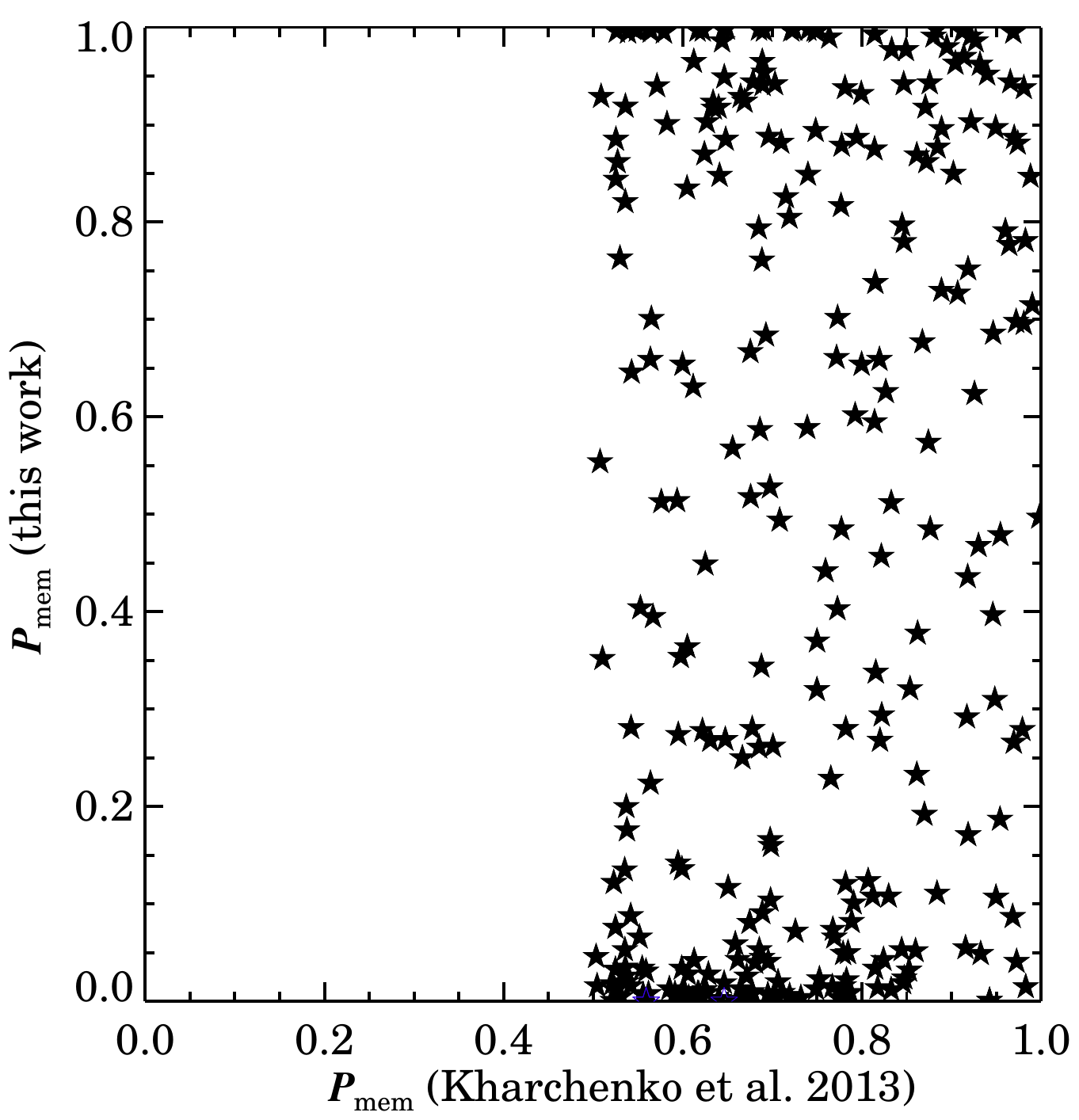}}
\caption{A comparison of membership probabilities calculated by \citet{kharchenko2013} and in this work. While both catalogs assign high \Pmem\ to stars near the cluster core that fall on the cluster's CMD sequence and proper-motion locus, the \citet{kharchenko2013} \Pmem\ calculation is more sensitive to field interlopers within the cluster's tidal radius, resulting in large numbers of non-members with artificially high \Pmem\ being listed in their catalog. } 
\label{memvsmem}
\end{figure}

The 258 cluster members are presented in Tables~\ref{tab:adam_members} and \ref{tab:not_adam_members}. A $J$ versus $(J-K)$ color-magnitude diagram (CMD) for NGC~752 is shown in Figure~\ref{cmd}.

\citet{kharchenko2013} investigated the membership of NGC 752 as part of a large-scale survey of Milky Way star clusters. We compared the \Pmem\ we derived for candidate members to those obtained by \citet{kharchenko2013} for 568 stars in their NGC 752 catalog; this included many stars for which we calculated a \Pmem~$< 50\%$ (i.e., not in our final cluster catalog) to make this comparison more meaningful. Figure~\ref{memvsmem} shows that both catalogs assign mutually high \Pmem\ to many candidates; these are stars near the cluster core that fall on the CMD sequence and proper-motion locus of the cluster. 

However, \citet{kharchenko2013} compute probabilities that capture spatial position with a step function, assigning \Pmem\ = 0\% for all stars outside the tidal radius and otherwise weighing all stars uniformly. Figure~\ref{memvsmem} therefore also contains a substantial population in the lower right corner, where we measure a \Pmem\ of near 0\% despite the high \Pmem\ estimated by \citet{kharchenko2013}. These stars are field interlopers that fall near the cluster sequence and proper-motion locus: since these interlopers should be uniformly distributed on the sky, most will be located at large radii from the cluster core (but still within its tidal radius) and will be down-weighted by our algorithm, which fits the radial-density profile, more effectively than the step function used by \citet{kharchenko2013}.

Finally, we note that our statistical approach to membership is bound to result in some contamination, with our catalog including stars with high \Pmem\ that would be excluded when additional information is included or becomes available. We expect that the forthcoming release of the second \textit{Gaia} data release (DR2) will be invaluable for improving the cluster census.

\subsection{Calculating masses for cluster members}\label{mass}
The availability of 2MASS photometry for nearly all of NGC 752's member stars---and for members of other clusters to which we wished to compare NGC 752---drove us to use these 2MASS magnitudes to estimate stellar masses, as in \cite{agueros11}. We calculated each star's absolute $K$ magnitude (M$_K$), using the source-specific distance modulus associated with the star's SED fit. Of the empirical absolute magnitude-mass relations identified by \cite{delfosse2000}, the M$_K$-mass relation is the best calibrated, and we used this relation for stars with M$_K > 5.5$ mag. 

For brighter stars, we used a theoretical relation for a 1.25 Gyr, [Fe/H] and [$\alpha$/H] = 0 population \citep[updated from the original version published in][]{dotter2008}.\footnote{Available from \url{http://stellar.dartmouth.edu/~models/}.} Systematic uncertainties in the \citet{delfosse2000} relation are of order $\approx$5-10\%, and we therefore adopt 10\% as the typical uncertainty in our derived masses. 

\section{Updating NGC 752's Age and Distance}\label{props}
\subsection{Previous efforts}
A critical step in establishing NGC~752 as a benchmark open cluster is accurately determining its age and distance. Main-sequence and red-giant-branch CMD modeling of NGC~752 have produced estimated ages ranging from 1 to 2 Gyr and DMs ranging from 7.7 to 8.5 mag \citep[e.g.,][]{Meynet1993,daniel1994,Dinescu1995,twarog2015}.\footnote{\citet{vanLeeuwen2009} used HIPPARCOS parallaxes and the photometric box method to derive a DM of 8.53$\pm$0.28 mag.} However, these ages and distances were usually derived using a by-eye comparison of model isochrones to various color-magnitude data sets, which does not provide statistically meaningful uncertainties on the output parameters. Furthermore, these isochrone fits generally used sub-solar metallicity isochrones, which are likely not appropriate for this cluster.

The two most recent and robust determinations of NGC 752's age and distance are those performed by \citet{barta2007,barta2011} and \citet{twarog2015}. \citet{barta2007} used a least-squares minimization to derive an isochrone age of $1.58$$\pm$$0.04$ Gyr and $(m-M)_V = 8.38$$\pm$$0.14$~mag for the upper main sequence of NGC~752. These authors' grid-search technique did provide a goodness-of-fit metric and solved for the best-fit age and DM. However, it did not fully account for correlated errors in colors and magnitudes, and the accuracy of the \citet{barta2007} results is limited to the spacing between isochrones in their model grid. In \citet{barta2011}, including newly identified photometric late-type candidate members led the authors to find an isochrone age of $1.41$ Gyr and a DM of $8.37$$\pm$$0.32$.  

\cite{twarog2015} obtained Str\"omgren photometry for the cluster using the WIYN 0.9-m telescope, achieving an internal precision of $\approx$0.005-0.01 mag. From their data for 68 F dwarfs near the cluster turnoff, \citet{twarog2015} inferred a reddening of $E(b-y)=$ 0.025$\pm$0.003, corresponding to $E(B-V)=$ 0.034$\pm$0.004, and [Fe/H] ranging from $-0.07$ to $-0.017$. Fitting these stars to isochrones computed for this metallicity and distance, \citet{twarog2015} derived an age of 1.45$\pm$0.05 Gyr and DM of 8.30 for NGC 752. These results are consistent with earlier results from the same group based on a re-analysis of the \cite{daniel1994} data \citep{twarog2009}, but do not attempt to quantify the potential systematic uncertainties associated with the choice of isochrones.

\begin{figure}[!ht] 
\centerline{\includegraphics[width=.99\columnwidth]{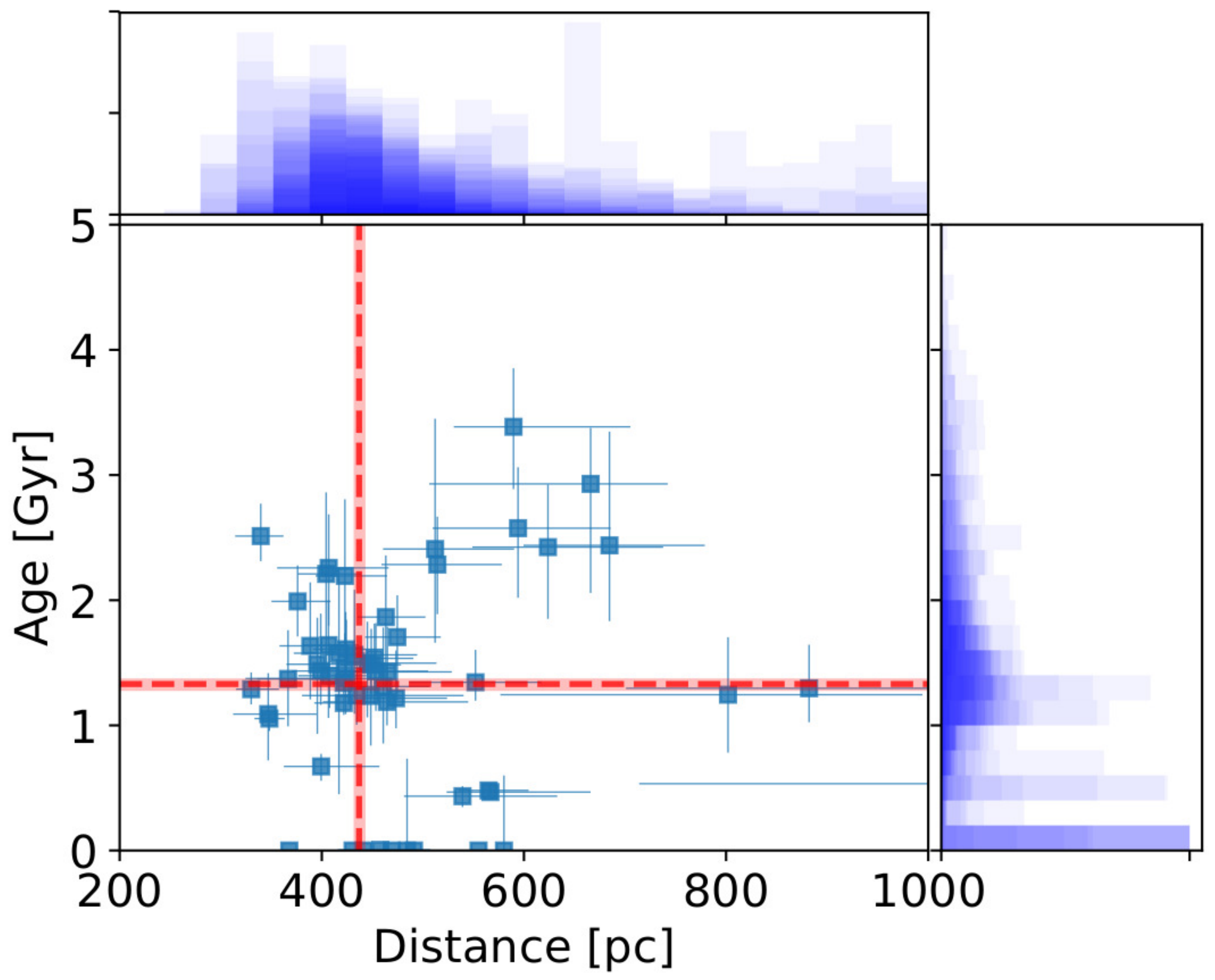}}
\caption{
\label{fig.MINESweeper}  
Age and distance estimates derived for 59 NGC 752 members with no evidence for a binary companion. Error bars indicate the characteristic 1$\sigma$ uncertainties associated with each age and distance estimate, with full probability distribution functions (PDFs) shown on the top and right sides of the main panel for each star. Individual PDFs are transparent, such that regions of parameter space favored by fits to multiple stars appear darker. The cluster's age (1.34$\pm$0.06 Gyr) and distance ($438^{+8}_{-6}$ pc) are derived by multiplying the individual stellar age/distance PDFs to identify the maximum likelihood values for the combined cluster population, and are highlighted in the main panel with dashed red lines.}
\end{figure}

\subsection{Our Bayesian approach and results}
We applied a Bayesian framework to cluster members with astrometric measurements in the Tycho-{\it Gaia} Astrometric Solution (TGAS) catalog \citep{Gaia2016a,gaia_DR1}. Our analysis used photometry from several publicly available surveys, typically Tycho-2 \citep{tycho-2} $BV$, \textit{Gaia} $G$, 2MASS $JHK$, and Wide-field Infrared Survey Explorer \citep{wise} $W1$, $W2$, and $W3$.  SED-based metallicities are inherently uncertain, and we therefore applied a Gaussian prior for metallicity for the cluster based on the \citet{guo2017} spectroscopic analysis. These authors measured metallicities for 36 candidate single members of the cluster using $R \approx 34,000$ spectra obtained with the Hectochelle multi-object spectrograph, finding that [Fe/H] = $-0.032$$\pm$$0.037$, a value consistent with that derived by \citet{twarog2015}.  We therefore adopt [Fe/H] $= -0.03$$\pm$$0.1$ as a prior, increasing the Gaussian width to account for potential systematic uncertainties in the individual stellar metallicities. Similarly, we applied a Gaussian prior of $A_V = 0.105\pm0.1$, based on the value derived by \citet{twarog2015}.

We then use MINESweeper, a newly developed Bayesian approach for determining stellar parameters using the newest MESA Isochrones \& Stellar Tracks (MIST) evolutionary models \citep{Choi2016,dotter2016} to infer probability distribution functions for the age and distance of each cluster member. A detailed description of MINESweeper will be given in Cargile et al.~(in prep); examples of its use include \citet{rodriguez2017}, \citet{temple2017}, and \citet{dotter2017}. MINESweeper provides full posterior distributions of all predicted stellar parameters from the MIST models, including ages, masses, and radii.


Since we are modeling each cluster member as a single star, unresolved binaries result in unreliable stellar parameters due to the influence of the binary on the stellar SED. There are 82 likely members in our catalog with \textit{Gaia} TGAS astrometric parallax measurements: of these, 23 have been identified as RV variables (see Tables~\ref{tab:adam_members} and \ref{tab:not_adam_members}), and we therefore derive estimates of the stellar parameters only for the 59 apparently single stars.\footnote{In practice, including the RV-variable stars does not change our results.}

To determine cluster-wide values for the stellar parameters inferred from the MINESweeper fits, we computed a kernel density estimation of the individual posterior distributions for the stellar parameters estimated for each star. The final combined posterior distributions provide the most probable age, distance, [Fe/H], and $A_V$ for NGC 752 given our priors and assuming that all of these stars are true cluster members. The maximum likelihood values for the distance and ages of individual stars are shown in Figure~\ref{fig.MINESweeper}, along with the super-positions of the individual age and distance probability distribution functions. The combined probability density functions imply the following maximum likelihood mean cluster parameters: an age $=1.34$$\pm$$0.06$ Gyr, a distance = $438^{+8}_{-6}$ pc (DM = 8.21$^{+0.04}_{-0.03}$), a [Fe/H] $=0.02$$\pm$$0.01$, and a $A_V = 0.198^{+0.008}_{-0.009}$. These cluster parameters are in agreement with those of \cite{barta2011} and \citet{twarog2015} and have more robust uncertainty estimates.

As a consistency check, we investigated the direct astrometric distances provided for 53 stars with accurate TGAS ($\sigma_{\pi}< 0.35$ mas) parallaxes.  For these cluster stars, we find a weighted mean parallax $\pi = 2.322$$\pm$$0.049$ mas, corresponding to a DM of 8.17$\pm$0.03~mag or $d = 431$$\pm$$6$~pc.\footnote{In this case we do not apply a binary cut, as binarity should not impact the parallax-derived distance.} These values are consistent with those we have determined using the MINESweeper analysis, and the RMS of the {\it Gaia} measurements is 0.324 mas, so the quoted uncertainties are consistent with the scatter. 

However, there are likely to be spatially correlated systematic uncertainties in the {\it Gaia} data at the level of this scatter \citep[$\approx$0.3 mas; e.g.,][]{gaia_DR1, stassun2016}, which implies corresponding systematic uncertainties of $\sigma_d =\ ^{+65}_{-50}$~pc and $\sigma_{DM} =\ ^{+0.30}_{-0.26}$~mag, respectively. The order-of-magnitude improvement in the precision of parallaxes and proper motions of \textit{Gaia} DR2 relative to its first data release will likely remove these potential systematic uncertainties, enabling proper motion selection to improve the cluster census and providing a precise and accurate distance measurement to this benchmark $>$1 Gyr open cluster.

\begin{figure}[!th] 
\centerline{\includegraphics[angle=90,width=1.04\columnwidth]{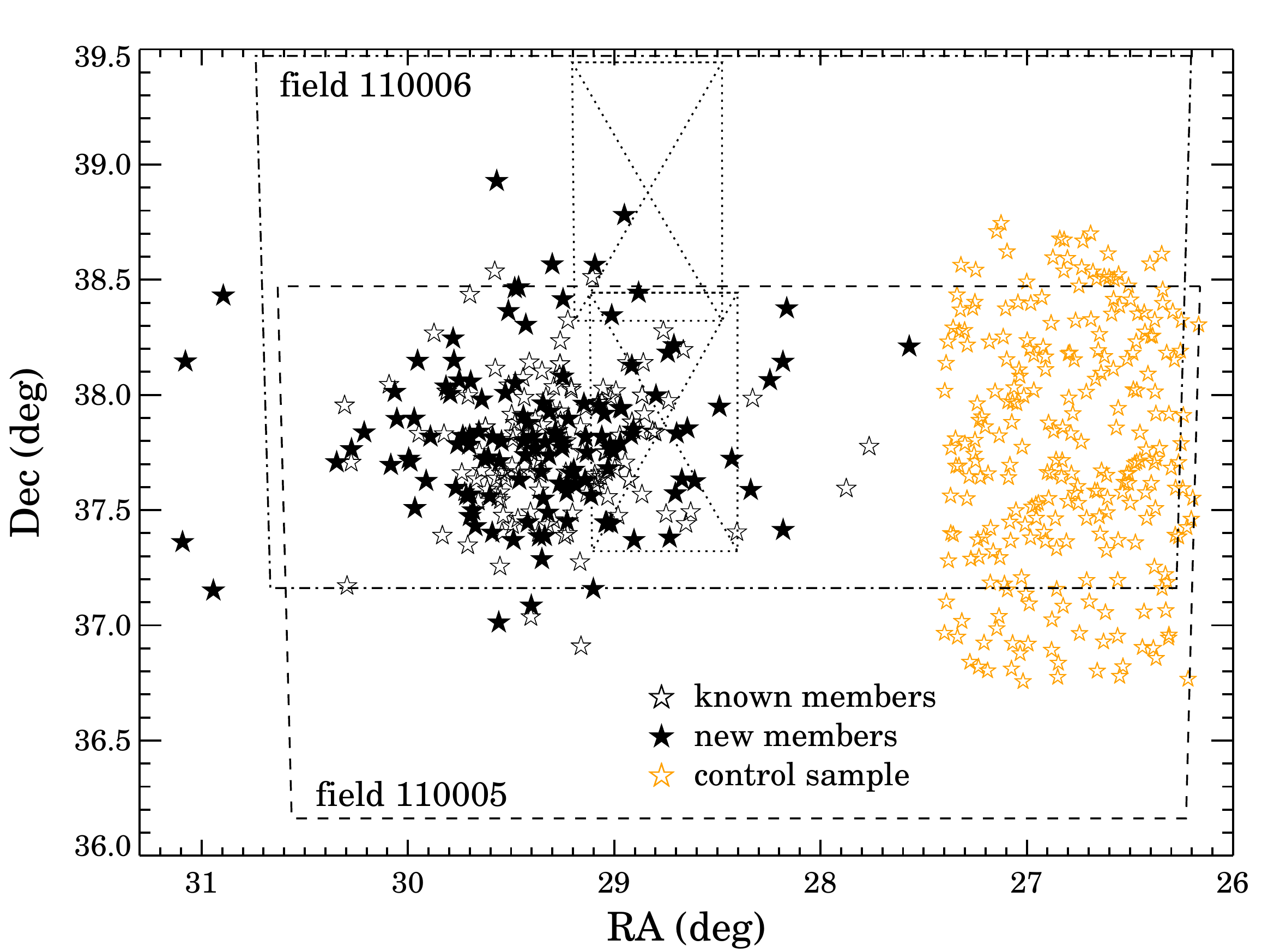}}
\caption{Spatial distribution of the 128 previously cataloged and the 130 newly identified NGC 752 members. The PTF fields are overlaid (110005: dashed lines; 110006: dot-dashed lines); the position of the dead chip in both fields is indicated by the dotted lines. The orange (empty) stars are selected for our control sample as part of the validation of our measured \Prot; see Section~\ref{periods}.} 
\label{fields}
\end{figure}

\begin{deluxetable}{lcc}
\tablewidth{0pt}
\tablecaption{PTF Observations of NGC 752 \label{obs}}
\tablehead{
\colhead{PTF Field}  & \colhead{Field Center}   & \colhead{Number of}   \\       
\colhead{Number} & \colhead{(J2000)} & \colhead{Observations}  
}
\startdata
110005 & 01:53:35 $+$37:19:00 & 377 \\
110006 & 01:53:53 $+$38:19:00 & 413 
\enddata
\end{deluxetable}

\section{Measuring Stellar Rotation at 1.3 Gyr}\label{rot}
\subsection{PTF observations and photometric data reduction}\label{ptf}
We monitored NGC 752 from 2010 Aug 22 to 2011 Jan 19 using time allocated to two PTF Key Projects: the PTF Open Cluster Survey \citep[POCS;][]{agueros11, Douglas2014, covey16, Kraus2017} and the PTF/M-dwarfs survey \citep{Law2011, Law2012}. The PTF infrastructure is described in \citet{nick2009}; of primary interest to us was one component, the robotic 48-inch Oschin (P48) telescope at Palomar Observatory, CA, which we used to conduct our imaging campaign. The P48 was equipped with the modified CFH12K mosaic camera, which had 11 working CCDs, 92 megapixels, 1$\arcsec$ sampling, and a 7.26 deg$^2$ field-of-view \citep{rahmer2008}. Under typical conditions (1$\farcs$1~seeing), it delivered 2\arcsec~full-width half-maximum images that reached a 5$\sigma$ limiting $R_{PTF} \approx21$ mag in 60~s \citep{nick2010}. 

We imaged two overlapping 3.5$^\circ$ $\times$ 2.31$^\circ$ fields covering the center of NGC 752. The fields were selected so that the bulk of the cluster members identified by \citet{daniel1994} and \citet{mermilliod1998} fell on one chip in each (see Figure~\ref{fields}). For most of the campaign, these fields were observed one to four times a night, weather permitting. There were gaps in our coverage each month when PTF conducted its $g$-band and/or H$\alpha$ surveys. Because we shared some of our observing time with the PTF/M-dwarfs survey, a transiting-planet search, there were multiple nights in our campaign when the cluster was observed with a higher frequency, resulting in $\approx$15 images per night. 

In total, we obtained close to 400 observations for each field (see Table~\ref{obs}).  After the standard PTF image calibrations were applied \citep[see][]{nick2009}, the photometric data reduction was done in the same manner as that described in \citet{Law2011}. We performed aperture photometry using SExtractor \citep{Bertin96} on each IPAC-processed PTF frame \citep{Laher2014}. After removing observations affected by e.g., bad pixels, diffraction spikes, or cosmic rays, the positions of single-epoch detections were matched using a 2\arcsec\ radius to produce multi-epoch light curves. This generated photometry for all objects at each epoch with approximate zero points determined on a chip-by-chip basis using USNO-B1.0 photometry of bright stars. The zero points were then refined by a downhill-simplex algorithm that minimized the median photometric variability over all bright non-variable stars in the images. 

\begin{figure}[!t]
\centerline{\includegraphics[width=\columnwidth]{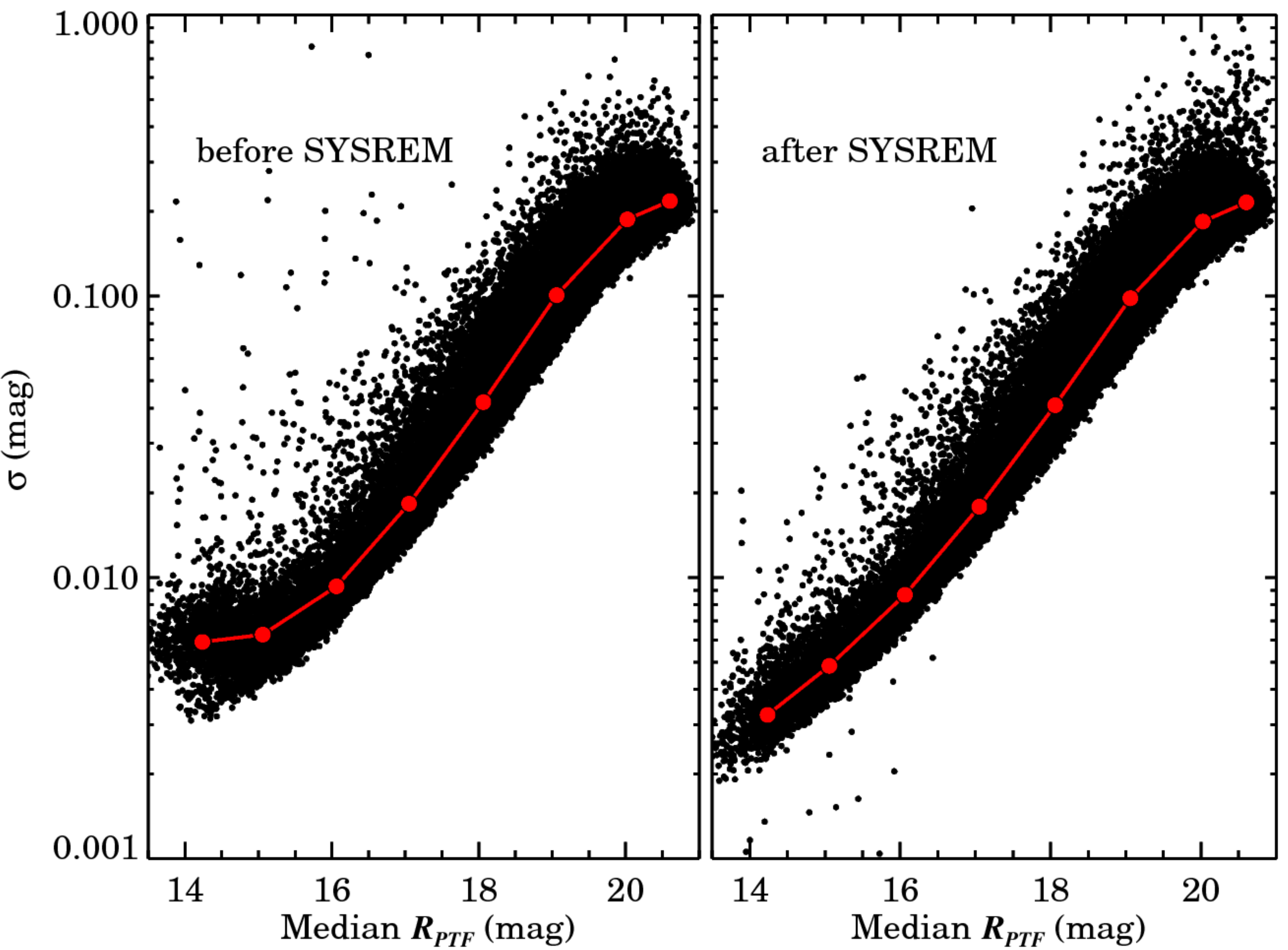}}
\caption{$\sigma$ versus median $R_{PTF}$ magnitude of the $\approx$53,000 objects detected in multiple epochs in field 110005. The median $\sigma$ when placing the objects in bins of width 1 mag is shown as the red points. The raw data are plotted in the left panel; at the bright end, the scatter exceeds the formal photometric errors by factors of a few, indicating that the precision is limited by systematic effects rather than by random photometric errors. As can be seen in the right panel, in addition to removing systematic trends in the data, applying the SYSREM algorithm significantly improves the photometric performance for $R_{PTF} < 16$ mag. } 
\label{sysrem}
\end{figure}

We then applied a version of the SYSREM algorithm to remove systematic trends from the data, e.g., those due to atmospheric extinction, detector efficiency, or point spread function changes \citep{sysrem}. Figure~\ref{sysrem} shows the impact this had on the photometry from field 110005, and in particular the resulting improved performance at the bright end. Applying SYSREM also allowed us to identify a few nights for which the overall photometric behavior of the chips differed significantly from the median over our entire observing campaign. 

\subsection{Period measurement}\label{periods}
To detect periodic signals in our light curves, we followed closely the methods developed for our Pleiades analysis \citep{covey16}. The 90 PTF light curves for NGC 752 members were first cleaned of unreliable data points---those with errors $>$0.5 mag or $>$6$\sigma$ removed from the mean magnitude---before computing a Lomb-Scargle periodogram \citep{Scargle1982, press1989} for 8000 candidate \Prot\ spaced logarithmically between 0.1 and 50 d. Each light curve was then phased on the period initially found to have the maximum power, and $4\sigma$ outliers from a smoothed, phase-folded light curve were clipped before generating an updated periodogram. This clipping and computing process was performed three times before a final period was assigned to the star.

The error on our \Prot\  measurements was estimated using the width of a Gaussian fit to the corresponding peak in the power spectrum \citep{lamm2004}. This width indicates a fundamental uncertainty in the period measurement that originates from the frequency resolution of the power spectrum and the time sampling of the data.

\subsection{Period validation}
\begin{figure}[t!]
\centerline{\includegraphics[width=\columnwidth]{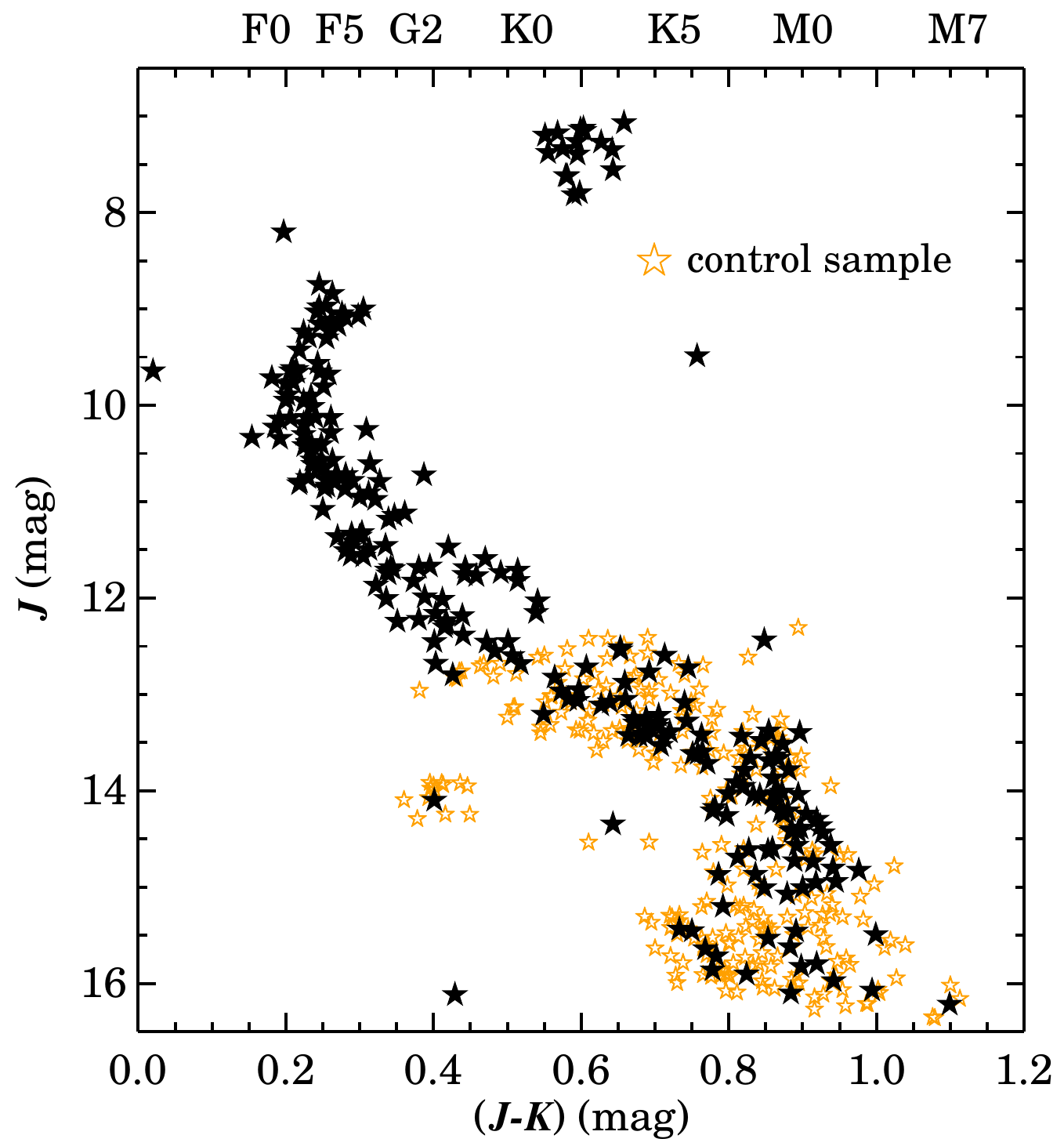}}
\caption{CMD for cluster members and control stars identified to test the robustness of our \Prot\ measurements. The control star sample has a color and magnitude distribution that mirrors that of NGC 752 stars with PTF light curves, but should be dominated by old field stars with little inherent variability.} 
\label{controlCMD}
\end{figure}

The process described above returns a \Prot\ for every light curve. We modified slightly the \citet{covey16} approach to select the significant and reliable \Prot\ measurements. We identified a sample of 254 field stars that have PTF light curves, high-quality 2MASS photometry, and $(J-K)$ colors and $J$ magnitudes similar to those of NGC 752 members (see Figure~\ref{controlCMD}). These stars' PTF photometry will exhibit the same instrumental signatures as those of the members, but because these stars should be older and have lower levels of magnetic activity, they should be less variable \citep[as expected based on the age-activity relation; e.g.,][]{hawley1999, soderblom2001, Douglas2014}. 

We then tested our ability to recover \Prot\ from data that reflect the cadence and noise properties of our targets' data by injecting artificial periodic signals into these quieter light curves and running the same period-detection algorithm as that applied to cluster members. We first removed every star in our control sample for which the periodogram included a peak with a power $>$20, thereby selecting a sample of 156 minimally variable stars. For each of these remaining stars, we then  generated 1500 periodic light curves in which a sine curve with an amplitude scaled relative to the light curve's $\sigma$\footnote{amplitude/$\sigma_{light\ curve} = $ 0.3, 0.6, 0.9, 1.2, or 1.5.} and a period randomly selected from a Gaussian distribution centered at 25$\pm$10 d was added to the PTF photometry while preserving the original light curve's timestamps.\footnote{We required that the injected \Prot\ be between 0.1 and 50 d. This choice of a Gaussian \Prot\ distribution for the simulations is the main difference with our approach in \citet{covey16}.}

By applying our period-detection algorithm to the resulting 234,000 artificial light curves, we measured the dependence of our recovery rate and accuracy of our \Prot\ measurement on the properties of the input light curve and of the output periodogram. We defined as a successful recovery any simulation in which the input and recovered \Prot\ agree to within 3\%. Our overall success rate was 73\%. This simulation allowed us to set a threshold power of 40 for the most significant peak in our periodograms as the one to use for identifying robust period measurements.  

To determine if the star exhibits a single, unambiguous period, we also cleaned the periodograms for cluster members of any aliases and beat periods between the candidate period and a one-day sampling frequency before searching for secondary peaks with power $\geq$60\% of the primary peak's. If no secondary peaks were found, the primary period was flagged as a secure detection (i.e., CLEAN = 1); sources with such secondary peaks were flagged as having an ambiguous period  (CLEAN = 0). In practice this step eliminated only three stars with peak periodogram powers $>$40.

\begin{figure*}[t!]
\centerline{\includegraphics[width=1.55\columnwidth]{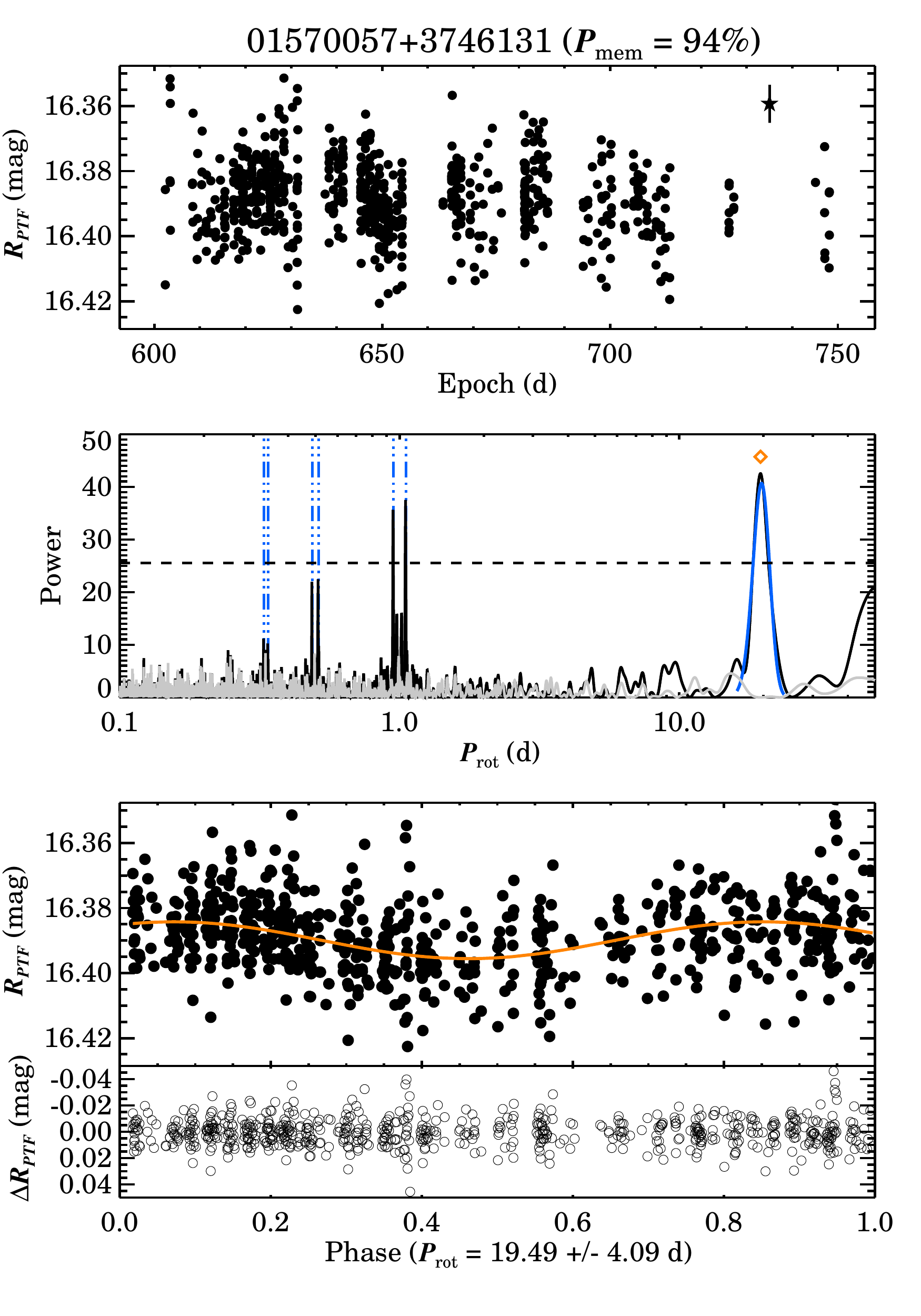}}
\caption{\textit{Top:} PTF light curve for a newly identified high-confidence NGC 752 member, 2MASS J01570057$+$3746131. The x-axis is the number of Julian days since 2009 Jan 1. The error bars on the star at the top right show the median photometric uncertainty. \textit{Middle:} The periodogram calculated via our iterative process (black line), with the peak power, corresponding to a period of 19.49~d, highlighted with an orange diamond. The blue Gaussian, with which we estimate the uncertainty on \Prot, is a fit to this peak. Beat periods between this \Prot\ and a one-day alias are flagged with vertical (blue) dot-dashed lines; the power threshold used to flag sources with ambiguous period detections (i.e., other periods with peaks with $\geq$60\% of the primary peak's power) is shown as a horizontal dashed line. \textit{Bottom:} The phase-folded light curve. A median-filtered version of this light curve, shown as an orange line, is subtracted to create a pre-whitened light curve, shown in the sub-panel at the bottom. The periodogram computed from this pre-whitened light curve is shown as a gray line in the middle panel. The primary peak and beat periods are not present in the periodogram of the pre-whitened light curve, indicating that the periodic signature removed during the pre-whitening accounts for all of the significant structure in the star's light curve. {\it The journal version of this article includes this figure for the other rotators we identify.}\\ } 
\label{fig:full}
\end{figure*}

The result of this analysis is a list of 12 NGC 752 stars for which we measured reliable rotations periods. Figure~\ref{fig:full} is an example of the outputs of the period-finding process described above for each of these stars, which are listed in Table~\ref{table:rotators}.

\begin{deluxetable*}{lccccccr@{$\pm$}l}
\tablewidth{0pt}
\tablecaption{NGC 752 Rotators \label{table:rotators}}
\tablehead{
  \colhead{2MASS ID} &
  \colhead{SpT} &
  \colhead{\Pmem} &
  \colhead{M$_K$} &
  \colhead{Mass} &
  \colhead{$R_{PTF}$} &
  \colhead{\# of Obs.} &
 \multicolumn{2}{c}{\Prot} \\
  \colhead{} &
  \colhead{} & 
  \colhead{(\%)} &
  \colhead{(mag)} & 
  \colhead{(\Msun)} & 
  \colhead{(mag)} & 
  \colhead{} & 
 \multicolumn{2}{c}{(d)}
}
\startdata
01525891$+$3803515 & K4.7 & 52.8 & 4.49 & {\bf 0.69} & $13.92$ & $303$ & $17.50$ & $2.49$ \\
01544738$+$3749590 & M1.8 & 57.2 & 5.82 & 0.49 & $17.28$ & $383$ & $18.85$ & $3.66$ \\
01553694$+$3722130 & K5.8 & 56.7 & 4.68 & {\bf 0.64} & $14.58$ & $389$ & $13.00$ & $1.54$ \\
01565531$+$3736463 & K5.2 & 95.2 & 4.54 & {\bf 0.68} & $14.83$ & $688$ & $14.90$ & $2.45$ \\
01570057$+$3746131 & M1.6 & 94.2 & 5.69 & 0.51 & $16.39$ & $695$ & $19.49$ & $4.09$ \\
01572074$+$3723159 & K5.9 & 84.8 & 4.73 & {\bf 0.63} & $14.87$ & $692$ & $14.03$ & $2.89$ \\
01572260$+$3732585 & K7.9 & 79.1 & 5.13 & {\bf 0.56} & $15.94$ & $690$ &  $5.27$ & $0.33$ \\
01581109$+$3747537 & K5.6 & 96.5 & 4.64 & {\bf 0.65} & $14.74$ & $690$ & $16.58$ & $2.28$ \\
01581346$+$3742456 & M0.5 & 94.1 & 5.31 & {\bf 0.53} & $16.66$ & $642$ & $17.53$ & $3.04$ \\
01582190$+$3724073 & K2.8 & 77.7 & 4.34 & {\bf 0.72} & $14.33$ & $583$ & $34.97$ & $25.10$ \\
01584873$+$3747010 & K4.7 & 96.2 & 4.47 & {\bf 0.69} & $14.94$ & $694$ & $13.92$ & $2.75$ \\
01591077$+$3800176 & K5.1 & 91.7 & 4.52 & {\bf 0.68} & $15.02$ & $693$ & $32.74$ & $9.77$
\enddata
\tablecomments{The formal uncertainty for the SpTs is 0.1 spectral classes. However, the systematic uncertainty in the underlying definition is $\approx$0.5 spectral classes for M dwarfs, and this systematic uncertainty will be reflected in the color-spectral type relations used for SED fits. M$_K$ is calculated for each star using the best-fit distance modulus determined from the SED fit, and in turn is used to obtain masses. Masses for sources brighter than M$_K =$ 5.5~mag are assigned using the theoretical model of \citet{dotter2008}, while masses for fainter sources are assigned using the empirical mass-luminosity relation measured by \citet{delfosse2000}. Although the \citet{delfosse2000} relation extends to stars with M$_K$ = 4.5 mag, the predicted mass values diverge by up to about $5\%$ from those of \citet{dotter2008} for stars brighter than M$_K$ = 5.5 mag. We provide the median $R_{PTF}$ magnitude of each light curve after filtering on flags and correcting for the (generally very small) photometric offset between fields for stars that were in both. The uncertainty on these magnitudes is of order 1\%.}
\end{deluxetable*}

\section{Measuring Chromospheric Activity at 1.3 Gyr}\label{spec}
We used the WIYN 3.5-m telescope on Kitt Peak, AZ, to obtain spectra for 96 stars; we used the MDM Hiltner 2.4-m telescope, also on Kitt Peak, to obtain spectra for 180 stars (see Table~\ref{specstats}). The resulting sample is $\approx$70\% complete for candidate cluster members with \Pmem~$\geq 50\%$ but that lacked spectra prior to this work. Our observational set-up and data reduction processes are described below. 

\begin{deluxetable}{lccc}[!t]
\tablewidth{0pt}
\tabletypesize{\scriptsize}
\tablecaption{Spectroscopic Statistics \label{specstats}}
\tablehead{
\colhead{} & \colhead{WIYN} & \colhead{Hiltner}    \\
\colhead{} & \colhead{(Hydra)} &  \colhead{(ModSpec)}      
}
\startdata
Targets				& 96 & 180  \\
... with \Pmem~$> 50$\%	& 60 & 106   \\
... with spectra in literature 	& 12 & 40   \\
... observed more than once     & 5  & 7  
\enddata
\tablecomments{\Pmem\ is the membership probability in our cluster catalog; see \S\ref{members}.}
\end{deluxetable}

\subsection{WIYN: Set-up and data reduction}\label{hyd}
We observed NGC~752 with the Hydra multi-object spectrograph during the nights of 2011 Feb 7 and 8. We used the bench-mounted spectrograph with the red fiber cables and an \'echelle grating with 600 lines mm$^{-1}$ set at a blaze angle of 13.9$^\circ$. This resulted in coverage from 6050 to 8950 \AA~with $\approx$1.4~$\AA$ sampling and a spectral resolution of $\approx$4000. We targeted two fields: a bright field (BF) centered at 01$^{\rm h}$56$^{\rm m}$34$^{\rm s}$, $+$37$^\circ$42$\amin$48$\asec$, and a faint field (FF) centered at 01$^{\rm h}$56$^{\rm m}$14$^{\rm s}$, $+$37$^\circ$34$\amin$50$\asec$ (J2000 coordinates). The two fields required exposure times of 1800 and 5400~s, respectively, which were split into three sub-exposures for cosmic-ray removal. We placed target fibers on 59 candidate cluster members in the BF and 42 candidate members in the FF; five stars were included in both fields, for a total of 96 individual targets.

\begin{figure}[h!]
\centerline{\includegraphics[width=1\columnwidth]{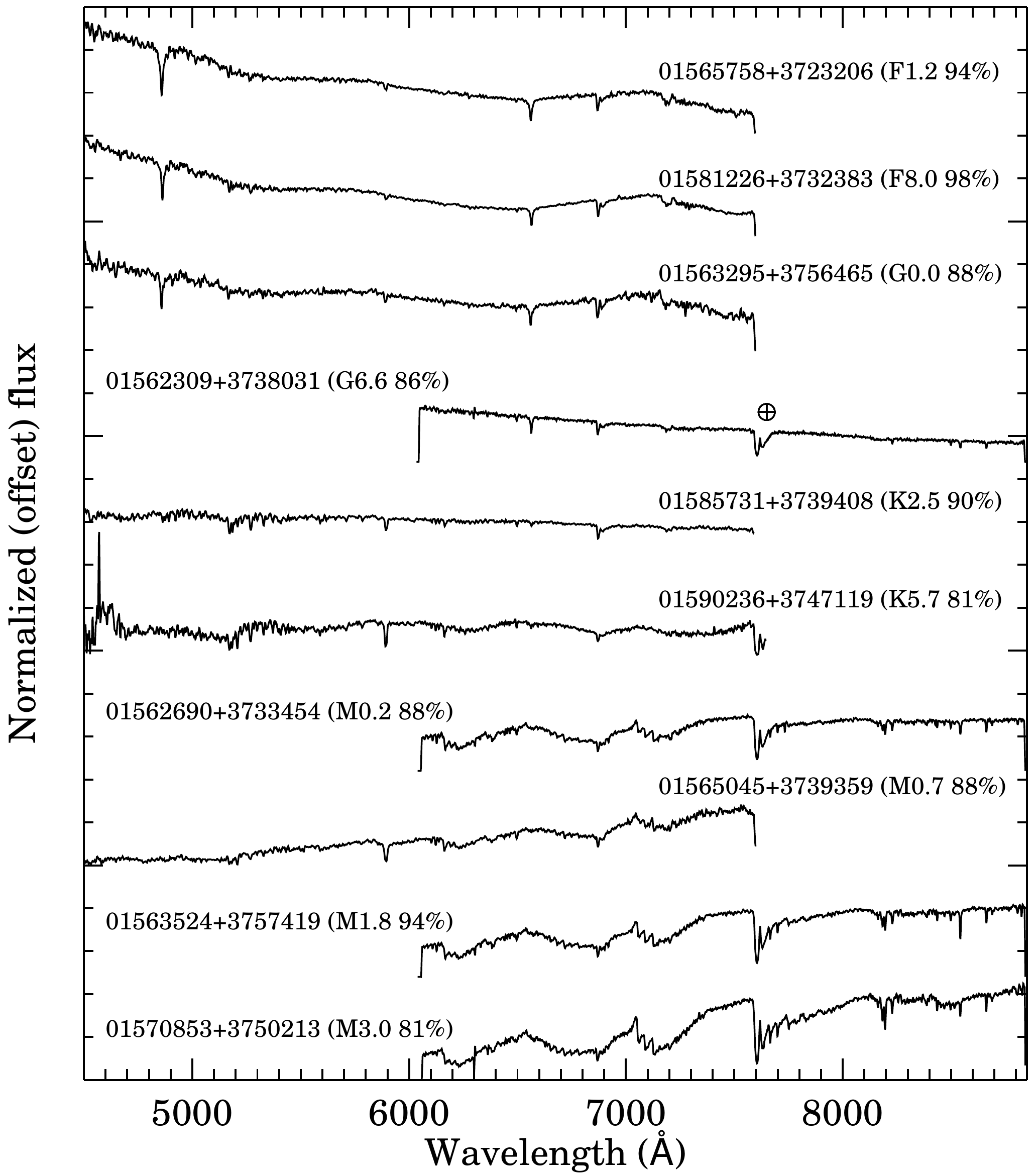}}
\caption{Sample Hydra and ModSpec data. Hydra provided coverage from 6050 to 8950~\AA, while ModSpec covered 4500 to 7500~\AA. Photometric SpT and \Pmem\ are indicated in parentheses. Imperfect flux calibrations are responsible for the structure seen in the continuum of the top three (ModSpec) spectra. The strong A-band telluric features are indicated by the $\oplus$.} 
\label{hyd_spec}
\end{figure}

The data were reduced using standard routines in the IRAF Hydra package.\footnote{Available from \url{http://iraf.noao.edu/tutorials/dohydra/dohydra.html}} Each image was trimmed and instrument biases were removed. The spectra for the individual fibers were extracted, flat-fielded, and dispersion-corrected. Sky spectra from $\approx$30 fibers placed across the field-of-view were combined and subtracted from our target  spectra. We throughput-corrected and flux-calibrated each spectrum using the flux standard G191-B2B, which was obtained using the same instrument set-up as our targets. We then combined the three sub-exposures for each object to form a single, high signal-to-noise spectrum for each candidate cluster member; four sample Hydra spectra are included in Figure \ref{hyd_spec}.

\subsection{MDM: Set-up and data reduction}\label{mdm}
We used the MDM Observatory Modular Spectrograph (ModSpec) on the 2.4 m to obtain spectra of 180 candidate cluster members over the course of five observing runs between 2010 Dec 1 and 2012 Feb 20. ModSpec provided coverage from 4500 to 7500~\AA~with $\approx$1.8~\AA~sampling and a spectral resolution of $\approx$3300. Using a PyRAF script, all the spectra were trimmed, overscan- and bias-corrected, cleaned of cosmic rays, flat-fielded, extracted, dispersion-corrected, and flux-calibrated using standard IRAF tasks.  

Wavelength shifts due to flexure were corrected using a custom IDL routine to measure the apparent wavelength of the 5577~\AA~[OI] sky emission line.  Measurements from lamp observations indicate that ModSpec's dispersion varies by $<$10\% across the full spectral range of these observations. Given this near constant dispersion and the lack of bright sky lines to provide a higher order solution, a simple linear offset was applied to each observation's wavelength solution to correct for the offset measured from the 5577 \AA~line. This line was too weak to provide accurate offset measurements for exposures shorter than 30 s, so the instrumental wavelength solution was preserved for spectra with short exposure times. The uncertainty in the measurement of the [OI] line center was typically $\approx$0.02-0.1~\AA, but occasionally as large as 0.2-0.3~\AA, depending on the exposure time and weather conditions. Uncertainties of a few 0.01-0.1~\AA~in each spectrum's wavelength solution limit the accuracy of the velocities that can be measured from these spectra to $\approx$2-15~km~s$^{-1}$. Six sample ModSpec spectra are shown in Figure~\ref{hyd_spec}. 

\subsection{Identifying chromospherically active members}\label{identifyActive}
\begin{figure}[h]
\centerline{\includegraphics[width=\columnwidth]{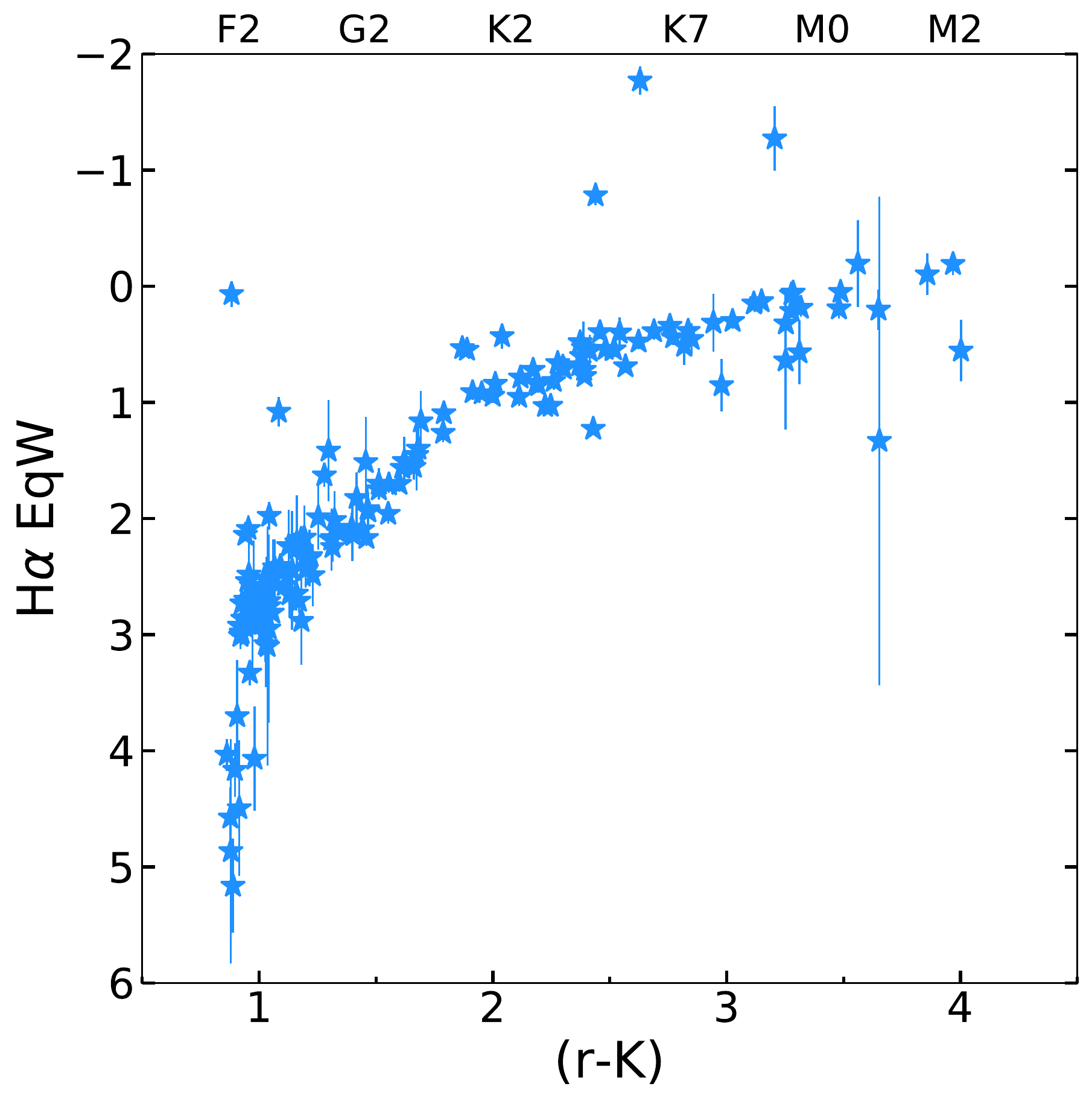}}
\caption{\Ha\ EqWs as a function of $(r-K)$ color for NGC 752 cluster members. Uncertainties in EqW are measured through a Monte Carlo process described in the text. The three emitters in the mid/late K regime are significant outliers from the remainder of the cluster population and have moderate membership probabilities: 50\% $<$~\Pmem~$<$ 80\%.}
\label{halpha_eqws}
\end{figure}

To identify chromospherically active cluster members, we measured the \Ha\ EqW for each spectrum. The measurement window used varied from spectrum to spectrum and was adjusted interactively. Ideally, we would always take the continuum flux to be the average between 6550-6560~\AA\ and 6570-6580~\AA. For spectra for which the \Ha\ line extended into these windows, the continuum flux was measured from 10~\AA\ windows on each side of the line. The resulting \Ha\ EqW measurements are shown in Figure~\ref{halpha_eqws} as a function of $(r-K)$. 

To estimate the human error in these interactive measurements, the same person measured each EqW twice, and we took the difference between the two measurements. We then used a Monte Carlo technique to determine the statistical significance of our \Ha\ measurements. Lacking noise spectra for these stars, we added noise drawn from a Gaussian distribution with a width equal to the $\sigma$ of the flux in the continuum region to each spectrum and remeasured the EqW 2500 times. The two error measurements were added in quadrature to produce the EqW uncertainties \citep[for details about this procedure, see][]{Douglas2014}. 

Using these EqW uncertainties, we identified magnetically active stars as those with \Ha ~EqW$+ 3\sigma <0$. We found only three stars that satisfy this criterion; we discuss this result in the next section, in the broader context of the age-rotation-activity relationship in NGC 752 and other open clusters. 

\begin{figure*}[t]
\centerline{\includegraphics[angle=-270,width=1.75\columnwidth]{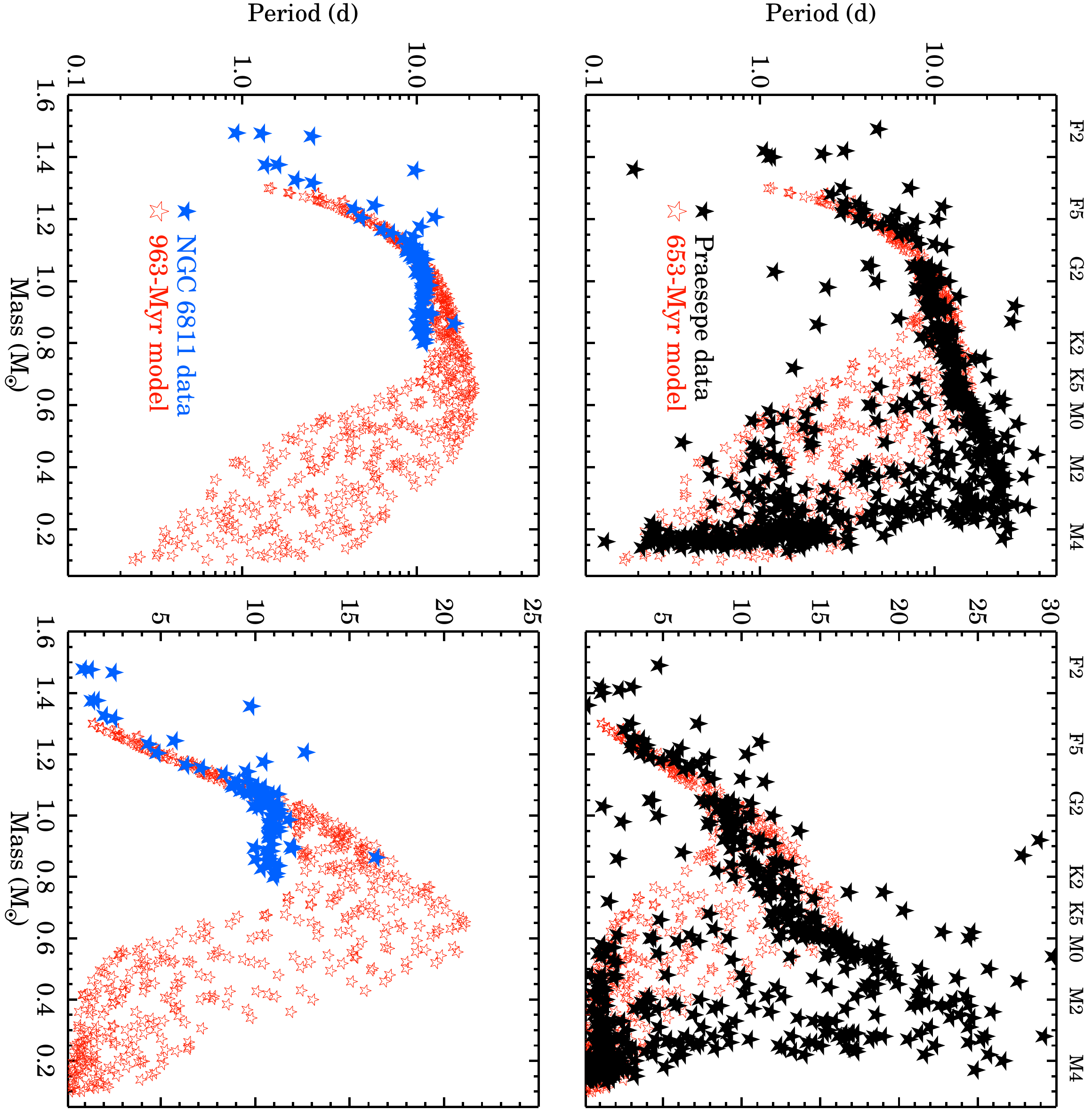}}
\caption{Comparison between the \citet{Matt2015} models (empty red stars) and the mass-period distributions for Praesepe \citep[black stars, top row; data from][]{Douglas2017} and NGC 6811 \citep[blue stars, bottom row; data from][]{Meibom2011}. The age of the \citet{Matt2015} model population is indicated in the right panel in each row. On the left, the periods are plotted logarithmically, and on the right linearly.} 
\label{mp}
\end{figure*}

\section{Placing Rotation and Activity at 1.3 Gyr in Context}\label{disc}
\subsection{Comparing the \Prot\ for NGC 752 to those for younger clusters and to the \citet{Matt2015} model for rotational evolution}\label{gyro}
As in previous papers, we placed our observations of rotators in NGC 752 in the context of the rotational evolution of low-mass stars \citep[cf.][]{agueros11,Douglas2016,Douglas2017}. Our empirical comparison was with the $\approx$650-Myr-old benchmark cluster Praesepe, for which extensive \Prot\ measurements exist for stars down to 0.2~\Msun, and with the $\approx$1-Gyr-old {\it Kepler} target NGC 6811, which is the only cluster close in age to NGC 752 with substantial rotational data. We also compared our data to the predictions from the \citet{Matt2015} model for stellar angular-momentum evolution. Initial conditions for this model are set by approximating the mass-period distribution observed in very young clusters. Angular momentum is then removed by winds using a prescription based on the solar wind described by \citet{kawaler1988} and \cite{Matt2012}, and the overall angular-momentum loss scales with mass and radius. 

Below, we extend our \citet{Douglas2017} test of the \citet{Matt2015} model. We then describe the process of constructing the mass-period sample for NGC 6811, examine evidence for rotational evolution between Praesepe, NGC 6811, and NGC 752's ages, and compare the NGC 752 data to the \citet{Matt2015} model for a 1.3-Gyr-old population.

\subsubsection{Comparing the Praesepe data to the \citet{Matt2015} model}
The top row of Figure~\ref{mp} replicates the comparison made in \cite{Douglas2017} between the \citet{Matt2015} model and mass-period data for the $\approx$650-Myr-old cluster Praesepe. The \citet{Matt2015} model reproduces the mass dependence of the slow-rotator sequence for $\gapprox$0.8 \Msun\ stars in Praesepe \citep[and in the Hyades, another $\approx$650-Myr-old cluster;][]{Douglas2016}, indicating that the \citet{Matt2015} stellar-wind prescription is correct for solar-type stars. 

However, the match between model and data is not as good for 0.8-0.3~\Msun\ stars. The model predicts more rapidly rotating $\lapprox$0.8~\Msun\ stars than are observed; $<$50\% of these stars are more efficient at spinning down than expected. The median rotation periods for 0.8-0.3~\Msun\ stars in the model and the data reflect this mismatch: the model predicts that the median rotator should have a \Prot~$= 4.5$ d, whereas the median observed rotator, when 26 known and candidate binaries are removed, has a \Prot~$= 14.0$~d.\footnote{Hereafter, we remove known and candidate binaries from our catalog when calculating median periods for Praesepe.} 

Furthermore, more than half of the cluster 0.6-0.3~\Msun\ stars have converged to the slow-rotator sequence, which extends from $\approx$1.2 to 0.3~\Msun, and more than half of the remaining rapid rotators are binaries. At 650 Myr, however, the \citet{Matt2015} model predicts that the slow-rotator sequence should end around 0.6~\Msun. If Praesepe is $\approx$650-Myr-old, early M dwarfs appear to brake more efficiently than predicted by \citet{Matt2015}. This may be because the \citet{Matt2015} model does not include any prescription for core-envelope decoupling; adding this to models may provide a better fit to stars in this mass range at this age \citep[e.g.,][and pers.~comm., S.~Matt]{gallet2015}.

\begin{figure*}[t]
\centerline{\includegraphics[angle=-270,width=1.75\columnwidth]{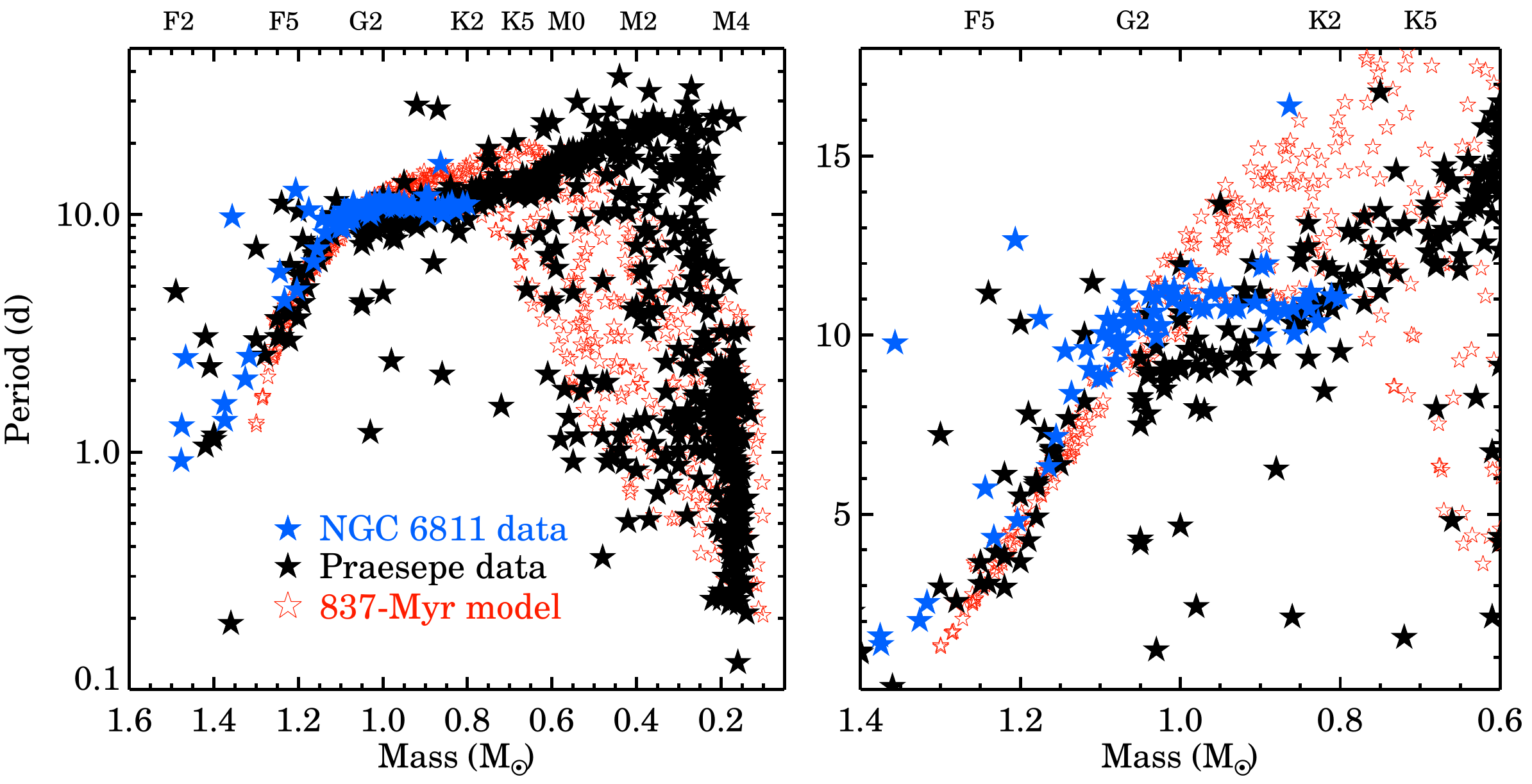}}
\caption{As in Figure~\ref{mp}, but now plotting Praesepe and NGC 6811 together with a \cite{Matt2015} 837-Myr-old population. The Praesepe and NGC 6811 slow-rotating sequences are well-matched, particularly when the \Prot\ are plotted logarithmically, and both appear to flatten for masses $\lapprox$1~\Msun, which is not predicted by the model. In the right (linear \Prot) panel, there is evidence of spin down for the 0.9-1.1~\Msun\ stars between the ages of Praesepe and NGC 6811, but the 0.8-0.9~\Msun\ stars do not appear to have spun down.} 
\label{mp2}
\end{figure*}

\subsubsection{Examining rotational evolution between Praesepe and NGC 6811}\label{surprise}
NGC 6811 \citep[1.00$\pm$0.17~Gyr;][]{janes2013} is one of four open clusters in the original {\it Kepler} field, and the only cluster close in age to NGC 752 for which \Prot\ have been obtained \citep{Meibom2011}. We matched the rotators listed in \cite{Meibom2011} to 2MASS and used the cluster properties determined by \citet{janes2013} ($E(B-V) = 0.074$, $(m-M)_V = 10.22$) and the 1 Gyr, [Fe/H] = 0.07 and [$\alpha$/H] = 0 (updated) \citet{dotter2008} model to calculate masses for these stars in the manner described in Section~\ref{mass}.\footnote{\citet{janes2013} find [Fe/H] = $-0.18$ for NGC 6811 based on isochrones fits, but an analysis of $R \approx 25,000$ spectra of individual members by \citet{molenda2014} finds a mean [Fe/H] = 0.04$\pm$0.01 and an overall abundance pattern for the cluster that is very close to solar.} In the bottom row of Figure~\ref{mp}, we show the resulting mass-period distribution for this cluster, along with the \cite{Matt2015} predictions for the distribution of a 963-Myr-old population. The \cite{Matt2015} model clearly overestimates the spin-down for $\lapprox$1~\Msun\ stars. Model stars with masses between 0.8 and 1.0~\Msun\ have a median \Prot~$= 14.2$ d; by contrast, the 26 NGC 6811 rotators in that mass bin have a median \Prot~$= 10.8$$\pm$0.4 d.

Indeed, the evolution for 0.8 and 1.0~\Msun\ stars is surprisingly small over the $\approx$350~Myr that should separate NGC 6811 from Praesepe: the 38 Praesepe stars in this mass range have a median \Prot~$=$ 9.9 d. In Figure~\ref{mp2}, we combine the data for Praesepe and NGC 6811 and show that the clusters' two slow-rotating sequences are very well-matched, especially considering that the bulk of the Praesepe data for stars $>$0.8~\Msun\ come from ground-based observations \citep{delorme2011,kovacs2014}. The quality of those data is not as high as those from {\it Kepler}, presumably contributing to the scatter in the periods for Praesepe stars between 0.8 and 1.2~\Msun\ relative to what is seen for NGC 6811. The combined cluster data are well described by the \citet{Matt2015} model population for 837 Myr, although the model continues to over-predict the spin down of stars between $\approx$0.95 and 0.6~\Msun~and under-predict the spin down of $\lapprox$0.6~\Msun\ stars.

\begin{figure*}[hbt!]
\centerline{\includegraphics[angle=-270,width=1.75\columnwidth]{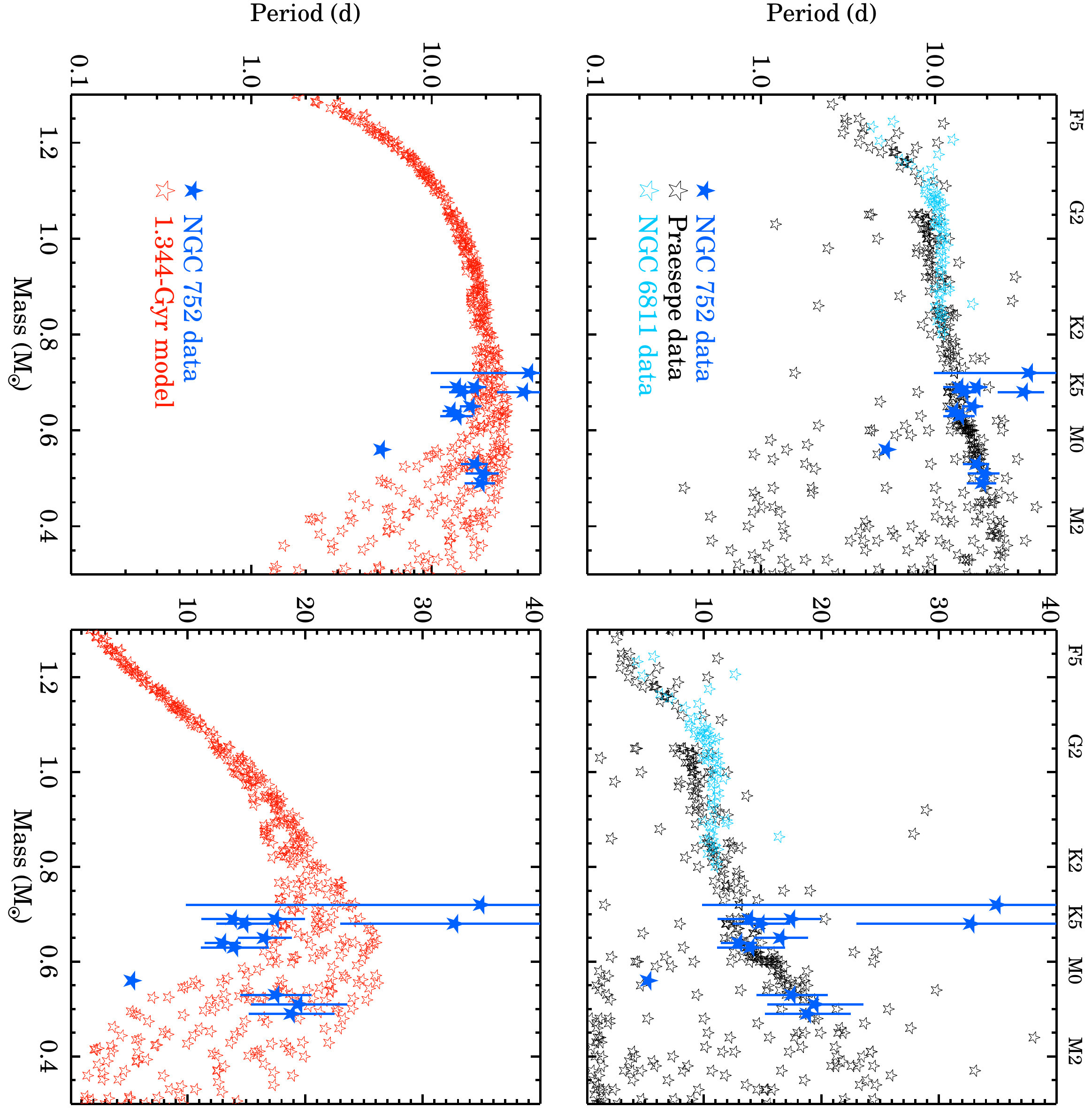}}
\caption{{\it Top ---} Comparison between the mass-period distribution for the joint Praesepe (empty black stars) and NGC 6811 (empty cyan stars) sample presented in Figure~\ref{mp2} and for NGC 752 (solid blue stars). {\it Bottom ---} Comparison between the \citet{Matt2015} model for 1.344 Gyr (empty red stars) and the NGC 752 data. As in previous figures, on the left, the periods are plotted logarithmically, and on the right linearly.} 
\label{mp3}
\end{figure*}

One can draw several possible conclusions from this comparison. If we assume that angular-momentum evolution is roughly constant with time, then at least one of the cluster ages is incorrect. NGC 6811's age could be younger than 1~Gyr, which we tested by comparing the cluster data to progressively younger \citet{Matt2015} model populations. While these comparisons do show that the cluster's mass-period sequence is better fit (by eye) when using $<$1 Gyr models, these younger populations show a spread in the \Prot\ at progressively higher masses (e.g., at 963 Myr, the single-valued mass-period sequence begins to fan out at $\approx$0.7~\Msun; at 837~Myr, at $\approx$0.8 \Msun; and at 653 Myr, at $\approx$0.9 \Msun). This spread is not seen in the NGC 6811 data, suggesting that stars with masses $\gapprox$0.8~\Msun~(the lowest mass for which we have {\it Kepler} data) have all had time to spin down to a slow-rotating sequence and setting a lower limit of $\approx$800 Myr for the cluster's age.

On the other hand, Praesepe could be older than previously thought, as argued by \cite{brandt2015}, who, by incorporating rotation into their evolutionary models, found that the cluster is closer to $\approx$800 Myr in age. Increasing Praesepe's age in this manner requires explaining the presence of fast rotators with masses between 0.5 and 1.1~\Msun, since these stars lie outside of the range of \Prot\ predicted by the \citet{Matt2015} model. The cluster of Praesepe stars at $\approx$0.6~\Msun\ and \Prot~$\approx1$~d in particular suggests that the 837-Myr-old model population is a poor fit to the data. However, as discussed in \citet{Douglas2016,Douglas2017}, many of these fast rotators are likely to be binaries. In the Hyades, all rapidly rotating $\gapprox$0.3~\Msun\ stars are binaries; in Praesepe, which has not been surveyed as extensively for binarity, half of the rapidly rotating $\gapprox$0.3~\Msun\ stars are confirmed or candidate binary systems, and the remaining $\gapprox$0.3~\Msun\ fast rotators are not confirmed single stars, because they have not been searched for companions.\footnote{The \citet{Meibom2011} periods are only for nominally single members of NGC 6811; these authors have extensive RV data for the cluster.}

Finally, the mass-period data for the two clusters may be suggesting that spin down progresses differently for solar-mass and lower-mass stars. The right panel of Figure~\ref{mp2}, where the periods are plotted linearly, shows that there is evidence for spin down for the 0.9-1.0~\Msun\ stars: for Praesepe, the 20 stars in this mass bin have a median \Prot = 9.4~d, while for NGC 6811, the 11 stars have a median \Prot~$=$ 10.8$\pm$0.3~d. That difference in the median \Prot\ is erased when considering 0.8-0.9~\Msun\ stars, however: the median for the 18 Praesepe stars is 10.8~d and for the 15 NGC~6811 members it is 10.8$\pm$0.4~d. 

Adding the \citet{Matt2015} model, which predicts a Skumanich-like, 1/$\sqrt{\rm age}$ spin down for these stars, strengthens the impression that spin down is stalling for these lower-mass stars: for 0.9-1.0~\Msun\ stars, the model predicts a median \Prot~$=12.3$~d, and for 0.8-0.9~\Msun\ stars, 13.7~d, at 837 Myr. The potential stalling of spin down observed for 0.8-0.9~\Msun\ stars needs to be tested with data from older clusters, with Ruprecht 147 a particularly promising cluster for this (J.~Curtis, pers.~comm.).  

\subsubsection{Comparing NGC 752 to the younger clusters and to the \citet{Matt2015} model}
In Figure~\ref{mp3}, we show a comparison of the combined mass-period data for Praesepe and NGC 6811 and for the 12 members of NGC 752 for which we have new \Prot\ measurements (top row). The sparseness of the data for NGC 752 make it difficult to draw strong conclusions from this comparison, although on average, the NGC 752 stars do appear to be rotating more slowly than their younger counterparts. The difference is not significant, with the lowest-mass stars in NGC 752 in particular being indistinguishable in the mass-period plane from their cousins in Praesepe. For the eight 0.6-0.8~\Msun\ stars in NGC 752, the median \Prot~=~16.6$\pm$2.8 d, while for Praesepe the 70 stars in this mass bin have a median \Prot~=~13.8~d. If we remove the two $\approx$0.7 \Msun\ longest-period rotators in NGC 752, which have associated large period uncertainties, the median \Prot~for cluster stars in this mass bin drops to 14.9$\pm$2.5~d, even closer to its cousin in Praesepe. 

Similarly, for the four 0.4-0.6~\Msun\ stars in NGC~752, the median \Prot~=~18.9$\pm$3.7~d, while in Praesepe, this \Prot~=~16.6~d for 83 stars. If we exclude the fast rotating Praesepe stars in this mass bin, which are likely binaries, so as to focus the comparison on the slow-rotating sequence only, the median Praesepe \Prot\ is 18.1 d for 59 stars.

The comparison to the \citet{Matt2015} model shown in Figure~\ref{mp3} illustrates the difficulty of calibrating gyrochronology models at these ages. Rather than a sequence of slow-rotating, $\approx$solar-mass stars as in Figure~\ref{mp}, we have a handful of lower-mass stars with which to anchor the comparison to the models. Still, it does appear that the model is significantly over-predicting the spin down for the 0.6-0.8~\Msun~stars, with the predicted median star in that mass range having a \Prot~=~21.0~d at 1.344~Gyr, $\approx$4.5~d more than what is observed.

The four lower-mass NGC 752 members have rotation periods that are more consistent with what is predicted by the \citet{Matt2015} model, with the median 0.4-0.6~\Msun\ star predicted to have a \Prot~=~17.2~d. One possible interpretation is that we are seeing the evolutionary stalling observed in the comparison of Praesepe and NGC 6811 for 0.8-0.9~\Msun\ stars shifted to lower masses, with the 0.6-0.8~\Msun~stars being the ones now rotating significantly faster than expected at this age.  

\subsection{Comparing magnetic activity in NGC 752 and in the Hyades and Praesepe}\label{act}
Studies of observational tracers of coronal or chromospheric activity have uncovered a mass-dependent transition between active and inactive stars in open clusters \citep[e.g.,][]{kafka2006,Douglas2014,Nunez2016,Nunez2017}. The dividing line between these two populations shifts to lower masses in older clusters, indicating that lower-mass stars possess longer activity lifetimes. For FGK stars, these lifetimes are $\leq$650 Myr, as calibrated by observations of open clusters younger than NGC 752 \citep{hawley1999,Douglas2014,Nunez2017}.

Extending such measurements to older open clusters is a primary motivation of this work. Our knowledge of the chromospheric activity lifetimes of lower-mass stars currently relies on indirect calibrations, such as modeling the vertical gradient in \Ha\ emission line strengths as a consequence of dynamical heating in the Galactic disk \citep[e.g.,][]{andy08}. 

\begin{figure}
\centerline{\includegraphics[width=1.\columnwidth]{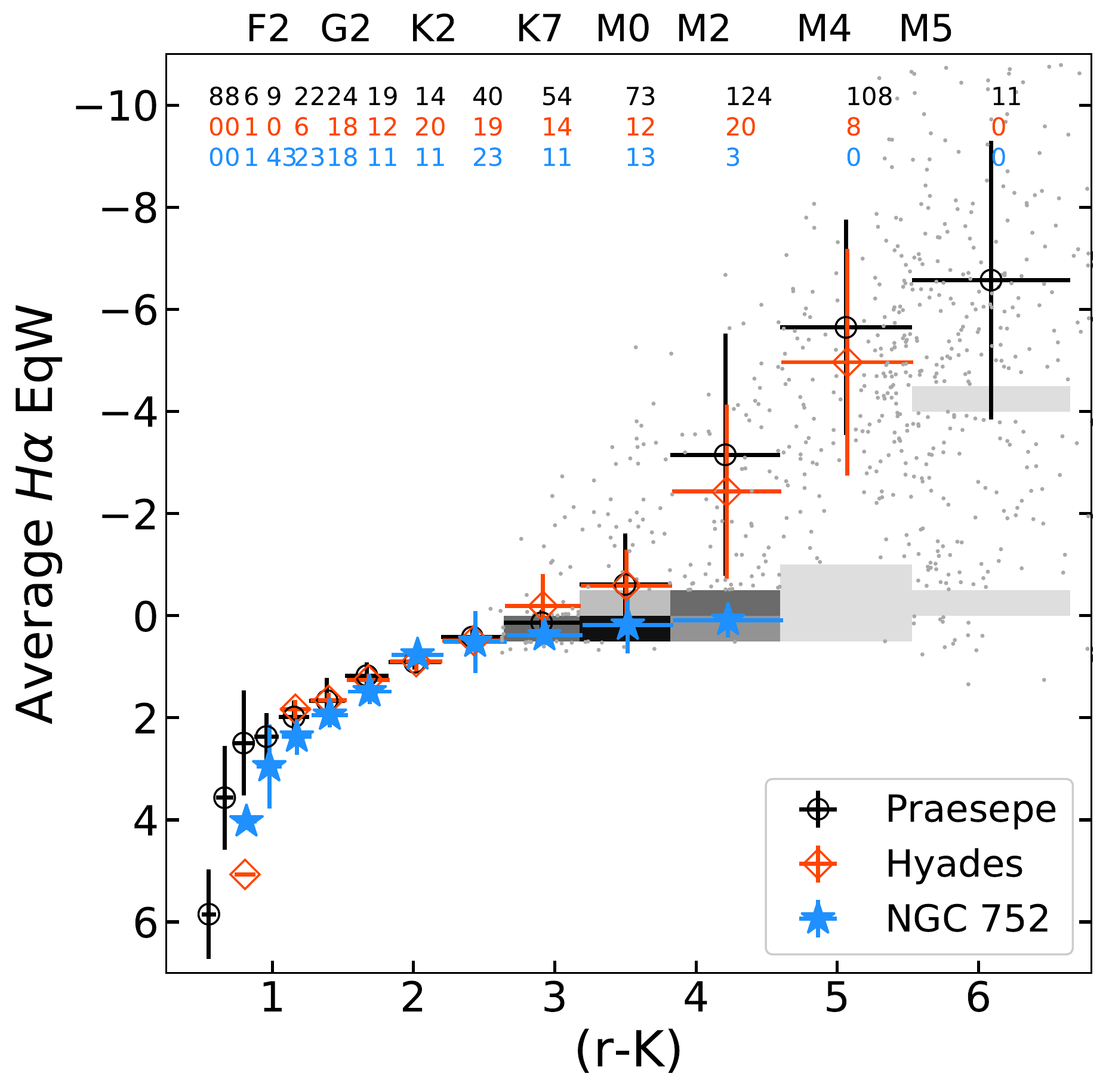}}
\caption{Average \Ha ~EqW as a function of logarithmically binned $(r-K)$ for stars in NGC 752 compared to the younger Praesepe and Hyades clusters. The number of stars in each bin is along the top; the vertical bars show the $\sigma$ for the bin, and the horizontal bars show the extent of the bin. Measurements of a comparison sample of $\approx$2800 M dwarf field stars with SDSS spectra and high-quality SDSS/2MASS photometry \citep[][]{covey2007,west2011} are shown as a grayscale histogram when more than 25 stars fell into a bin, and as points otherwise. The inactive region of the histogram includes 2059 stars. The stars in NGC 752 are clearly less active than those in Praesepe and the Hyades, with EqWs consistent with the field star distribution.} 
\label{halpha_spec}
\end{figure}

Our spectroscopic campaign confirms that the boundary between active and inactive stars has shifted well into the M dwarf regime in NGC 752.  As noted above, there are three stars in our spectroscopic sample with formal detections of \Ha\ emission, but we do not consider these stars indicative of the location of the active/inactive boundary in this cluster. As Figure~\ref{halpha_eqws} demonstrates, even the modest activity signatures measured from these stars (EqW $>$ 2~\AA) make them clear outliers from the dominant cluster locus, and their moderate membership probabilities (50 $<$~\Pmem~$<$ 80\%) indicate that these stars may not be {\it bona fide} cluster members. Calculating the mean \Ha\ EqW for NGC 752 members in bins of $(r-K)$, as shown in Figure \ref{halpha_spec}, indicates that there is no transition to activity in NGC 752, at least within the domain of our spectroscopic survey, which includes stars with spectral types as late as $\approx$M2.  

Indeed, the comparison of EqW loci in Figure~\ref{halpha_spec} demonstrates that the activity properties of NGC 752's early type (SpT $<$ M2) members are fully consistent with those of the largely inactive field star population. Comparing the NGC 752 stars with those in Praesepe and the Hyades, which exhibit a clear transition to active populations at $(r-K) \approx 4$, also indicates that the location of the active/inactive boundary shifts to lower masses as stars age from $<$1~Gyr to 1.3~Gyr. Our measurements are thus consistent with, but cannot fully test, the prediction based on the activity lifetimes calculated by \citet{andy08} that the active/inactive boundary in NGC 752 should lie at a spectral type of $\approx$M3.

\section{Conclusions}\label{concl}
We present an updated list of likely cluster members for NGC 752, one of the only nearby open clusters significantly older than the Hyades. Our catalog is constructed by supplementing the catalogs of \citet[][]{daniel1994, mermilliod1998} with candidates identified using updated photometric and proper motion criteria, and refined via radial velocity measurements. We produce a list of 258 probable cluster members, a $>$50\% increase over previous catalogs, and in particular provide the first high confidence list of late K- and M-dwarf members of the cluster. 

Using a Bayesian framework to fit MIST isochrones to literature photometry and the {\it Gaia} TGAS astrometry available for 59 NGC 752 members, we derive maximum likelihood mean parameters for the cluster. We find an age $=1.34$$\pm$$0.06$ Gyr, a distance = $438^{+8}_{-6}$ pc (DM = 8.21$^{+0.04}_{-0.03}$), a [Fe/H] $=0.02$$\pm$$0.01$, and an $A_V = 0.198^{+0.008}_{-0.009}$. These cluster parameters are in agreement with those of \cite{barta2011} and \citet{twarog2015} and have more robust uncertainty estimates.


We report on the results of our optical monitoring of the cluster. We targeted NGC 752 with PTF for five months in 2010-2011, producing light curves with 400-700 $R_{PTF}$ measurements for 90 cluster members. We use these these light curves to identify 12 high-confidence K and M cluster members with reliable \Prot\ measurements. These are the first periods measured for such low-mass stars with a well-constrained age $>$1 Gyr. 

We use data from the younger clusters Praesepe ($\approx$650~Myr) and NGC 6811 ($\approx$1 Gyr), and the \citet{Matt2015} models for angular-momentum loss, to place these new mass-period data in the broader context of stellar spin down. Our comparison of the mass-period data for Praesepe and NGC 6811 suggest that there may be a mass-dependent stalling of spin down, with $\approx$solar-mass stars losing angular momentum as predicted by a Skumanich-type spin-down law, whereas 0.8-0.9~\Msun\ stars do not appear to have spun down significantly over the $\approx$350 Myr that separate the two clusters. An alternative interpretation is that at least one of the ages for these two clusters is incorrect, as has already been argued for Praesepe by \citet{brandt2015}, who find its age to be closer to 800 Myr. 

The sparseness of the NGC 752 \Prot\ data make it difficult to draw strong conclusions from a comparison to the data for the younger clusters or to the \citet{Matt2015} model. Although it does seem that, on average, the NGC 752 stars are rotating more slowly than their younger counterparts, the difference is not significant, and in particular the lowest-mass stars in NGC 752 for which we measure \Prot\ are indistinguishable from their cousins in Praesepe. Comparisons with the \citet{Matt2015} model data suggest that the model over-predicts the angular momentum lost by K and early M stars over their first 1.3 Gyr; this excess in the predicted spin down for these stars was also observed when comparing the model predictions to the data for the younger clusters. On the other hand, the \citet{Matt2015} model systematically under-predicts the spin down of 0.4-0.6~\Msun\ stars at Praesepe's age, but the model \Prot\ are consistent with the \Prot\ measured for these stars at NGC 752's age. There are only four $<$0.6~\Msun\ NGC 752 stars for which we have these measurements, however. 

Finally, we discuss spectroscopic observations of over 270 candidate cluster members with the MDM 2.4-m and WIYN 3.5-m telescopes. Based on our measurements of \Ha, we find that NGC 752's stars are magnetically inactive at spectral types of $\approx$M2 and earlier, and indeed that these stars' activity properties are fully consistent with those of the largely inactive field-star population. Comparing the NGC 752 stars with those in Praesepe and the Hyades also indicates that the location of the active/inactive boundary shifts to lower masses as stars age from $<$1 Gyr to 1.3 Gyr. Our measurements are consistent with, but cannot fully test, the prediction of \citet{andy08} that the active/inactive boundary in NGC 752 should lie at a spectral type of $\approx$M3.

The fraction of NGC 752 members for which we measured \Prot, 13\%, is smaller than that we obtained in our PTF Pleiades campaign \citep[19\%;][]{covey16}, but is higher than that in our Praesepe campaign \cite[7\%;][]{agueros11}. This highlights the challenge in defining appropriate metrics for identifying robust \Prot\ measurements. These efforts are essential, however: while yields from satellite observations are much higher \citep[i.e., essentially 100\% for the Pleiades with {\it K2};][]{Rebull2016a}, an analysis by \citet{Douglas2017} of the properties of rotators in Praesepe suggested that {\it K2} was not uncovering rotators with smaller amplitudes than those identified from the ground. Even in the era of {\it K2} and (soon) the {\it Transiting Exoplanet Survey Satellite}, ground-based surveys of rotation in clusters still have an important role to play. And forthcoming {\it Gaia} data should solidify and extend the membership of NGC 752 to lower masses, thereby increasing its importance for studies of low-mass stars.

\acknowledgments 
We thank Eran Ofek for his help scheduling PTF observations. David Fierroz participated in the WIYN and MDM observing runs, and we thank him for his help in collecting the spectra presented here. We thank Jules Halpern and John Thorstensen for their help with the MDM observations, and the WIYN observing specialists for their assistance with the Hydra observations. Caty Pilachowski very generously shared the results of her RV monitoring of NGC 752, and Stanislava Barta\v si\={u}t\.{e} and Justas Zdanavi\v cius sent us their photometric catalog of candidate NGC 752 members, for which we thank them. We are grateful to Sean Matt for running many iterations of his model for us, enabling the comparisons presented here, and for his comments on our results. We than Jason Curtis and Bruce Twarog for providing detailed comments on a draft, and the anonymous referee for comments that improved the final paper. 

M.A.A.\ acknowledges support provided by the NSF through grant AST-1255419. P.A.C.\ acknowledges support provided by the NSF through grant AST-1109612. 

The MDM Observatory is operated by Dartmouth College, Columbia University, Ohio State University, Ohio University, and the University of Michigan. 

This paper is based on observations obtained with the Samuel Oschin Telescope as part of the Palomar Transient Factory project, a scientific collaboration between the California Institute of Technology, Columbia University, Las Cumbres Observatory, the Lawrence Berkeley National Laboratory, the National Energy Research Scientific Computing Center, the University of Oxford and the Weizmann Institute of Science.

This research has made use of NASA's Astrophysics Data System Bibliographic Services, the SIMBAD database, operated at CDS, Strasbourg, France, the NASA/IPAC Extragalactic Database, operated by the Jet Propulsion Laboratory, California Institute of Technology, under contract with the National Aeronautics and Space Administration, and the VizieR database of astronomical catalogs \citep{Ochsenbein2000}. 
 
IRAF is distributed by the National Optical Astronomy Observatories, which are operated by the Association of Universities for Research in Astronomy, Inc., under cooperative agreement with the National Science Foundation. PyRAF is a product of the Space Telescope Science Institute, which is operated by AURA for NASA.
 
The Two Micron All Sky Survey was a joint project of the University of Massachusetts and the Infrared Processing and Analysis Center (California Institute of Technology). The University of Massachusetts was responsible for the overall management of the project, the observing facilities and the data acquisition. The Infrared Processing and Analysis Center was responsible for data processing, data distribution and data archiving.  

\appendix

\begin{deluxetable}{lrrrr}
\tablewidth{0pt}
\tablecaption{SED Templates for USNO-B1.0 Photometry \label{SEDs} }
\tablehead{
\colhead{SpT} & \colhead{M$_B$} & \colhead{M$_{R}$} & \colhead{M$_{I}$} & \colhead{M$_{bol}$}
\\
\colhead{} & \colhead{(mag)} & \colhead{(mag)} & \colhead{(mag)} & \colhead{(mag)}
}
\startdata
B8 & $-$0.29 & $-$0.20 & $-$0.07 & $-$1.00\\
A0 & 0.57 & 0.56 & 0.63 & 0.30\\
A2 & 1.32 & 1.23 & 1.24 & 1.10\\
A5 & 1.98 & 1.79 & 1.74 & 1.75\\
A7 & 2.31 & 2.07 & 1.99 & 2.08\\
F0 & 2.87 & 2.50 & 2.38 & 2.61\\
F2 & 3.20 & 2.76 & 2.63 & 2.89\\
F5 & 4.00 & 3.39 & 3.28 & 3.61\\
F8 & 4.70 & 3.92 & 3.85 & 4.24\\
G0 & 5.05 & 4.08 & 4.03 & 4.47\\
G2 & 5.29 & 4.20 & 4.13 & 4.60\\
G5 & 5.68 & 4.46 & 4.33 & 4.89\\
G8 & 6.34 & 5.04 & 4.86 & 5.30\\
K0 & 6.72 & 5.29 & 5.09 & 5.69\\
K2 & 7.31 & 5.74 & 5.43 & 6.08\\
K4 & 8.02 & 6.27 & 5.89 & 6.55\\
K5 & 8.41 & 6.52 & 6.03 & 6.68\\
K7 & 9.05 & 6.93 & 6.34 & 6.89\\
M0 & 10.45 & 7.95 & 7.22 & 7.60\\
M1 & 11.07 & 8.42 & 7.51 & 7.97\\
M2 & 12.01 & 9.15 & 8.07 & 8.44\\
M3 & 13.07 & 10.11 & 8.73 & 9.09\\
M4 & 14.30 & 11.28 & 9.72 & 9.92\\
M5 & 16.02 & 12.91 & 10.95 & 11.01\\
M6 & 17.41 & 14.18 & 12.12 & 12.06\\
M7 & 18.72 & 15.35 & 13.03 & 12.70\\
M8 & 20.35 & 16.69 & 13.87 & 13.13\\
M9 & 20.93 & 17.23 & 14.37 & 13.43\\
L0 & 22.03 & 17.47 & 14.79 & 13.69
\enddata
\end{deluxetable}

\section*{SED Templates}\label{appendix}
We based our SED fitting procedures on those described by KH07, but since NGC 752 is not in the SDSS footprint, we extended the SED templates to use USNO-B1.0 photometry. We calculated the absolute magnitudes in the USNO-B1 filters (photographic $BRI$) by bootstrapping from our highly probable members of Praesepe and Coma Berenices, which span spectral types of A0-M7. For each star, we already had a measurement of m$_{bol}$ and SpT based on SED fits to SDSS and 2MASS photometry. We then downloaded the USNO-B1.0 magnitudes for those stars and computed the $(B-$m$_{bol})$, $(R-$m$_{bol})$, and $(I-$m$_{bol})$ colors. Finally, we calculated the average value for these colors for SpT bins of cluster members (i.e., G4.0-G6.4 to correspond to G5 stars, or M0.6-M1.5 to correspond to M1 stars) and combined them with the M$_{bol}$ absolute values from KH07 to compute the absolute magnitudes M$_{B}$, M$_{R}$, and M$_{I}$. 

For B8 stars, we linearly extrapolated the color-SpT relations of early A stars with respect to similar SDSS filters---($g'-B$), ($r'-R$), and ($i'-I$)---to compute absolute magnitudes from KH07. For the latest-type stars (M8-L0), we conducted a similar extrapolation on the colors of mid-M stars, then verified them by conducting SED fits on a sample of bright ultracool dwarfs \citep[from][]{leggett2002} that had photometry in both USNO-B1.0 and SDSS. There were too few ultracool dwarfs with photometry in USNO-B1.0 to justify fitting color relations to those data, but the measurements sufficed to confirm that the extrapolation from mid-M stars was valid. We give M$_B$, M$_R$, M$_I$, and M$_{bol}$ as a function of SpT in Table~\ref{SEDs}.

Based on the scatter in colors between very similar filters (i.e., ($i'-I$) and ($r'-R$)) in color-SpT relations for our sample of open cluster members, we estimate that the typical photometric uncertainty for USNO-B1.0 magnitudes is $\sigma \approx 0.25$ mag. Differences in the emulsions used for the original photographic plates also will introduce some color terms; for example, POSS-I conducted $B$ ``filtered'' observations with a Kodak 103a-O emulsion and no filter, while POSS-II used Kodak IIIaJ emulsions with a GG385 filter. The corresponding southern surveys that contribute to USNO-B1.0 (which are not relevant to our survey, but could be interpreted using the same SEDs) also used Kodak IIIaJ emulsions, but with a slightly redder GG385 filter. The color terms appear to be small compared to the uncertainty for individual stars, so we computed a single calibration for all versions of $B$, $R$, and $I$. However, the color terms could introduce small systematic uncertainties in SED fits for stellar populations. 

\begin{ThreePartTable}
\begin{longtable*}{lcr@{$\pm$}lr@{$\pm$}lcccccc}
\tablecaption{\Pmem\ Selected NGC 752 Members}
\tablehead{
 \colhead{2MASS ID} &
 \colhead{Platais ID} &
 \multicolumn{2}{c}{$J$} &
 \multicolumn{2}{c}{$K$} &
 \colhead{Mass} &
 \colhead{SpT} &
 \colhead{DM} &
 \colhead{M$_{bol}$} &
 \colhead{Binary?\tnote{a}} &
 \colhead{\Pmem}\\
 \colhead{} & 
 \colhead{} &
 \multicolumn{2}{c}{(mag)} & 
 \multicolumn{2}{c}{(mag)} & 
 \colhead{(\Msun)} & 
 \colhead{} & 
 \colhead{(mag)} & 
 \colhead{(mag)} &
 \colhead{} &   
 \colhead{(\%)} 
}
01501676$+$3812369 & \nodata & 9.96 & 0.02 & 9.73 & 0.02 & 1.24 & F3.4 & 7.38$\pm$0.22 & 10.60$\pm$0.10 & \nodata & 54.6 \\
01523927$+$3822334 & \nodata & 10.79 & 0.02 & 10.53 & 0.02 & 1.06 & F7.4 & 7.58$\pm$0.20 & 11.70$\pm$0.08 & \nodata & 54.4 \\
01524348$+$3724497 & \nodata & 10.57 & 0.02 & 10.3 & 0.02 & 1.04 & F8.0 & 7.28$\pm$0.16 & 11.52$\pm$0.05 & \nodata & 93.7 \\
01524372$+$3808381 & \nodata & 11.59 & 0.02 & 11.12 & 0.02 & 0.82 & G9.3 & 7.17$\pm$0.04 & 12.73$\pm$0.05 & \nodata & 52.7 \\
01525891$+$3803515 & \nodata & 12.51 & 0.02 & 11.86 & 0.02 & 0.69 & K4.7 & 7.37$\pm$0.07 & 14.01$\pm$0.02 & \nodata & 52.8 \\
01531903$+$3759057 & \nodata & 11.75 & 0.02 & 11.41 & 0.02 & 0.96 & G3.5 & 8.02$\pm$0.07 & 12.76$\pm$0.04 & \nodata & 51.4 \\
01532120$+$3735162 & \nodata & 10.86 & 0.02 & 10.58 & 0.02 & 1.17 & F4.9 & 8.01$\pm$0.22 & 11.59$\pm$0.10 & \nodata & 97.7 \\
01533728$+$3724173 & \nodata & 11.45 & 0.02 & 11.11 & 0.03 & 0.82 & G8.3 & 7.19$\pm$0.08 & 12.55$\pm$0.11 & \nodata & 61.1 \\
01534317$+$3743224 & \nodata & 12.95 & 0.03 & 12.35 & 0.03 & 0.73 & K2.9 & 8.03$\pm$0.08 & 14.32$\pm$0.07 & \nodata & 57.4 \\
01535762$+$3756556 & \nodata & 14.56 & 0.04 & 13.67 & 0.04 & 0.51 & M1.0 & 8.23$\pm$0.14 & 16.20$\pm$0.04 & \nodata & 73.0 \\
01542643$+$3737321 & \nodata & 14.13 & 0.03 & 13.27 & 0.04 & 0.58 & \nodata & 8.30$\pm$0.14 & 15.61$\pm$0.05 & \nodata & 78.0 \\
01543105$+$3729316 & \nodata & 10.34 & 0.02 & 10.15 & 0.02 & 1.17 & F4.6 & 7.59$\pm$0.05 & 11.10$\pm$0.03 & \nodata & 98.7 \\
01543493$+$3751183 & \nodata & 14.24 & 0.03 & 13.33 & 0.03 & 0.57 & K7.8 & 8.28$\pm$0.14 & 15.73$\pm$0.05 & \nodata & 59.5 \\
01543660$+$3726262 & \nodata & 11.82 & 0.02 & 11.45 & 0.02 & 0.81 & G9.1 & 7.48$\pm$0.04 & 12.99$\pm$0.06 & \nodata & 65.9 \\
01544130$+$3737590 & \nodata & 14.73 & 0.03 & 13.81 & 0.04 & 0.51 & \nodata & 8.36$\pm$0.18 & 16.33$\pm$0.04 & \nodata & 76.1 \\
01544738$+$3749590 & \nodata & 14.68 & 0.04 & 13.87 & 0.04 &  0.49  & M1.8 & 8.06$\pm$0.17 & 16.41$\pm$0.03 & \nodata & 57.2 \\
01544887$+$3734216 & \nodata & 11.81 & 0.02 & 11.3 & 0.02 & 0.72 & K2.8 & 6.96$\pm$0.08 & 13.23$\pm$0.08 & \nodata & 55.4 \\
01545054$+$3812578 & \nodata & 10.81 & 0.02 & 10.59 & 0.02 & 1.11 & F5.9 & 7.84$\pm$0.21 & 11.64$\pm$0.10 & \nodata & 93.2 \\
01545549$+$3722500 & \nodata & 14.03 & 0.03 & 13.14 & 0.03 & 0.55 & M0.0 & 7.99$\pm$0.07 & 15.59$\pm$0.03 & \nodata & 69.6 \\
01545672$+$3758289 & 215 & 10.61 & 0.02 & 10.37 & 0.02 & 1.15 & F5.2 & 7.73$\pm$0.21 & 11.38$\pm$0.10 & \nodata & 99.3 \\
01545773$+$3811013 & \nodata & 10.22 & 0.02 & 10.04 & 0.02 & 1.23 & F3.4 & 7.67$\pm$0.22 & 10.90$\pm$0.10 & \nodata & 99.1 \\
01545965$+$3728599 & \nodata & 10.6 & 0.02 & 10.29 & 0.02 & 1.07 & F7.6 & 7.35$\pm$0.21 & 11.51$\pm$0.08 & \nodata & 89.8 \\
01550283$+$3816383 & \nodata & 11.36 & 0.02 & 11.09 & 0.02 & 1.03 & F8.6 & 8.00$\pm$0.21 & 12.31$\pm$0.06 & \nodata & 54.5 \\
01551140$+$3759524 & \nodata & 11.47 & 0.02 & 11.18 & 0.02 & 1.06 & F7.5 & 8.24$\pm$0.49 & 12.37$\pm$0.19 & \nodata & 92.4 \\
01552488$+$3755032 & \nodata & 12.96 & 0.02 & 12.39 & 0.02 & 0.72 & K3.0 & 8.04$\pm$0.08 & 14.36$\pm$0.07 & \nodata & 63.1 \\
01552616$+$3808222 & 300 & 12.18 & 0.02 & 11.74 & 0.02 & 0.78 & K0.6 & 7.64$\pm$0.13 & 13.44$\pm$0.07 & \nodata & 72.7 \\
01553162$+$3826370 & \nodata & 11.86 & 0.02 & 11.54 & 0.02 & 0.82 & G8.6 & 7.59$\pm$0.03 & 13.01$\pm$0.04 & \nodata & 51.8 \\
01553694$+$3722130 & \nodata & 13.21 & 0.02 & 12.51 & 0.02 & 0.64 & K5.8 & 7.83$\pm$0.06 & 14.67$\pm$0.02 & \nodata & 56.7 \\
01553777$+$3750567 & \nodata & 13.43 & 0.02 & 12.61 & 0.02 & 0.58 & K7.6 & 7.63$\pm$0.11 & 14.95$\pm$0.04 & \nodata & 91.8 \\
01553948$+$3807379 & \nodata & 14.01 & 0.03 & 13.14 & 0.03 & 0.59 & K7.5 & 8.22$\pm$0.11 & 15.47$\pm$0.04 & \nodata & 81.7 \\
01554062$+$3749375 & \nodata & 13.07 & 0.02 & 12.44 & 0.02 & 0.70 & K4.1 & 8.01$\pm$0.07 & 14.57$\pm$0.05 & \nodata & 88.7 \\
01554493$+$3808213 & 363 & 9.62 & 0.02 & 9.41 & 0.02 & 1.15 & F5.1 & 6.78$\pm$0.21 & 10.41$\pm$0.10 & P & 98.9 \\
01554817$+$3846508 & \nodata & 9.77 & 0.02 & 9.57 & 0.02 & 1.22 & F3.9 & 7.13$\pm$0.22 & 10.47$\pm$0.10 & \nodata & 87.0 \\
01555154$+$3756270 & \nodata & 15.49 & 0.05 & 14.49 & 0.07 &  0.44  & \nodata & 8.43$\pm$0.22 & 17.19$\pm$0.04 & \nodata & 54.8 \\
01555302$+$3747219 & \nodata & 13.71 & 0.03 & 12.94 & 0.03 & 0.66 & \nodata & 8.34$\pm$0.23 & 15.15$\pm$0.05 & \nodata & 94.2 \\
01555313$+$3756368 & \nodata & 13.29 & 0.02 & 12.58 & 0.02 & 0.68 & K5.1 & 8.07$\pm$0.09 & 14.77$\pm$0.02 & \nodata & 84.7 \\
01555351$+$3749267 & 391 & 12.45 & 0.02 & 11.98 & 0.02 & 0.78 & K0.6 & 7.90$\pm$0.12 & 13.70$\pm$0.08 & \nodata & 88.7 \\
01555801$+$3801028 & 406 & 13.11 & 0.02 & 12.48 & 0.02 & 0.70 & K4.0 & 8.04$\pm$0.07 & 14.59$\pm$0.05 & \nodata & 78.1 \\
01555942$+$3740486 & 413 & 11.18 & 0.02 & 10.84 & 0.02 & 0.99 & G1.0 & 7.59$\pm$0.45 & 12.13$\pm$0.17 & \nodata & 87.0 \\
01560091$+$3745139 & \nodata & 14.02 & 0.03 & 13.23 & 0.03 & 0.57 & K7.6 & 8.22$\pm$0.14 & 15.53$\pm$0.05 & \nodata & 72.8 \\
01560288$+$3820453 & \nodata & 13.04 & 0.02 & 12.46 & 0.02 & 0.71 & K3.6 & 8.05$\pm$0.07 & 14.51$\pm$0.06 & \nodata & 73.8 \\
01560368$+$3759224 & 435 & 10.57 & 0.02 & 10.32 & 0.02 & 1.27 & F2.9 & 8.06$\pm$0.21 & 11.16$\pm$0.09 & D & 99.6 \\
01560380$+$3726267 & \nodata & 13.08 & 0.02 & 12.34 & 0.02 & 0.67 & K5.4 & 7.78$\pm$0.09 & 14.54$\pm$0.02 & \nodata & 66.7 \\
01560381$+$3745595 & \nodata & 14.86 & 0.04 & 14.08 & 0.05 &  0.52  & \nodata & 8.41$\pm$0.15 & 16.57$\pm$0.03 & \nodata & 90.0 \\
01560469$+$3746478 & \nodata & 13.95 & 0.03 & 13.13 & 0.03 & 0.58 & K7.6 & 8.14$\pm$0.14 & 15.46$\pm$0.05 & \nodata & 94.3 \\
01560705$+$3740529 & \nodata & 15 & 0.05 & 14.1 & 0.05 &  0.40  & M2.8 & 7.80$\pm$0.23 & 16.76$\pm$0.03 & \nodata & 70.2 \\
01560754$+$3733285 & 444 & 13.06 & 0.02 & 12.46 & 0.02 & 0.71 & K3.5 & 8.07$\pm$0.07 & 14.50$\pm$0.07 & \nodata & 85.0 \\
01560879$+$3726484 & \nodata & 14.95 & 0.04 & 14.03 & 0.05 &  0.52  & \nodata & 8.38$\pm$0.15 & 16.63$\pm$0.03 & \nodata & 80.8 \\
01560895$+$3739526 & 455 & 9.66 & 0.02 & 9.44 & 0.01 & 1.14 & F5.3 & 6.78$\pm$0.20 & 10.45$\pm$0.10 & P & 99.7 \\
01561009$+$3726459 & \nodata & 13.66 & 0.03 & 12.79 & 0.02 & 0.56 & K7.8 & 7.72$\pm$0.13 & 15.17$\pm$0.04 & \nodata & 64.6 \\
01561109$+$3745114 & \nodata & 10.41 & 0.02 & 10.19 & 0.02 & 1.21 & F4.0 & 7.74$\pm$0.21 & 11.10$\pm$0.10 & D & 99.5 \\
01561141$+$3755057 & \nodata & 12.59 & 0.02 & 11.88 & 0.02 & 0.66 & K5.6 & 7.25$\pm$0.09 & 14.06$\pm$0.02 & \nodata & 92.9 \\
01561287$+$3801433 & 472 & 10.12 & 0.02 & 9.87 & 0.02 & 1.17 & F4.8 & 7.28$\pm$0.21 & 10.85$\pm$0.10 & D & 99.6 \\
01561427$+$3758144 & \nodata & 10.01 & 0.02 & 9.78 & 0.02 & 1.30 & F2.2 & 7.62$\pm$0.19 & 10.55$\pm$0.08 & \nodata & 99.6 \\
01561447$+$3749071 & \nodata & 12.72 & 0.02 & 11.97 & 0.02 & 0.64 & K5.9 & 7.30$\pm$0.11 & 14.17$\pm$0.02 & \nodata & 94.4 \\
01561789$+$3757110 & \nodata & 16.11 & 0.09 & 15.68 & 0.2 &  0.23  & M3.8 & 8.26$\pm$0.41 & 18.02$\pm$0.06 & \nodata & 58.5 \\
01561863$+$3737393 & 505 & 9.9 & 0.02 & 9.669 & 0.02 & 1.22 & F3.7 & 7.26$\pm$0.21 & 10.56$\pm$0.09 & DP & 99.5 \\
01562250$+$3833549 & \nodata & 14.59 & 0.03 & 13.74 & 0.04 &  0.50  & M1.7 & 8.00$\pm$0.16 & 16.30$\pm$0.03 & \nodata & 60.2 \\
01562255$+$3739180 & \nodata & 12.68 & 0.02 & 12.16 & 0.02 & 0.74 & K2.3 & 7.88$\pm$0.09 & 14.03$\pm$0.07 & \nodata & 87.7 \\
01562309$+$3738031 & \nodata & 11.68 & 0.02 & 11.34 & 0.02 & 0.88 & G6.6 & 7.63$\pm$0.22 & 12.73$\pm$0.02 & \nodata & 86.4 \\
01562427$+$3709287 & \nodata & 14.42 & 0.03 & 13.52 & 0.04 & 0.52 & M0.9 & 8.13$\pm$0.06 & 16.06$\pm$0.03 & \nodata & 72.4 \\
01562547$+$3830461 & \nodata & 10.77 & 0.02 & 10.51 & 0.02 & 1.19 & F4.5 & 7.99$\pm$0.21 & 11.48$\pm$0.09 & \nodata & 98.1 \\
01562690$+$3733454 & \nodata & 14.29 & 0.03 & 13.37 & 0.03 & 0.54 & M0.2 & 8.17$\pm$0.07 & 15.85$\pm$0.03 & \nodata & 87.5 \\
01563180$+$3744582 & \nodata & 14.04 & 0.03 & 13.2 & 0.03 & 0.57 & K7.7 & 8.18$\pm$0.11 & 15.57$\pm$0.04 & \nodata & 94.4 \\
01563295$+$3756465 & 555 & 10.77 & 0.02 & 10.48 & 0.02 & 0.99 & G0.0 & 7.26$\pm$0.56 & 11.73$\pm$0.22 & DP & 88.2 \\
01563337$+$3749050 & \nodata & 14.04 & 0.03 & 13.2 & 0.03 & 0.56 & K7.8 & 8.11$\pm$0.11 & 15.57$\pm$0.04 & \nodata & 96.3 \\
01563390$+$3737502 & \nodata & 13.29 & 0.02 & 12.62 & 0.02 & 0.69 & K4.9 & 8.12$\pm$0.08 & 14.78$\pm$0.02 & \nodata & 92.9 \\
01563524$+$3757419 & \nodata & 14.2 & 0.03 & 13.32 & 0.03 &  0.49  & M1.8 & 7.55$\pm$0.14 & 15.89$\pm$0.03 & \nodata & 94.0 \\
01563686$+$3745127 & 575 & 12.38 & 0.02 & 11.94 & 0.02 & 0.79 & K0.4 & 7.88$\pm$0.12 & 13.64$\pm$0.09 & \nodata & 88.8 \\
01563867$+$3654342 & \nodata & 10.64 & 0.02 & 10.4 & 0.02 & 1.06 & F7.7 & 7.42$\pm$0.06 & 11.60$\pm$0.04 & \nodata & 89.4 \\
01563921$+$3751411 & \nodata & 9.63 & 0.02 & 9.416 & 0.02 & 1.23 & F3.6 & 7.02$\pm$0.21 & 10.29$\pm$0.09 & \nodata & 99.9 \\
01563985$+$3716293 & \nodata & 12.15 & 0.02 & 11.75 & 0.02 & 0.73 & K2.2 & 7.43$\pm$0.03 & 13.56$\pm$0.03 & \nodata & 75.2 \\
01564576$+$3736496 & \nodata & 14.61 & 0.03 & 13.78 & 0.04 & 0.51 & \nodata & 8.36$\pm$0.17 & 16.25$\pm$0.04 & \nodata & 93.9 \\
01564701$+$3740305 & \nodata & 13.42 & 0.02 & 12.75 & 0.03 & 0.66 & K5.4 & 8.15$\pm$0.09 & 14.92$\pm$0.02 & \nodata & 96.5 \\
01564860$+$3729114 & 622 & 9.71 & 0.02 & 9.533 & 0.02 & 1.27 & F2.8 & 7.26$\pm$0.21 & 10.34$\pm$0.09 & \nodata & 99.7 \\
01565045$+$3739359 & \nodata & 13.63 & 0.02 & 12.77 & 0.02 & 0.52 & M0.7 & 7.40$\pm$0.16 & 15.26$\pm$0.04 & \nodata & 88.1 \\
01565339$+$3753502 & \nodata & 13.33 & 0.03 & 12.62 & 0.03 & 0.68 & K5.1 & 8.11$\pm$0.09 & 14.81$\pm$0.02 & \nodata & 94.9 \\
01565432$+$3723521 & 648 & 10.95 & 0.02 & 10.65 & 0.02 & 1.27 & \nodata & 8.39$\pm$0.22 & 11.50$\pm$0.10 & \nodata & 98.2 \\
01565491$+$3727055 & \nodata & 13.91 & 0.03 & 13.1 & 0.03 & 0.52 & M0.5 & 7.75$\pm$0.20 & 15.54$\pm$0.05 & \nodata & 68.1 \\
01565530$+$3734416 & \nodata & 14.43 & 0.03 & 13.5 & 0.03 & 0.55 & \nodata & 8.37$\pm$0.16 & 15.97$\pm$0.05 & \nodata & 88.2 \\
01565531$+$3736463 & \nodata & 13.38 & 0.02 & 12.66 & 0.02 & 0.68 & K5.2 & 8.13$\pm$0.09 & 14.85$\pm$0.02 & \nodata & 95.2 \\
01565758$+$3723206 & 667 & 10.13 & 0.02 & 9.925 & 0.02 & 1.34 & F1.2 & 7.87$\pm$0.19 & 10.65$\pm$0.09 & \nodata & 93.8 \\
01565952$+$3824559 & \nodata & 12.82 & 0.02 & 12.25 & 0.02 & 0.74 & K2.3 & 7.99$\pm$0.10 & 14.14$\pm$0.08 & \nodata & 58.7 \\
01565981$+$3804466 & \nodata & 15.9 & 0.08 & 15.07 & 0.11 &  0.30  & M3.6 & 8.15$\pm$0.40 & 17.73$\pm$0.05 & \nodata & 62.6 \\
01570025$+$3747243 & \nodata & 15.61 & 0.07 & 14.73 & 0.08 &  0.39  & \nodata & 8.41$\pm$0.27 & 17.37$\pm$0.04 & \nodata & 68.4 \\
01570057$+$3746131 & \nodata & 14.02 & 0.03 & 13.16 & 0.03 &  0.51  & M1.6 & 7.47$\pm$0.08 & 15.72$\pm$0.03 & \nodata & 94.2 \\
01570249$+$3753077 & 682 & 10.33 & 0.02 & 10.17 & 0.02 & 1.17 & F4.5 & 7.61$\pm$0.22 & 11.10$\pm$0.10 & \nodata & 99.5 \\
01570276$+$3748143 & \nodata & 13.42 & 0.02 & 12.65 & 0.02 & 0.67 & K5.6 & 8.07$\pm$0.10 & 14.87$\pm$0.02 & \nodata & 52.4 \\
01570317$+$3755447 & \nodata & 10.83 & 0.02 & 10.58 & 0.02 & 1.20 & F4.3 & 8.09$\pm$0.21 & 11.53$\pm$0.09 & \nodata & 99.6 \\
01570352$+$3736549 & \nodata & 15.97 & 0.08 & 15.02 & 0.11 &  0.32  & M3.5 & 8.27$\pm$0.39 & 17.78$\pm$0.05 & \nodata & 73.4 \\
01570362$+$3805118 & \nodata & 10.72 & 0.02 & 10.43 & 0.02 & 1.22 & F3.8 & 8.03$\pm$0.25 & 11.35$\pm$0.11 & \nodata & 99.5 \\
01570546$+$3750428 & 701 & 11.68 & 0.02 & 11.24 & 0.02 & 0.77 & K0.9 & 7.11$\pm$0.13 & 12.98$\pm$0.06 & \nodata & 69.3 \\
01570780$+$3749279 & \nodata & 13.78 & 0.02 & 12.96 & 0.02 & 0.58 & K7.6 & 7.98$\pm$0.14 & 15.29$\pm$0.05 & \nodata & 98.0 \\
01570853$+$3750213 & \nodata & 14.23 & 0.03 & 13.36 & 0.03 &  0.37  & M3.0 & 6.94$\pm$0.28 & 16.03$\pm$0.03 & \nodata & 81.4 \\
01571048$+$3802065 & 720 & 11.32 & 0.02 & 11.01 & 0.02 & 1.06 & F7.6 & 8.06$\pm$0.25 & 12.21$\pm$0.08 & \nodata & 78.6 \\
01571052$+$3727267 & \nodata & 12.03 & 0.02 & 11.48 & 0.02 & 0.72 & K2.9 & 7.13$\pm$0.07 & 13.42$\pm$0.07 & \nodata & 86.2 \\
01571205$+$3834016 & \nodata & 12.24 & 0.02 & 11.89 & 0.02 & 0.89 & G6.3 & 8.23$\pm$0.45 & 13.30$\pm$0.08 & \nodata & 52.3 \\
01571216$+$3756048 & 731 & 10.72 & 0.02 & 10.33 & 0.02 & 0.94 & G5.0 & 6.86$\pm$0.32 & 11.75$\pm$0.06 & DP & 90.1 \\
01571517$+$3755377 & \nodata & 14.4 & 0.03 & 13.51 & 0.04 &  0.51  & M1.6 & 7.84$\pm$0.15 & 16.09$\pm$0.03 & \nodata & 92.9 \\
01571577$+$3744134 & \nodata & 15.43 & 0.05 & 14.7 & 0.08 &  0.37  & M2.9 & 8.25$\pm$0.32 & 17.27$\pm$0.04 & \nodata & 87.9 \\
01571727$+$3729210 & \nodata & 15.63 & 0.06 & 14.87 & 0.1 &  0.34  & M3.2 & 8.23$\pm$0.32 & 17.49$\pm$0.05 & \nodata & 64.3 \\
01571941$+$3759235 & 768 & 11.08 & 0.02 & 10.83 & 0.02 & 1.12 & F5.8 & 8.10$\pm$0.24 & 11.88$\pm$0.11 & \nodata & 97.0 \\
01572038$+$3747404 & \nodata & 15.79 & 0.07 & 14.87 & 0.09 &  0.27  & M3.9 & 7.76$\pm$0.63 & 17.60$\pm$0.05 & \nodata & 85.9 \\
01572074$+$3723159 & \nodata & 13.43 & 0.02 & 12.74 & 0.02 & 0.63 & K5.9 & 8.02$\pm$0.05 & 14.89$\pm$0.02 & \nodata & 84.8 \\
01572260$+$3732585 & \nodata & 13.86 & 0.03 & 13 & 0.03 & 0.56 & K7.9 & 7.88$\pm$0.14 & 15.41$\pm$0.04 & \nodata & 79.1 \\
01572291$+$3757466 & \nodata & 14.61 & 0.03 & 13.76 & 0.04 &  0.39  & M2.8 & 7.45$\pm$0.23 & 16.41$\pm$0.03 & \nodata & 84.3 \\
01572379$+$3752119 & 790 & 11.11 & 0.02 & 10.75 & 0.02 & 0.94 & G4.5 & 7.31$\pm$0.15 & 12.15$\pm$0.05 & \nodata & 91.9 \\
01572409$+$3717151 & \nodata & 12.53 & 0.02 & 11.88 & 0.02 & 0.63 & K5.9 & 7.16$\pm$0.05 & 14.03$\pm$0.02 & \nodata & 86.2 \\
01572493$+$3740013 & \nodata & 15.45 & 0.05 & 14.56 & 0.07 &  0.43  & \nodata & 8.44$\pm$0.25 & 17.20$\pm$0.04 & \nodata & 82.6 \\
01572599$+$3743197 & 798 & 9.57 & 0.02 & 9.322 & 0.02 & 1.17 & F4.9 & 6.74$\pm$0.21 & 10.32$\pm$0.10 & DP & 99.9 \\
01572615$+$3739202 & 799 & 10.41 & 0.02 & 10.18 & 0.02 & 1.28 & F2.7 & 7.94$\pm$0.39 & 11.00$\pm$0.20 & \nodata & 99.0 \\
01572738$+$3723085 & \nodata & 14.82 & 0.04 & 13.84 & 0.04 &  0.49  & M2.0 & 8.06$\pm$0.20 & 16.50$\pm$0.04 & \nodata & 61.3 \\
01572746$+$3735104 & 806 & 9.95 & 0.02 & 9.748 & 0.02 & 1.16 & F4.9 & 7.15$\pm$0.21 & 10.73$\pm$0.10 & \nodata & 99.7 \\
01573184$+$3753407 & 824 & 10.72 & 0.02 & 10.45 & 0.02 & 1.32 & \nodata & 8.33$\pm$0.39 & 11.24$\pm$0.18 & \nodata & 99.8 \\
01573256$+$3742058 & \nodata & 12.45 & 0.03 & 12.05 & \nodata & 0.78 & K0.2 & 7.95$\pm$0.16 & 13.68$\pm$0.10 & D & 75.7 \\
01573280$+$3746007 & \nodata & 16.06 & 0.09 & 15.07 & 0.11 &  0.34  & \nodata & 8.42$\pm$0.16 & 17.76$\pm$0.05 & \nodata & 90.1 \\
01573476$+$3749063 & \nodata & 14.56 & 0.04 & 13.62 & 0.03 &  0.51  & M1.6 & 7.95$\pm$0.19 & 16.21$\pm$0.04 & \nodata & 94.2 \\
01573646$+$3705082 & \nodata & 11.76 & 0.02 & 11.31 & 0.02 & 0.73 & K2.2 & 7.01$\pm$0.10 & 13.14$\pm$0.07 & \nodata & 51.8 \\
01573777$+$3749505 & 859 & 11.73 & 0.02 & 11.24 & 0.02 & 0.74 & K2.2 & 6.97$\pm$0.13 & 13.10$\pm$0.09 & \nodata & 90.3 \\
01573875$+$3808303 & 864 & 11.71 & 0.02 & 11.37 & 0.02 & 0.87 & G6.7 & 7.64$\pm$0.50 & 12.76$\pm$0.12 & \nodata & 58.9 \\
01573944$+$3752259 & 868 & 9.68 & 0.02 & 9.418 & 0.02 & 1.16 & F5.0 & 6.82$\pm$0.22 & 10.43$\pm$0.10 & \nodata & 99.7 \\
01574051$+$3726489 & \nodata & 13.59 & 0.02 & 12.83 & 0.02 & 0.62 & K7.1 & 8.06$\pm$0.12 & 15.02$\pm$0.04 & \nodata & 93.8 \\
01574053$+$3752441 & \nodata & 13.39 & 0.02 & 12.49 & 0.02 &  0.49  & M1.9 & 6.68$\pm$0.19 & 15.07$\pm$0.03 & \nodata & 71.8 \\
01574283$+$3818167 & \nodata & 12.43 & 0.02 & 11.58 & 0.02 & 0.57 & K7.7 & 6.56$\pm$0.14 & 13.94$\pm$0.05 & \nodata & 83.1 \\
01574529$+$3748024 & \nodata & 16.21 & 0.1 & 15.11 & 0.12 &  0.30  & M3.7 & 8.25$\pm$0.41 & 17.92$\pm$0.06 & \nodata & 60.2 \\
01574568$+$3754297 & \nodata & 14.21 & 0.03 & 13.34 & 0.03 & 0.59 & \nodata & 8.44$\pm$0.26 & 15.68$\pm$0.07 & \nodata & 94.4 \\
01574604$+$3804284 & 897 & 9.653 & 0.02 & 9.407 & 0.01 & 1.13 & F5.7 & 6.70$\pm$0.21 & 10.46$\pm$0.10 & DP & 99.8 \\
01574713$+$3747303 & 901 & 10.12 & 0.02 & 9.897 & 0.02 & 1.24 & F3.4 & 7.55$\pm$0.21 & 10.78$\pm$0.09 & \nodata & 99.9 \\
01575049$+$3737559 & \nodata & 14.72 & 0.04 & 13.83 & 0.04 &  0.36 & M3.1 & 7.33$\pm$0.25 & 16.50$\pm$0.03 & \nodata & 64.3 \\
01575125$+$3828010 & \nodata & 13.25 & 0.02 & 12.57 & 0.02 & 0.69 & K4.9 & 8.08$\pm$0.08 & 14.74$\pm$0.02 & \nodata & 71.5 \\
01575140$+$3739525 & \nodata & 12.59 & 0.02 & 12.08 & 0.02 & 0.78 & K0.9 & 7.98$\pm$0.12 & 13.85$\pm$0.06 & \nodata & 93.8 \\
01575512$+$3803041 & \nodata & 14.38 & 0.03 & 13.48 & 0.03 & 0.56 & \nodata & 8.39$\pm$0.16 & 15.92$\pm$0.05 & \nodata & 90.3 \\
01575517$+$3752461 & \nodata & 10.75 & 0.02 & 10.52 & 0.02 & 1.23 & F3.6 & 8.13$\pm$0.21 & 11.40$\pm$0.09 & \nodata & 99.8 \\
01575572$+$3827482 & \nodata & 13.31 & 0.02 & 12.62 & 0.02 & 0.67 & K5.2 & 8.07$\pm$0.09 & 14.79$\pm$0.02 & \nodata & 79.7 \\
01575643$+$3750011 & 941 & 9.81 & 0.02 & 9.556 & 0.02 & 1.18 & F4.7 & 7.01$\pm$0.21 & 10.54$\pm$0.10 & \nodata & 99.8 \\
01575711$+$3722144 & \nodata & 13.41 & 0.02 & 12.72 & 0.02 & 0.60 & K7.2 & 7.87$\pm$0.25 & 14.90$\pm$0.05 & \nodata & 87.0 \\
01575777$+$3748224 & \nodata & 10.39 & 0.02 & 10.16 & 0.02 & 1.10 & F6.4 & 7.36$\pm$0.20 & 11.26$\pm$0.10 & D & 99.9 \\
01575883$+$3741269 & 953 & 11.14 & 0.02 & 10.79 & 0.02 & 0.96 & G3.1 & 7.43$\pm$0.17 & 12.13$\pm$0.06 & DP & 83.5 \\
01580275$+$3802305 & 964 & 11.66 & 0.02 & 11.27 & 0.02 & 0.85 & G7.7 & 7.44$\pm$0.06 & 12.70$\pm$0.06 & \nodata & 88.5 \\
01580314$+$3821510 & \nodata & 14.03 & 0.03 & 13.17 & 0.03 & 0.51 & M1.0 & 7.71$\pm$0.15 & 15.68$\pm$0.04 & \nodata & 65.9 \\
01580629$+$3738068 & \nodata & 11.75 & 0.02 & 11.31 & 0.02 & 0.81 & G9.7 & 7.32$\pm$0.12 & 12.95$\pm$0.12 & \nodata & 83.3 \\
01580647$+$3800323 & \nodata & 14.34 & 0.03 & 13.7 & 0.04 &  0.39 & K4.8 & 7.39$\pm$1.20 & 14.04$\pm$0.45 & \nodata & 82.8 \\
01580769$+$3739571 & \nodata & 10.14 & 0.02 & 9.95 & 0.02 & 1.30 & F2.1 & 7.79$\pm$0.19 & 10.71$\pm$0.08 & D & 99.8 \\
01581063$+$3724058 & 999 & 12.23 & 0.02 & 11.81 & 0.02 & 0.82 & G9.0 & 7.87$\pm$0.13 & 13.37$\pm$0.12 & \nodata & 51.2 \\
01581109$+$3747537 & \nodata & 13.24 & 0.02 & 12.55 & 0.02 & 0.65 & K5.6 & 7.92$\pm$0.09 & 14.73$\pm$0.02 & \nodata & 96.5 \\
01581140$+$3739334 & 1000 & 10.56 & 0.02 & 10.33 & 0.02 & 1.20 & F4.2 & 7.85$\pm$0.21 & 11.26$\pm$0.10 & \nodata & 99.7 \\
01581226$+$3732383 & 1003 & 10.25 & 0.02 & 9.94 & 0.02 & 1.05 & F8.0 & 6.95$\pm$0.11 & 11.19$\pm$0.03 & \nodata & 98.4 \\
01581269$+$3734405 & \nodata & 10.18 & 0.02 & 9.96 & 0.02 & 1.28 & F2.6 & 7.74$\pm$0.20 & 10.77$\pm$0.09 & \nodata & 99.8 \\
01581293$+$3715202 & \nodata & 11.33 & 0.02 & 11.04 & 0.02 & 0.94 & G4.0 & 7.59$\pm$0.15 & 12.38$\pm$0.05 & \nodata & 67.8 \\
01581346$+$3742456 & \nodata & 14.42 & 0.03 & 13.53 & 0.03 & 0.53 & M0.5 & 8.23$\pm$0.20 & 16.02$\pm$0.05 & \nodata & 94.1 \\
01581427$+$3700453 & \nodata & 13.05 & 0.02 & 12.39 & 0.02 & 0.69 & K4.7 & 7.89$\pm$0.07 & 14.53$\pm$0.02 & \nodata & 67.7 \\
01581667$+$3855431 & \nodata & 11.35 & 0.02 & 11.05 & 0.02 & 1.01 & F9.4 & 7.90$\pm$0.27 & 12.31$\pm$0.08 & \nodata & 84.4 \\
01581685$+$3738160 & 1023 & 10.4 & 0.02 & 10.15 & 0.02 & 1.10 & F6.4 & 7.35$\pm$0.21 & 11.26$\pm$0.10 & D & 99.9 \\
01581840$+$3806540 & 1027 & 11.49 & 0.02 & 11.18 & 0.02 & 0.94 & G4.4 & 7.71$\pm$0.06 & 12.54$\pm$0.02 & \nodata & 77.7 \\
01581897$+$3832137 & \nodata & 10.3 & 0.02 & 10.07 & 0.02 & 1.14 & F5.4 & 7.41$\pm$0.21 & 11.10$\pm$0.10 & \nodata & 98.6 \\
01582147$+$3748568 & \nodata & 15.71 & 0.08 & 14.92 & 0.1 &  0.34 & \nodata & 8.31$\pm$0.12 & 17.56$\pm$0.05 & \nodata & 76.3 \\
01582190$+$3724073 & \nodata & 13.2 & 0.02 & 12.65 & 0.02 & 0.72 & \nodata & 8.32$\pm$0.08 & 14.59$\pm$0.07 & \nodata & 77.7 \\
01582469$+$3733380 & \nodata & 15.52 & 0.06 & 14.67 & 0.08 &  0.42  & \nodata & 8.50$\pm$0.09 & 17.20$\pm$0.04 & \nodata & 70.9 \\
01582651$+$3743421 & \nodata & 14.8 & 0.04 & 13.86 & 0.04 &  0.49  & M1.9 & 8.06$\pm$0.20 & 16.46$\pm$0.03 & \nodata & 95.4 \\
01582759$+$3735222 & 1083 & 10.97 & 0.02 & 10.65 & 0.02 & 1.21 & F4.2 & 8.20$\pm$0.42 & 11.62$\pm$0.24 & \nodata & 99.8 \\
01583120$+$3743202 & \nodata & 15.45 & 0.06 & 14.7 & 0.08 &  0.34  & M3.2 & 8.07$\pm$0.32 & 17.33$\pm$0.04 & \nodata & 62.8 \\
01583407$+$3758490 & \nodata & 15.19 & 0.05 & 14.4 & 0.06 &  0.42  & M2.5 & 8.25$\pm$0.25 & 17.01$\pm$0.04 & \nodata & 79.0 \\
01583743$+$3750269 & \nodata & 15.85 & 0.08 & 15.07 & 0.11 &  0.28  & M3.6 & 8.05$\pm$0.40 & 17.64$\pm$0.05 & \nodata & 80.5 \\
01583811$+$3732157 & 1123 & 10.64 & 0.02 & 10.39 & 0.02 & 1.29 & F2.5 & 8.20$\pm$0.19 & 11.21$\pm$0.08 & \nodata & 99.7 \\
01584149$+$3725492 & \nodata & 12.45 & 0.02 & 11.94 & 0.02 & 0.72 & K2.7 & 7.61$\pm$0.03 & 13.85$\pm$0.03 & \nodata & 66.1 \\
01584419$+$3729563 & \nodata & 13.61 & 0.03 & 12.86 & 0.02 & 0.64 & K5.9 & 8.18$\pm$0.22 & 15.05$\pm$0.05 & \nodata & 92.3 \\
01584696$+$3803299 & \nodata & 14.2 & 0.03 & 13.43 & 0.03 &  0.54 & M1.1 & 7.89$\pm$0.17 & 15.91$\pm$0.04 & \nodata & 54.2 \\
01584750$+$3734036 & \nodata & 15.82 & 0.08 & 14.92 & 0.1 &  0.33 & M3.3 & 8.26$\pm$0.36 & 17.60$\pm$0.05 & \nodata & 68.6 \\
01584788$+$3728278 & \nodata & 13.47 & 0.03 & 12.63 & 0.03 & 0.53 & M0.4 & 7.33$\pm$0.30 & 15.08$\pm$0.08 & \nodata & 87.9 \\
01584814$+$3749252 & \nodata & 14.86 & 0.04 & 14.02 & 0.05 &  0.45 & M2.3 & 8.00$\pm$0.09 & 16.63$\pm$0.03 & \nodata & 76.5 \\
01584873$+$3747010 & \nodata & 13.45 & 0.02 & 12.74 & 0.03 & 0.69 & K4.7 & 8.27$\pm$0.07 & 14.91$\pm$0.03 & \nodata & 96.2 \\
01584997$+$3759465 & \nodata & 12.71 & 0.02 & 12.1 & 0.02 & 0.68 & K4.9 & 7.57$\pm$0.03 & 14.24$\pm$0.02 & \nodata & 77.7 \\
01585042$+$3720520 & \nodata & 9.643 & 0.02 & 9.44 & 0.02 & 1.13 & F5.6 & 6.73$\pm$0.22 & 10.47$\pm$0.10 & \nodata & 97.7 \\
01585206$+$3733522 & \nodata & 12.67 & 0.02 & 12.27 & 0.02 & 0.77 & K0.6 & 8.14$\pm$0.04 & 13.95$\pm$0.03 & \nodata & 70.9 \\
01585393$+$3734424 & 1178 & 11.71 & 0.02 & 11.19 & 0.02 & 0.71 & K3.0 & 6.82$\pm$0.06 & 13.14$\pm$0.06 & DP & 55.1 \\
01585614$+$3749184 & \nodata & 14.94 & 0.04 & 13.99 & 0.04 & 0.50 & \nodata & 8.50$\pm$0.13 & 16.52$\pm$0.04 & \nodata & 76.8 \\
01585731$+$3739408 & 1196 & 12.15 & 0.02 & 11.61 & 0.02 & 0.73 & K2.5 & 7.30$\pm$0.08 & 13.50$\pm$0.07 & \nodata & 89.6 \\
01585985$+$3801189 & 1204 & 10.79 & 0.02 & 10.57 & 0.02 & 1.26 & \nodata & 8.30$\pm$0.21 & 11.41$\pm$0.09 & \nodata & 99.6 \\
01590030$+$3803467 & \nodata & 12.76 & 0.02 & 12.07 & 0.02 & 0.67 & K5.3 & 7.52$\pm$0.10 & 14.26$\pm$0.02 & \nodata & 88.5 \\
01590236$+$3747119 & \nodata & 13.27 & 0.02 & 12.53 & 0.02 & 0.65 & K5.7 & 7.89$\pm$0.11 & 14.72$\pm$0.02 & \nodata & 80.9 \\
01590427$+$3735475 & \nodata & 13.42 & 0.02 & 12.76 & 0.02 & 0.70 & \nodata & 8.30$\pm$0.07 & 14.89$\pm$0.04 & \nodata & 91.8 \\
01590640$+$3808537 & \nodata & 14.25 & 0.03 & 13.45 & 0.03 & 0.57 & \nodata & 8.43$\pm$0.14 & 15.75$\pm$0.05 & \nodata & 79.3 \\
01590742$+$3814427 & \nodata & 11.98 & 0.02 & 11.6 & 0.02 & 0.84 & G7.7 & 7.75$\pm$0.18 & 13.01$\pm$0.08 & \nodata & 79.2 \\
01591077$+$3800176 & \nodata & 13.52 & 0.02 & 12.82 & 0.03 & 0.68 & \nodata & 8.30$\pm$0.10 & 15.00$\pm$0.02 & \nodata & 91.7 \\
01591569$+$3802102 & \nodata & 13.61 & 0.02 & 12.85 & 0.02 & 0.65 & K5.7 & 8.21$\pm$0.12 & 15.04$\pm$0.02 & \nodata & 65.4 \\
01591802$+$3749494 & \nodata & 12.55 & 0.02 & 12.06 & 0.02 & 0.76 & K1.3 & 7.89$\pm$0.12 & 13.84$\pm$0.06 & \nodata & 51.3 \\
01592960$+$3816042 & \nodata & 10.48 & 0.02 & 10.25 & 0.02 & 1.15 & F5.0 & 7.63$\pm$0.22 & 11.24$\pm$0.10 & \nodata & 99.1 \\
01593379$+$3749086 & \nodata & 16.09 & 0.08 & 15.21 & 0.16 &  0.28 & M3.8 & 8.15$\pm$0.42 & 17.90$\pm$0.06 & \nodata & 62.4 \\
01593875$+$3737369 & \nodata & 13.51 & 0.03 & 12.64 & 0.03 & 0.55 & M0.0 & 7.47$\pm$0.16 & 15.07$\pm$0.05 & \nodata & 82.1 \\
01594725$+$3749538 & \nodata & 12 & 0.02 & 11.66 & 0.02 & 0.92 & G5.3 & 8.12$\pm$0.05 & 13.06$\pm$0.03 & \nodata & 56.8 \\
01594872$+$3808543 & \nodata & 13.78 & 0.02 & 12.89 & 0.03 & 0.52 & M0.7 & 7.52$\pm$0.19 & 15.38$\pm$0.05 & \nodata & 84.9 \\
01595186$+$3730317 & \nodata & 12.79 & 0.02 & 12.37 & 0.02 & 0.82 & \nodata & 8.43$\pm$0.26 & 13.91$\pm$0.19 & \nodata & 56.8 \\
01595269$+$3753502 & \nodata & 15.07 & 0.04 & 14.19 & 0.06 &  0.44 & M2.4 & 8.13$\pm$0.24 & 16.83$\pm$0.04 & \nodata & 53.0 \\
01595715$+$3742387 & \nodata & 12.87 & 0.03 & 12.21 & 0.02 & 0.70 & K4.4 & 7.76$\pm$0.07 & 14.36$\pm$0.04 & \nodata & 69.8 \\
01595909$+$3743239 & \nodata & 13.66 & 0.03 & 12.83 & 0.03 & 0.52 & M0.7 & 7.43$\pm$0.16 & 15.29$\pm$0.04 & \nodata & 70.1 \\
02001281$+$3753452 & \nodata & 14.17 & 0.03 & 13.39 & 0.04 & 0.57 & \nodata & 8.37$\pm$0.14 & 15.69$\pm$0.05 & \nodata & 58.8 \\
02001529$+$3800479 & \nodata & 15 & 0.04 & 14.15 & 0.06 &  0.50  & \nodata & 8.40$\pm$0.18 & 16.70$\pm$0.04 & \nodata & 80.9 \\
02001988$+$3741466 & \nodata & 14.36 & 0.04 & 13.43 & 0.04 & 0.53 & M0.6 & 8.16$\pm$0.08 & 15.98$\pm$0.03 & \nodata & 73.9 \\
02002194$+$3802409 & \nodata & 9.89 & 0.03 & 9.692 & 0.02 & 1.25 & F3.2 & 7.36$\pm$0.22 & 10.54$\pm$0.09 & \nodata & 98.6 \\
02005086$+$3750145 & \nodata & 13.38 & 0.03 & 12.52 & 0.03 & 0.55 & M0.0 & 7.34$\pm$0.12 & 14.94$\pm$0.04 & \nodata & 79.4 \\
02010562$+$3745515 & \nodata & 12.22 & 0.03 & 11.84 & 0.02 & 0.86 & G7.2 & 8.06$\pm$0.18 & 13.26$\pm$0.04 & \nodata & 63.0 \\
02011366$+$3757177 & \nodata & 10.78 & 0.03 & 10.46 & 0.02 & 1.10 & F6.5 & 7.66$\pm$0.21 & 11.58$\pm$0.10 & \nodata & 95.4 \\
02012287$+$3742278 & \nodata & 13.68 & 0.03 & 12.83 & 0.03 &  0.54  & M1.2 & 7.27$\pm$0.32 & 15.33$\pm$0.08 & \nodata & 65.4 \\
02033474$+$3825529 & \nodata & 10.12 & 0.03 & 9.88 & 0.02 & 1.19 & F4.3 & 7.37$\pm$0.22 & 10.82$\pm$0.10 & \nodata & 83.4 \\
02034628$+$3709039 & \nodata & 10.28 & 0.03 & 10.02 & 0.02 & 1.11& F6.4 & 7.23$\pm$0.22 & 11.14$\pm$0.10 & \nodata & 71.9 \\
02041884$+$3808429 & \nodata & 9.76 & 0.03 & 9.55 & 0.02 & 1.17 & F4.7 & 6.98$\pm$0.22 & 10.52$\pm$0.10 & \nodata & 68.4 \\
02042216$+$3721411 & \nodata & 10.85 & 0.02 & 10.6 & 0.02 & 1.14 & F5.5 & 7.93$\pm$0.21 & 11.64$\pm$0.10 & \nodata & 57.9 \\
WISE 015743.22$$+$$374417.6\tnote{b} & \nodata & \multicolumn{2}{c}{\nodata} & \multicolumn{2}{c}{\nodata} & \nodata & M3.7 & 6.84$\pm$2.52 & 16.51$\pm$0.88 & \nodata & 89.5\\
\end{longtable*}
\begin{tablenotes}
\item[a]{Based on RV measurements published by \citet{daniel1994} (``D'') or \citet{mermilliod1998}, or collected by C.~Pilachowski (``P'').}
\item[b]{ 2MASS photometry does not exist for WISE 015743.22$+$374417.6 due to confusion with a persistence artifact in the 2MASS image. Based on a fit to its optical SED, we classify the source as a probable (\Pmem~=~89.5\%) cluster member; the source does not factor in to our subsequent analysis, so we do not attempt to derive a separate non-$K$-band based mass estimate.}
\end{tablenotes}
\end{ThreePartTable}

\begin{ThreePartTable}
\begin{longtable*}{lcr@{$\pm$}lr@{$\pm$}lccr@{$\pm$}lccc}
\tablecaption{Other NGC 752 Members}
\tablehead{
  \colhead{2MASS ID} &
  \colhead{Platais ID} &
 \multicolumn{2}{c}{$J$} &
 \multicolumn{2}{c}{$K$} &
  \colhead{Mass} &
  \colhead{SpT} &
 \multicolumn{2}{c}{RV$_D$} &
  \colhead{RV$_P$} &
  \colhead{Binary?\tnote{a}} \\
  \colhead{} & 
  \colhead{} &
 \multicolumn{2}{c}{(mag)} & 
 \multicolumn{2}{c}{(mag)} & 
  \colhead{(\Msun)} & 
  \colhead{} & 
 \multicolumn{2}{c}{(km s$^{-1}$)} & 
  \colhead{(km s$^{-1}$)} &
  \colhead{}
 }
01510351$+$3746343  &  \nodata  &    7.35 & 0.02  &   6.70 & 0.03  & \nodata &\nodata &  \multicolumn{2}{c}{\nodata}  & \nodata &      \nodata  \\
01513012$+$3735380  &  \nodata  &    7.34 & 0.02  &   6.76 & 0.02  & \nodata &\nodata &  \multicolumn{2}{c}{\nodata}  & \nodata &      \nodata  \\
01543966$+$3811455  &  \nodata  &    7.13 & 0.01  &   6.52 & 0.02  & \nodata &\nodata &  \multicolumn{2}{c}{\nodata}  & \nodata &      \nodata  \\
01551261$+$3750145  &  \nodata  &    7.82 & 0.02  &   7.22 & 0.02  & \nodata &\nodata &  \multicolumn{2}{c}{\nodata}  & \nodata &      \nodata  \\
01551528$+$3750312  &  \nodata  &    7.80 & 0.02  &   7.19 & 0.02  & \nodata &\nodata &   4.5 & 0.5  & \nodata &      \nodata  &  \\
01552765$+$3759551  &  \nodata  &    7.62 & 0.02  &   7.03 & 0.02  & \nodata &\nodata &   4.9 & 0.3  & \nodata &      \nodata  & \\
01552769$+$3734046  &    305    &    9.29 & 0.01  &   9.06 & 0.02  & \nodata &  F5.6  &  \multicolumn{2}{c}{\nodata}  & 6.25$^{\pm0.09}_{\pm0.19}$  &     \nodata  \\
01552928$+$3750262  &    313    &    9.00 & 0.02  &   8.69 & 0.02  & \nodata &\nodata &   4.6 & 0.5  &               $-$11.0$^{\pm1.97}_{\pm36.7}$  &          DP  \\
01553936$+$3752525  &  \nodata  &    7.15 & 0.02  &   6.54 & 0.02  & \nodata &\nodata &  \multicolumn{2}{c}{\nodata} & \nodata &      \nodata  \\
01554239$+$3737546  &  \nodata  &    7.40 & 0.01  &   6.80 & 0.02  & \nodata &\nodata &  \multicolumn{2}{c}{\nodata} & \nodata &      \nodata  \\
01554740$+$3742265  &    372    &    8.97 & 0.02  &   8.71 & 0.02  & \nodata &\nodata &   7.1 & 1.5  &               7.95$^{\pm0.06}_{\pm0.16}$  &     \nodata   \\
01555552$+$3728333  &    397    &    8.84 & 0.02  &   8.57 & 0.02  & \nodata &\nodata &   7.4 & 2.0  &               8.17$^{\pm0.13}_{\pm0.54}$  &     \nodata   \\
01561029$+$3745000  &  \nodata  &    9.24 & 0.02  &   9.01 & 0.02  & 1.26 &   F3.2  &   5.4 & 0.2  & \nodata &            D   \\
01561550$+$3738416  &  \nodata  &    9.12 & 0.02  &   8.85 & 0.02  & \nodata &\nodata &  \multicolumn{2}{c}{\nodata} & \nodata &      \nodata \\
01561890$+$3758005  &  \nodata  &    7.17 & 0.01  &   6.60 & 0.02  & \nodata &\nodata &   4.5 & 0.1  & \nodata &           DM   \\
01562163$+$3736084  &  \nodata  &    7.56 & 0.02  &   6.91 & 0.02  & \nodata &\nodata &  \multicolumn{2}{c}{\nodata} & \nodata &      \nodata   \\
01563204$+$3734223  &  \nodata  &   11.69 & 0.02  &  11.35 & 0.02  & 0.95 &   G3.7  &   5.9 & 0.4  & \nodata &            D   \\
01564975$+$3801216  &    626    &    8.20 & 0.02  &   8.00 & 0.02  & \nodata &\nodata &   7.1 & 1.7  &               7.39$^{\pm0.24}_{\pm0.98}$  &     \nodata \\
01565043$+$3801581  &  \nodata  &    7.38 & 0.02  &   6.82 & 0.02  & \nodata &\nodata &   4.9 & 0.2  & \nodata &           DM   \\
01565395$+$3819272  &  \nodata  &   14.09 & 0.02  &  13.69 & 0.04  & \nodata &\nodata &  \multicolumn{2}{c}{\nodata} & \nodata &      \nodata   \\
01565576$+$3747594  &    654    &   10.38 & 0.02  &  10.15 & 0.02  & \nodata &\nodata &  \multicolumn{2}{c}{\nodata} &               6.11$^{\pm0.25}_{\pm0.37}$  &     \nodata  \\
01565633$+$3739514  &  \nodata  &    9.22 & 0.01  &   8.95 & 0.02  & \nodata &  F5.1  &  \multicolumn{2}{c}{\nodata} & \nodata &      \nodata   \\
01570279$+$3814035  &    684    &   11.41 & 0.02  &  11.12 & 0.02  & \nodata &\nodata &  \multicolumn{2}{c}{\nodata} &               2.81$^{\pm0.15}_{\pm7.12}$  &           P   \\
01570311$+$3808026  &  \nodata  &    7.14 & 0.02  &   6.53 & 0.02  & \nodata &\nodata &  \multicolumn{2}{c}{\nodata} & \nodata &      \nodata   \\
01571425$+$3746510  &    745    &    9.03 & 0.02  &   8.79 & 0.02  & \nodata &  F4.0  &   $-$7.6 & 18.8  &               22.33$^{\pm0.40}_{\pm12.8}$  &          DP  \\
01571709$+$3726089  &  \nodata  &    9.30 & 0.01  &   9.05 & 0.02  & 1.18 &   F4.6  & \multicolumn{2}{c}{\nodata} & \nodata &      \nodata   \\
01572399$+$3806104  &  \nodata  &   11.56 & 0.02  &  11.26 & 0.02  & \nodata &\nodata &   4.5 & 1.1  & \nodata &      \nodata  \\
01572825$+$3724026  &    814    &    9.43 & 0.02  &   9.20 & 0.02  & 1.21 &   F4.1  & \multicolumn{2}{c}{\nodata} &               6.55$^{\pm0.25}_{\pm12.3}$  &           P \\
01573621$+$3745101  &    849    &    9.05 & 0.02  &   8.77 & 0.02  & \nodata &\nodata &   4.5 & 0.9  &               9.16$^{\pm2.13}_{\pm8.39}$  &          DP    \\
01573696$+$3702110  &  \nodata  &    9.49 & 0.02  &   8.73 & 0.02  & \nodata &\nodata &  \multicolumn{2}{c}{\nodata} & \nodata &      \nodata  \\
01573735$+$3729276  &  \nodata  &    9.64 & 0.01  &   9.62 & 0.02  & \nodata &\nodata &  \multicolumn{2}{c}{\nodata} & \nodata &      \nodata  \\
01573760$+$3739380  &  \nodata  &    7.07 & 0.02  &   6.41 & 0.02  & \nodata &\nodata &   5.9 & 0.8  & \nodata &           DM  \\
01573767$+$3749008  &    857    &    9.07 & 0.02  &   8.77 & 0.02  & \nodata &\nodata &   5.7 & 9.3  &               8.88$^{\pm0.24}_{\pm7.62}$  &          DP  \\
01573895$+$3746123  &  \nodata  &    7.26 & 0.02  &   6.66 & 0.03  & \nodata &\nodata &   6.1 & 0.5  & \nodata &            M  \\
01574472$+$3759184  &  \nodata  &    9.17 & 0.02  &   8.90 & 0.02  & \nodata &\nodata &  \multicolumn{2}{c}{\nodata} & \nodata &      \nodata \\
01575198$+$3727460  &  \nodata  &   11.50 & 0.01  &  11.22 & 0.02  & \nodata &\nodata &   5.0 & 1.0  & \nodata &      \nodata  \\
01575822$+$3726064  &    952    &   11.55 & 0.01  &  11.26 & 0.02  & \nodata &\nodata &  \multicolumn{2}{c}{\nodata} &               5.44$^{\pm0.10}_{\pm0.10}$  &     \nodata   \\
01575934$+$3754540  &  \nodata  &    9.08 & 0.02  &   8.80 & 0.02  & \nodata &\nodata &   4.8 & 0.3  & \nodata &            D  \\
01580924$+$3728355  &    993    &   12.27 & 0.02  &  11.85 & 0.02  & 0.82 &  G8.8  &   5.3 & 0.7  &               5.77$^{\pm0.12}_{\pm0.17}$  &     \nodata \\
01581531$+$3733196  &   1017    &   12.01 & 0.02  &  11.60 & 0.02  & 0.77 &  K0.7  &   5.8 & 0.3  &               5.78$^{\pm0.08}_{\pm0.13}$  &     \nodata  \\
01582981$+$3751374  &  \nodata  &    7.62 & 0.02  &   7.04 & 0.02  & \nodata &\nodata &   5.2 & 0.4  & \nodata &      \nodata   \\
01583440$+$3740152  &   1107    &   12.29 & 0.02  &  11.88 & 0.02  & 0.81  &   G9.1  &   5.5 & 0.6  &               5.45$^{\pm0.14}_{\pm0.13}$  &     \nodata \\
01583687$+$3745106  &   1117    &    8.75 & 0.02  &   8.50 & 0.02  & \nodata &\nodata &   5.0 & 0.5  &               4.69$^{\pm34.0}_{\pm4.11}$  &          DP   \\
01584005$+$3738051  &   1129    &   10.91 & 0.02  &  10.59 & 0.02  & \nodata &\nodata &   3.4 & 2.9  &               6.05$^{\pm0.16}_{\pm0.21}$  &           D   \\
01584793$+$3826078  &  \nodata  &    9.16 & 0.02  &   8.91 & 0.02  & \nodata &  F4.8  &  \multicolumn{2}{c}{\nodata} & \nodata &      \nodata   \\
01585290$+$3748572  &  \nodata  &    7.27 & 0.01  &   6.64 & 0.02  & \nodata &\nodata &   5.6 & 0.4  & \nodata &      \nodata    \\
01591479$+$3800554  &  \nodata  &    7.20 & 0.01  &   6.64 & 0.02  & \nodata &\nodata &  \multicolumn{2}{c}{\nodata} & \nodata &      \nodata   \\
01591990$+$3723230  &  \nodata  &   11.50 & 0.02  &  11.05 & 0.02  & \nodata &\nodata &  \multicolumn{2}{c}{\nodata} & \nodata &      \nodata \\
02010594$+$3742235  &  \nodata  &    8.98 & 0.02  &   8.73 & 0.02  & \nodata &\nodata &  \multicolumn{2}{c}{\nodata} & \nodata &      \nodata  \\
02011082$+$3710164  &  \nodata  &   11.68 & 0.02  &  11.30 & 0.02  & 0.85 & G7.5  & \multicolumn{2}{c}{\nodata} & \nodata &      \nodata   \\
\end{longtable*}
\begin{tablenotes}
\item[a]{Based on RV measurements published by \citet{daniel1994} (``D'') or \citet{mermilliod1998}, or collected by C.~Pilachowski (``P'').}
\end{tablenotes}
\end{ThreePartTable}



\end{document}